\pdfoutput=1
\documentclass[
pdflatex,sn-mathphys,
Numbered]{sn-jnl}

\usepackage{graphicx}%
\usepackage{multirow}%
\usepackage{amsmath,amssymb,amsfonts}%
\usepackage{amsthm}%
\usepackage{mathrsfs}%
\usepackage[title]{appendix}%
\usepackage{xcolor}%
\usepackage{textcomp}%
\usepackage{manyfoot}%
\usepackage{booktabs}%
\usepackage{listings}%
\usepackage{mySymbols}
\usepackage{myPreamble}
\usepackage{myColors}
\usepackage{myPgfplots}

\newtheorem{remark}{Remark}%

\newcommand{\myeqref}[1]{\added{Eq.}~(\ref{#1})}
\newcommand{\myeqrefs}[2]{\added{Eqs.}~(\ref{#1})-(\ref{#2})}
\newcommand{\myeqrefsa}[2]{\added{Eqs.}~(\ref{#1}) and (\ref{#2})}

\begin{document}

\title[A consistent diffuse-interface model for two-phase flow problems with rapid evaporation]{A consistent diffuse-interface model for two-phase flow problems with rapid evaporation}

\author*[1]{\fnm{Magdalena}  \sur{Schreter-Fleischhacker}}\email{magdalena.schreter@tum.de}
\author[2,3]{\fnm{Peter} \sur{Munch}}\email{peter.munch@it.uu.se}%
\author[1]{\fnm{Nils} \sur{Much}}\email{nils.much@tum.de}%
\author[2,4]{\fnm{Martin} \sur{Kronbichler}}\email{martin.kronbichler@rub.de}%
\author[1]{\fnm{Wolfgang} A. \sur{Wall}}\email{wolfgang.a.wall@tum.de}%
\author[1]{\fnm{Christoph} \sur{Meier}}\email{christoph.anton.meier@tum.de}%

\affil*[1]{\orgdiv{Institute for Computational Mechanics}, \orgname{Technical University of Munich}, \orgaddress{\street{Boltzmannstrasse 15}, \city{Garching}, \postcode{85748}, \country{Germany}}}
\affil[2]{\orgdiv{Institute for High-Performance Scientific Computing}, \orgname{University of Augsburg}, \orgaddress{\street{Universitätsstraße 12a}, \city{Augsburg}, \postcode{86159}, \country{Germany}}}
\affil[3]{\orgdiv{Department of Information Technology}, \orgname{Uppsala University}, \orgaddress{\street{Box 337}, \city{Uppsala}, \postcode{75105}, \country{Sweden}}}
\affil[4]{\orgdiv{Faculty of Mathematics}, \orgname{Ruhr University Bochum}, \orgaddress{\street{Universitätsstraße 150}, \city{Bochum}, \postcode{44780}, \country{Germany}}}

\abstract{
	We present accurate and mathematically consistent formulations of a diffuse-interface model for two-phase flow problems involving rapid evaporation. 
	The model addresses challenges including discontinuities in the density field by several orders of magnitude, leading to high velocity and pressure jumps across the liquid-vapor interface,  along with dynamically changing interface topologies. 	To this end,  we integrate an incompressible Navier--Stokes solver combined with a conservative level-set formulation and a regularized, i.e.,  diffuse, representation of discontinuities into a matrix-free adaptive finite element framework. The achievements are three-fold: First, \replaced{we propose mathematically consistent definitions for the level-set transport velocity in the diffuse interface region by extrapolating the velocity from the liquid or gas phase. They exhibit superior prediction accuracy for the evaporated
	mass and the resulting interface dynamics compared to a local velocity evaluation, especially for strongly curved interfaces.}{First, this work proposes mathematically consistent definitions for the level-set transport velocity in the diffuse interface region by extrapolating the velocity from the liquid or gas phase, which exhibit superior prediction accuracy for the evaporated
	mass and the resulting interface dynamics compared to a local velocity evaluation, especially for highly curved interfaces.}
	Second, we show that accurate prediction of the evaporation-induced pressure jump requires a consistent, namely a reciprocal, density interpolation across the interface, which satisfies local mass conservation. 
	Third, the combination of diffuse interface models for evaporation with standard Stokes-type constitutive relations for viscous flows leads to significant pressure artifacts in the diffuse interface region. To mitigate these, we propose \replaced{to introduce a correction term}{a modification} for such constitutive model types.
	Through selected analytical and numerical examples, the aforementioned properties are validated. The presented model promises new insights in simulation-based prediction of melt-vapor interactions in thermal multiphase flows such as in laser-based powder bed fusion of metals.	
	}


\maketitle

\section{Introduction}\label{sec1}

\subsection{Background and challenges}
Phase change phenomena in immiscible two-phase flows, specifically evaporation, play a crucial role in various industrial and environmental processes such as spray combustion, boiling of water in power plants, heat exchangers/cooling systems and, notably, laser-based powder-bed fusion additive manufacturing of metals (\pbfam{}). The focus of our investigation is primarily motivated by the latter area, where evaporation plays a central role, as discussed in the following.
 
Metal additive manufacturing by \pbfam{} is a promising technology offering
unique capabilities for the on-demand production of high-performance metal parts with nearly unlimited freedom of
design. 
Considering the mesoscale\added{, on the order of micrometers}, metal powder particles begin to melt in the vicinity of the laser, i.e., a solid-liquid phase transition is induced \replaced{that ideally results in the formation of a continuous melt pool.}{, leading to the formation of a continuous melt pool in the ideal case.} However, the rapid transition from molten metal to metal vapor and their interaction under typical process conditions can cause local process instabilities \cite{ly2017metal,kiss2019laser,cunningham2019keyhole,bitharas2022interplay}, resulting in quality-degrading defects. \deleted{such as pores, spatter
, denudation
, and lack of fusion.}
 The main thermo-physical effects of evaporation affecting the melt pool behavior are summarized as follows: 
Evaporation from a liquid surface leads to  (1) the release of vapor into the surrounding environment \added{and thus a movement of the melt pool surface}, resulting in  (2) a potentially strong vapor flow \added{and thus a velocity jump at the liquid-vapor interface}. This induces (3) a force or pressure \added{jump} in the opposite direction of the vapor flow, which is known in the literature as ``recoil pressure". Furthermore, (4) evaporative cooling due to absorption of the latent heat of evaporation as well as (5) the convective heat transfer due to the vapor flow may influence the thermal field.
	\added{Existing numerical melt pool models typically consider evaporation only through simplified models by taking into account
solely the evaporation-induced pressure jump via a phenomenological model \replaced{(i.e., 3)}{(3)} and evaporative cooling \replaced{(i.e., 4)}{(4)}, neglecting the other effects (i.e., 1, 2 and 5), e.g., \cite{khairallah2016laser,chen2020spattering,meier2020meltpool,fuchs2022versatile}.}
 
\deleted{ \replaced{E}{While e}xperimental investigations 
	have hypothesized that the majority of undesired defects are associated with e\-va\-po\-ra\-tion-induced material redistribution dynamics \deleted{(related to (2) and (3))}, e.g., powder particle entrainment or spattering\added{. However,}
 	existing numerical thermal-multiphase flow models typically consider evaporation only through simplified models by taking into account
 	solely the evaporation-induced pressure jump \replaced{(i.e., 3)}{(3)} and evaporative cooling \replaced{(i.e., 4)}{(4)}, neglecting the other effects.}
 
\deleted{
 The additional consideration of the exchange in mass\added{ and thus phase-change dependent movement of the liquid-vapor interface} (\added{i.e., }1), \added{the velocity jump (normal to the interface) and the} resultant vapor flow (\added{i.e., }2), and the convective heat transfer induced by the vapor velocity (\added{i.e., }5) as well as their cumulative effects on the melt pool dynamics may be important modeling aspects, which have been attempted to be solved only a few times in the literature 
 .
 }
  \deleted{Our long-term goal is to incorporate all the mentioned effects \replaced{(i.e., 1-5)}{(1)-(5)} into a high-fidelity model of melt pool thermo-hydrodynamics. On the way to this goal, however, there are several important and challenging modeling aspects to be addressed.}

\replaced{
	Our long-term goal is to incorporate all of these effects into a high-fidelity model of melt pool thermo-hydrodynamics. To this end, we present a key submodel for modeling of two-phase flow with evaporative phase change across curved, dynamically changing interfaces in this contribution. Specifically, model components are developed to accurately predict evaporated mass (i.e., 1), velocity jump (i.e, 2), and recoil pressure (i.e., 3), and the resulting flow dynamics. 
}
{
 This contribution sheds light on a specific aspect of \added{two-phase flow with phase change }modeling by targeting the mentioned effects \replaced{(i.e., 1-3)}{(1)-(3)}\added{, i.e., the mass flux, the velocity jump and the recoil pressure}. Our primary objective is the development of a robust, efficient, and accurate model for two-phase flow with rapid evaporative phase change across curved dynamically changing interfaces. 
 In this endeavor, particular emphasis is put on the accurate prediction of the resulting flow dynamics, the motion of the liquid surface and the associated evaporated mass. 
}
For this purpose, to investigate the latter in an isolated manner, isothermal conditions are assumed by prescribing the evaporative mass flux as an analytical field. 
The extension to anisothermal conditions \added{and thus to incorporate effects (4) and (5)} is part of our future work.

\subsection{Related work about computational modeling of two-phase flows with moving interfaces}
The behavior of multiphase flows with phase change is inherently complex because  mass, momentum and energy are exchanged across an a-priori unknown moving, deformable interface transported with the flow. Typically, the thickness of this interface region is orders of magnitude smaller than the scale of the flow characteristics.  A numerical modeling framework for two-phase flow with phase change requires two essential components: (i) a method for representing and tracking the motion of the interface, and (ii) a method for modeling discontinous changes of parameters, variables and coupling (jump) conditions  between primary variables of the phases.
A non-exhaustive overview of existing methods for modeling multi-phase flows without/with phase change is provided below, while the interested reader is referred to reviews e.g. provided in~\cite{scardovelli1999direct,tryggvason2011direct}.

\emph{Moving grid methods} \added{\cite{hirt1974arbitrary,tang2005moving}} explicitly resolve the evolving interface through alignment with an element or cell boundary and appropriate coupling conditions at the interface, yielding high accuracy of the obtained solution. However, once the interface undergoes a complex deformation, frequent remeshing is required by these schemes, which is a computationally demanding task, especially in 3D. Additionally, the solution fields need to be remapped from the old mesh to the new mesh, which may additionally introduce undesired diffusive effects \cite{anderson1998diffuse}. Those frameworks are frequently formulated in an Arbitrary-Lagrangian-Eulerian (ALE) setting. 
To the authors' best knowledge, only a few ALE-based frameworks for two-phase flow with phase change are available such as \cite{anjos20143d,jafari2016,gros2018,zhang2019locally}, all limited to scenarios of low density ratios and restricted to non-topology changing geometries. 


In contrast,  in \emph{fixed grid methods}~\cite{tryggvason2011direct} the interface intersects the grid. 
	\emph{Interface-tracking} schemes can be employed to follow the motion of the interface by explicitly describing the surface using a Lagrangian mesh or marker points~\cite{unverdi1992front}. Alternatively, \emph{interface-capturing} schemes describe the interface implicitly via an auxiliary function. Commonly employed interface-capturing schemes are the volume of fluid (VOF) method~\cite{Hirt1981VolumeOF}, the level-set method~\cite{Osher1988,sussman1994level}, and the phase field method, adapted in~\cite{lowengrub1998quasi} for fluid mechanics. The major advantage of the VOF method over the level-set method is its inherent mass-conserving property, while in the level-set method mass conservation can be reobtained via reinitialization of the level-set function \cite{olsson2007conservative}.
The reconstruction of the interface geometry, e.g. curvature for calculating surface tension, is not accurately possible from the VOF function~\cite{popinet2009accurate} and needs special treatment. The mathematical framework of the level-set method and the phase-field method is similar, but the level-set method is a strictly mathematical approach while the phase-field method can be consistently derived from thermodynamics~\cite{jacqmin1999}.

For fixed grid methods, additional numerical effort is required to cope with coupling conditions at the interface.
\emph{Sharp interface methods}~\added{\cite{chessa2002extended,chessa2003enriched}} maintain the discontinuity of the solution fields at the interface. Thus, they enable a highly accurate representation of the original multi-phase problem. 
For sharp modeling of discontinuities at the interface and interface conditions, \replaced{these methods rely on extended discrete solution spaces. Examples include}{they are based on extended discrete solution spaces, e.g.,} 
the extended finite element method (XFEM)~\cite{rasthofer2011extended,sauerland2013stable,schottrasthofer2015face}, the cut finite element method (CutFEM)~\cite{hansbo2014cut,massing2018stabilized,claus2019cutfem,frachon2019cut,frachon2023cut} or the extended discontinuous Galerkin method~\cite{henneaux2023} together with level-set schemes, and other methods such as the ghost-fluid method~\cite{fechter2015discontinuous,fedkiw1999non,fechter2018approximate} or the immersed interface method~\cite{lee2003immersed}. Typically, the accuracy gains of sharp interface methods require complex modifications of the numerical schemes, such as stabilization\replaced{ to account for the small cut-cell problem~\cite{burman2015cutfem}. This stabilization is a necessary ingredient to ensure robustness of the numerical scheme in terms of a priori error estimates and applicability to complex surface-coupled problems~\cite{massing2015nitsche,schottrasthofer2015face}}{~\cite{schottrasthofer2015face}}. \deleted{ which, in turn, can restrict their practical applicability to challenging real-world problems.}

Alternatively, for a more straightforward and robust implementation by regularizing discontinuities, \emph{diffuse interface methods}~\added{\cite{brackbill1992,anderson1998diffuse}}, have been introduced. 
There, a smooth transition of the properties between the fluids may be assumed over a finite but small thickness of the interface region. Overall, diffuse interface methods yield a less accurate solution compared to sharp interface methods. Nevertheless, they are mathematically consistent such that the solution converges to the one of a sharp model with decreasing interface thickness. 
\replaced{Their most important features include the inherent ability to handle topology changes, and the use of a standard computational fluid dynamics solver that supports variable density and viscosity parameters, as detailed in the review article by Gibou et al.~\cite{gibou2018review}.}
{Their most important features are that they are inherently capable of handling topology changes, and that a standard computational fluid dynamics solver supporting variable density and viscosity parameters can be directly used for the simulation of real-world evaporation problems,  
as elaborated in the review article by Gibou et al.~\cite{gibou2018review}.}

In this work, we have adopted a level-set based diffuse fixed grid method to model the liquid-vapor interaction. This choice was primarily motivated by the imperative requirement for robustness, particularly in view of the intended application to mesoscale modeling of \pbfam{}\added{. Mesoscale models typically resolve individual powder particles and melt pool thermo-hydrodynamics in order to study defect generation mechanisms that lead to porosity.~\cite{meier2017thermophysical}.}
 The melt pool undergoes significant deformation and also topological changes as soon as e.g. spattering of melt drops or pore formation occurs. This is particularly challenging considering the significant differences in material properties between molten metal and metal vapor, as well as the rapid evaporation rates leading to substantial velocity jumps at the liquid-vapor interface.

\subsection{Related work about computational models for two-phase flows with phase change}
 In the following, some of the most important previous research endeavors related to the advancement of fixed grid methods for two-phase flows with phase change are summarized.

In \cite{juric1998computations} a finite difference scheme combined with front-tracking for studying two-phase boiling flows in 2D was presented. Similarly, in \cite{welch2000volume} simulations of film boiling were performed based on the VOF method combined with a MAC staggered scheme, where general applicability of their method to 3D was demonstrated.
In \cite{nguyen2001boundary,gibou2007level} a level-set framework was combined with the ghost-fluid method to employ sharp jump conditions by extracting an extension velocity from the liquid/vapor phases to advect the level-set transport. The authors applied it also for studying film boiling in the presence of immersed solid bodies \cite{son2007level}. A similar framework was presented in \cite{tanguy2007level} and employed to investigate evaporation of droplets based on a 2D studies. 

Only a few works deal with diffuse interface methods that consider a smeared representation of the evaporation-induced volume expansion and consequently the evaporation-induced velocity as well as pressure jumps across a finite interface region: 
In \cite{hardt2008evaporation}, a continuous field for the evaporation source term in the continuity equation was derived by solving an inhomogeneous Helmholtz equation and combined with a VOF framework, which seemed to work well for flat interfaces. 
\added{A numerical framework }for modeling high density ratios and evaporation of curved surfaces \replaced{was proposed}{,} in \cite{lee2017direct} \deleted{a hybrid cframework was proposed}\replaced{. This hybrid framework}{ that} involves modeling aspects from both --- sharp and diffuse models\added{ --- by employing a smooth treatment of the mass flux and a sharp treatment for jumps in the velocity, pressure and temperature gradient.} It is based on the level-set method using discretization by the marker-and-cell staggered grid method and was successfully applied to study droplet evaporation with a density ratio of up to 1000. It is one of the few works (together with \cite{tanguy2014benchmarks}), where the issues associated with the computation of the level-set transport velocity in presence of diffuse velocity jumps at the interface and associated misleading mass predictions are mentioned. In \cite{scapin2020volume}, a VOF-based finite difference framework involving a
smeared evaporation-induced velocity jump was presented. The authors proposed to transport the VOF function with a combination of a divergence-free extension of the liquid velocity computed from the solution of a Poisson equation and an irrotational term due to phase change. Their framework was verified by studying the droplet evaporation at relatively low density ratios (less than 100). 

It is evident that modeling evaporative jump conditions in a diffuse manner is very attractive, mainly because of its seamless integration into various numerical frameworks, including finite element methods. This approach comes at the price of requiring a sufficiently fine numerical mesh to resolve the interface region. However, it provides a robust solution, without the need for additional stabilization techniques. 
\added{When combined with an interface-tracking or interface-capturing scheme, the interface transport velocity is a crucial model component to accurately predict the liquid surface motion including the evaporated mass. However, its computation poses a significant challenge due to the smeared velocity jump at the liquid-vapor interface.}
\deleted{Nevertheless, when combined with an interface-tracking or interface-capturing scheme, the computation of the transport velocity, which governs the liquid surface motion including the evaporated mass, poses a significant challenge due to the smeared velocity jump at the liquid-vapor interface. }
As shown above, only a few related frameworks have been presented in the literature\replaced{. Additionally,}{, while} a comprehensive discussion of the difficulties associated with the precise modeling of diffuse evaporative jump conditions for curved surfaces subject to rapid evaporation is lacking. In particular, to the best of the authors' knowledge, none of the works discuss the undesirable effect of \added{numerical} pressure artifacts in the interface region\replaced{. This issue typically arises}{
as typically occurring} when diffuse interface models for evaporation are combined with Stokes-type constitutive laws for viscous flow.

\subsection{Our contributions}
We present accurate and mathematically consistent formulations of a level-set-based diffuse-interface model for two-phase viscous flow problems involving rapid evaporation. For incorporating evaporative phase change into a diffuse framework, modifications in terms of (i) the level-set transport velocity, (ii) the density interpolation function to ensure mass conservation, and (iii) \added{adding an evaporation-induced correction term in the Stokes-type }constitutive relation are needed, which are detailed in the following objectives:
\begin{itemize}
	\item We elucidate the difficulties associated with modeling of diffuse jump conditions for rapid evaporation of curved surfaces based on analytical benchmark examples. Since most existing verification examples consider problem setups, where the velocity in one phase is zero, we propose a new verification example, where both phases are subject to a velocity field. 
	\item We propose mathematically consistent formulations for the level-set transport velocity to accurately predict the evaporated mass and the resulting
	interface topology.	 For this purpose, two fundamental approaches are considered: one based on a local velocity evaluation, and the other based on a velocity extrapolation from the
	liquid or gas phase in the diffuse interface region. For a given interface thickness, it is	demonstrated that the latter approach typically fufills the requirement of a divergence-free
	level set transport velocity with higher accuracy, in particular for highly curved interfaces. 
	\item We show that the evaporation-induced pressure jump can only be
	accurately predicted if a consistent, i.e., a reciprocal, density interpolation across the interface is chosen that satisfies local mass conservation.	
	\item For mitigating pressure artifacts in the interface region induced by non-physical deformations resulting from the regularized treatment of evaporative dilation in viscous flows, we introduce a \replaced{correction term in the}{modified} Stokes-type constitutive relation.
\end{itemize} 
 All these aspects are important in view of the further development of this model for studying melt pool thermo-hydrodynamics of \pbfam{} in the future. 
 
 Furthermore, in contrast to the existing literature, we exploit \replaced{recent developments}{novel aspects} related to high-performance computing by considering a matrix-free operator evaluation and an adaptive refinement of the finite element mesh in the interface region by building our implementation on the finite element library \texttt{deal.II}~\cite{arndt2022deal}. 
 However, since the focus of this article is on the development of modeling techniques for two-phase flow with phase change, we will omit the performance analysis in this article. The interested reader may be referred to corresponding articles, where this topic is discussed in detail \cite{kronbichler2012generic,Kronbichler18multiphase,proell2023highly,munch2023}. 

The remainder of this article is structured as follows: The governing equations, a comprehensive theoretical discussion related to regularized modeling of the evaporation-induced velocity discontinuity and the derivation of mathematically consistent level-set transport velocity approaches as well as \replaced{the correction term in the }{a modified} Stokes-type constitutive relation are presented in Section~\ref{sec2}. In Section~\ref{sec3}, the proposed methods are employed to several numerical benchmark examples and verified by comparison to analytical solutions. Conclusions are drawn in Section~\ref{sec4}. 

\section{Methods}\label{sec2}

\subsection{Preliminaries}
In the following, \replaced{a consistent}{a novel} model for two-phase flow with phase change, which we interchangeably refer to as ``evaporation" throughout this publication, is presented. The focus lies on accurately predicting the evaporation-induced velocity and pressure change across the interface as well as the evaporation-induced dynamics of the liquid surface resulting from the phase transition from liquid to vapor/gas phase. To this end, we propose the mathematical model based on the following assumptions:

\begin{itemize}
	\item The flow is incompressible and viscous (Newtonian) at a moderate Reynolds number.
	\item  Spatially resolved vapor phase as well as liquid–vapor phase transition to explicitly resolve evaporation-induced recoil pressure and gas/vapor flows and thereby induced material redistribution dynamics.
	\item Isothermal conditions to investigate the evaporation-induced discontinuities arising in the mass/momentum equation of the two-phase flow framework in an isolated manner. Thus, for the present work, the evaporative mass flux, representing the evaporated mass rate per unit area, is prescribed as an analytical field. The extension of the framework to anisothermal conditions will be considered in our future work. 
	\item Diffuse interface capturing scheme with a finite but small interface thickness. 
\end{itemize}

\subsection{Governing equations of isothermal two-phase flow with evaporative phase change}
\label{sec:governing_equations}
\begin{figure}[!t]
	\centering
	\includegraphics{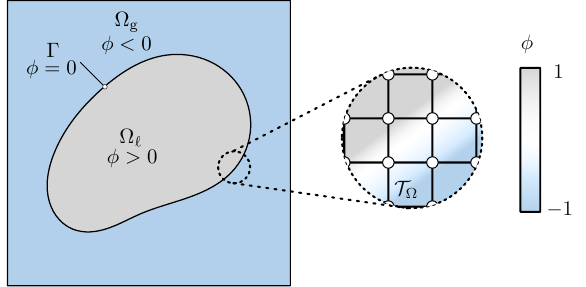}
	\caption{Physical domain of interest for the two-phase flow with phase change problem. The domain is decomposed into a liquid and a gaseous phase, represented by $\liqP{\Omega}$ and $\gasP{\Omega}$, respectively, separated by an interface $\Gamma$. The spatial discretization of the domain is performed by a finite element mesh $\mathcal{T}_\Omega$. Based on the level-set function $\phi$, the two phases are implicitly distinguished.}
	\label{fig:multi_field_problem}
\end{figure}

We assume that the Eulerian domain of interest $\Omega=\OmegaG\cup\OmegaL\in\mathbb{R}^{n}$ with $n\in\{1,2,3\}$ is occupied by a liquid phase $\OmegaL$ and a gaseous (vapor) phase $\OmegaG$, both modeled as incompressible and immiscible fluids, illustrated in Fig.~\ref{fig:multi_field_problem}. 
Irreversible phase transition between liquid and gaseous (vapor) phase, i.e., evaporation, along the liquid-gaseous interface $\Gamma\in\mathbb{R}^{n-1}$ may occur. 
By employing a level-set based diffuse interface capturing scheme for the position of the interface between the gaseous and the liquid phase, a single set of equations for the entire multi-phase domain can be formulated. 

\subsubsection{Flow field}
The velocity field $\boldsymbol{u}(\Bx,t)$ and the pressure field $p(\Bx,t)$ for point $\Bx\in\Omega$ and at time~$t\,\in[0,\tEnd]$ are governed by the incompressible, isothermal Navier--Stokes equations formulated in an Eulerian setting, consisting of the continuity equation and the momentum balance equation:
\begin{subequations}
	\begin{align}
		\label{eq:continuity}	
		\nabla\cdot\boldsymbol{u} &= \evaporDilationRate
		\quad&&\text{ in }\Omega\times[0,\tEnd]\,,\\
		\label{eq:momentum_balance}
		\rhoEff\left(\fracPartial{\boldsymbol{u}}{t}
		+(\boldsymbol{u}\cdot\nabla)\,\boldsymbol{u}
		\right) &= -\nabla p + \diver\viscousStress+\rhoEff\,\boldsymbol{g}+\surfaceTensionForce&&\text{ in }\Omega\times[0,\tEnd]\,.
		\end{align}
	\end{subequations}
	Within each phase, thermo-physical properties are assumed to be constant but vary smoothly across the interface region. They are designated as \emph{effective} properties by the subscript $\eff{(\bullet)}$ to refer to the two-phase (liquid/gas) mixture.
	These effective properties, the density $\rhoEff$ and the dynamic viscosity $\muEff$, are specified in Section~\ref{sec:matProps}. The dynamic viscosity influences the viscous stress tensor $\viscousStress$, as discussed in Section~\ref{sec:const_law}, and $\eff{\rho}\,\boldsymbol{g}$ denotes gravitational forces. Variables indicated by a tilde $\tilde{(\bullet)}$ represent diffuse interface \replaced{contributions}{fluxes} and consist of the evaporative dilation rate $\evaporDilationRate$ and surface tension $\surfaceTensionForce$ which are specified in Section~\ref{sec:interface_fluxes}. 
	\myeqrefs{eq:continuity}{eq:momentum_balance} are supplemented by a suitable initial condition
	\begin{equation}
	\Bu=\Bu^{(0)}\quad\text{in }\Omega\times\{t=0\}
	\end{equation}
	where the superscript $(\bullet)^{(0)}$  denotes an initial field function.
	 Dirichlet and Neumann boundary conditions are imposed according to
	\begin{align}
		\Bu &= \bar{\Bu}&&\text{on }\partial\Omega_{\text{D},u}\subset\partial\Omega\times[0,\tEnd], \\
	 \boldsymbol{\sigma}\cdot\normalDomain&= \bar{\Bt}\text{ if }\Bu\cdot\hat{\Bn} > 0 \text{ (outflow)}&&\text{on }\partial\Omega_{\text{N},u}\subset\partial\Omega\times[0,\tEnd]
	\end{align}
	with the Cauchy stress tensor $\boldsymbol{\sigma}=\viscousStress - p\,\mathcal{I}$, where $\mathcal{I}$ represents the second-order identity tensor, and the outward-pointing unit normal vector  $\normalDomain$ to the domain boundary $\partial\Omega=\partial\Omega_{\text{D},u}\cup\partial\Omega_{\text{N},u}$ with $\partial\Omega_{\text{D},u}\cap\partial\Omega_{\text{N},u}=\emptyset$.
	
	\subsubsection{Level-set field}
The temporal evolution of the interface $\GammaLevelSet$, represented by the zero-isosurface of a level-set function chosen as a regularized function $-1\leq\phi(\boldsymbol{x},t)\leq1$ according to \cite{olsson2007conservative}, is obtained by solving the advection equation
			\begin{align}
		\label{eq:transport}
		\fracPartial{\phi}{t}+\uGamma\nabla\phi&=0  \quad\text{ in }\Omega\times[0,\tEnd]\,.
	\end{align}
	We denote $\phi>0$ as inside the liquid phase and $\phi<0$ as inside the gaseous phase. 
In \myeqref{eq:transport}, $\uGamma$ represents the level-set transport velocity.
\added{
	At this point it should be noted that the formulation of the level-set transport velocity $\uGamma$ is a key modeling component in presence of rapid evaporation and is part of a detailed discussion in Section~\ref{sec:level_set_transport}. Without phase change, it is typically assumed that the latter corresponds to the local fluid velocity $\uGamma(\Bx,t):=\Bu(\Bx,t)$. }
 \replaced{For the initial level set we assume the regularized characteristic function
\begin{equation}
	\label{eq:initial_phi}
	\phi(\Bx) = 
	\tanh\left(\frac{d(\Bx)}{2\epsilon}\right)
\end{equation}
}{\myeqref{eq:transport} is subject to the initial condition 
$	\phi^{(0)} = 
	\tanh\left(\frac{d(\Bx)}{2\epsilon}\right)\quad\text{in }\Omega\times\{t=0\}\,$
}
depending on the interface thickness parameter $\epsilon$ and a signed distance function $d(\Bx)$. For a given level-set function, \myeqref{eq:initial_phi} can be inverted to obtain an expression for the signed distance function 
\begin{equation}
	\label{eq:distance}
	d(\phi) = \epsilon\log\left({\frac{1+\phi}{1-\phi}}\right)\,.
\end{equation}
Dirichlet and Neumann boundary conditions are prescribed
\begin{align}
	\phi&=\bar{\phi}&&\text{on }\partial\Omega^{\text{inflow}}_{\text{D},\phi}\subset\partial\Omega\times[0,\tEnd]&&& \nabla\phi\cdot\hat{\Bn}&=0&&&&\text{on }\partial\Omega_{\text{N},\phi}\subset\partial\Omega\times[0,\tEnd]
\end{align}
along the domain boundary $\partial\Omega=\partial\Omega^{\text{inflow}}_{\text{D},\phi}\cup\partial\Omega_{\text{N},\phi}$ with $\partial\Omega=\partial\Omega^{\text{inflow}}_{\text{D},\phi}\cap\partial\Omega_{\text{N},\phi}=\emptyset$.

\deleted{For keeping the shape of the profile of the level-set function constant as the interface moves, additionally, reinitialization is performed according to \cite{olsson2007conservative}\deleted{, as described in Appendix \ref{app:level_set}}.}

\deleted{
At this point it should be noted that the formulation of the level-set transport velocity $\uGamma$ is a key modeling component in presence of rapid evaporation and is part of a detailed discussion in Section~\ref{sec:level_set_transport}. Without phase change, it is typically assumed that the latter corresponds to the local fluid velocity $\uGamma(\Bx,t):=\Bu(\Bx,t)$. }

{
Subsequent to solving the advection equation \eqref{eq:transport} of the level-set function $\phi$ at time~$t$, a reinitialization step \cite{olsson2007conservative} is performed to preserve the shape of the regularized indicator function as the interface moves. For this purpose, we solve
\begin{align}
	\label{eq:reinitialization}
	\fracPartial{\psi}{\tau}+\nabla\cdot\left(\frac{1-\psi^2}{2}\nGamma\right) & =\epsilon\nabla\cdot((\nabla\psi\cdot\nGamma)\nGamma) \qquad  \text{in }\Omega\times[0,\tauEnd]
\end{align}
for the pseudo-time $\tau\in[0,\tauEnd]$ with initial condition $\psi(\Bx,\tau=0)=\phi(\Bx,t)$ and homogeneous Neumann boundary conditions $\nabla\psi\cdot\hat{\Bn}=0\text{ on }\partial\Omega\times[0,\tauEnd]$
until steady state is obtained at $\tau=\tauEnd$. Here, $\psi$ represents an auxiliary field, which is transferred to the level-set field, $\phi(\boldsymbol{x},t) = \psi(\boldsymbol{x},\tau=\tauEnd)$, at the end of the reinitialization pseudo-time stepping scheme. The parameter $\epsilon$ is the interface thickness parameter, and $\nGamma$ represents the interface normal vector, evaluated at time $t$ (or pseudo-time $\tau=0$) and assumed as constant over the pseudo-time. The determination of the latter is described below.
For discretization in time of the reinitialization equation, we employ a semi-implicit Euler time stepping scheme, considering an explicit scheme for the nonlinear compressive flux term $\left(\added{(1-\psi^2)/2}\,\nGamma\right)$ in order to obtain a linear system of equations. 
The pseudo-time step size is chosen as $\Delta\tau=\min\left(\epsilon, \Delta t\right)$. 

As proposed in \cite{olsson2007conservative}, the interface normal vector is computed from a projection step of the normalized level-set gradient 
\begin{equation}
	\label{eq:normal_vector}
	\bar{\Bn}_\Gamma=\frac{\nabla\phi}{|\nabla\phi|}\qquad\text{ in }\Omega\,
\end{equation}
to the level-set space
\begin{equation}
	\label{eq:normal_vector_regularized}
	\nGamma-\eta_{n}\,h^2\Delta\nGamma=\bar{\Bn}_\Gamma\qquad\text{ in }\Omega\,
\end{equation}
subject to homogeneous Neumann boundary conditions $\nabla\nGamma\cdot\hat{\Bn}=\boldsymbol{0}\text{ on }\partial\Omega$. The filter parameter $\eta_{n}\,h^2$ is determined from the element edge length $h$ and the constant $\eta_n$ and represents the radius of nonlocal interaction. 
The mean curvature is defined as
\begin{equation}
	\label{eq:curvature}
	\bar{\kappa}=-\nabla\cdot\nGamma\qquad\text{ in }\Omega\,.
\end{equation}
In order to avoid spurious high-frequency oscillations of the curvature, likewise to the projected interface normal vector, we compute a regularized curvature $\kappa$ as proposed in \cite{olsson2007conservative}
\begin{equation}
	\label{eq:curvature_regularized}
	\kappa-\eta_{\kappa}\,h^2\Delta\kappa=\bar{\kappa}\qquad\text{ in }\Omega\,.
\end{equation}
with the filter parameter $\eta_{\kappa}\,h^2$ from the element edge length $h$ and a constant $\eta_{\kappa}$. 
We use homogeneous Neumann boundary conditions $\nabla\kappa\cdot\hat{\Bn}=\boldsymbol{0}\text{ on }\partial\Omega$. 
\begin{remark}For quadrilateral or hexahedral elements, we compute the element edge length $h$ as $h=\max(d)/\sqrt{dim}$, where $\max(d)$ is the largest diagonal of the element and $dim\in\{1,2,3\}$ is the considered dimension. The influence of the filter parameter was investigated in \cite{zahedi2012spurious,cenanovic2020finite}. In our simulations we consider $\eta_{\kappa}=2$ and $\eta_{\Bn}=2$ as default values.
\end{remark}
}

\subsubsection{Effective material properties}
\label{sec:matProps}
\begin{figure}[tb!]
	\centering
	\includegraphics{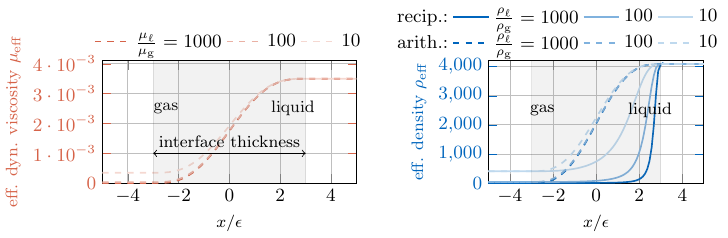}
	\caption{\added{Distribution of (left) the effective dynamic viscosity (\myeqref{eq:muEff}) and (right) the effective density using a (standard) arithmetic phase-fraction weighted average versus the employed reciprocal interpolation function (\myeqref{eq:rhoEff}).} The values for the liquid phase are chosen to represent \tiSixFour{}, i.e. $\liqP{\rho}=\SI{4133}{kg/m^3}$ and $\liqP{\mu}=\SI{3.5e-3}{Pa\cdot s}$. The ratios between the phases are artificially varied for the sake of demonstration.}
	\label{fig:effective_mat_props}
\end{figure}
From the level-set function $\phi$, a localized, indicator-like representation can be constructed by employing the smoothed approximation of the Heaviside function \cite{sussman1994level,peskin2002immersed}
\begin{equation}
	\label{eq:heaviside}
	H_\phi(\phi) = 
	\begin{cases} 	
			0 & d(\phi) \leq -3\epsilon\\
			\frac{1}{2} + \frac{d(\phi)}{6\,\epsilon} + \frac{1}{2\,\pi}\,\sin\left(\frac{\pi\,d(\phi)}{3\,\epsilon}\right) & -3\epsilon < d(\phi) < 3\epsilon\\
			1 & d(\phi) \geq 3\epsilon
	\end{cases} \,.
\end{equation}
This function is used to interpolate quantities between the two phases. \added{In addition, the gradient of \myeqref{eq:heaviside} is employed to compute a smoothed approximation of the Dirac delta function, allowing for a diffuse representation of interface contributions with non-zero support localized to a finite interface region (see Section~\ref{sec:interface_fluxes}).} For example, the effective dynamic viscosity $\muEff$ is evaluated as arithmetic phase-fraction weighted average of the values for the liquid and the gaseous phase, i.e., $\liqP{\mu}$ and $\gasP{\mu}$, respectively: 
\begin{equation}
	\eff{\mu}\left(\phi\right) = H_\phi(\phi)\,\liqP{\mu}
	+ \left(1-H_\phi\left(\phi\right)\right)\,\gasP{\mu}\,.
	\label{eq:muEff}
\end{equation}
While the type of interpolation function used for the effective viscosity is arbitrary \added{(see discussion in Appendix~\ref{app:viscosity})}, this is not the case for the density in the presence of phase change. We employ a reciprocal interpolation function of the density between the two phases
\begin{equation}
\frac{1}{\func{\rhoEff}{\phi}} = \frac{\Hphi(\phi)}{\rhoL} + \frac{1-\Hphi(\phi)}{\rhoG}
\label{eq:rhoEff}
\end{equation}
considering the density of the liquid phase $\rhoL$ and the one of the vapor phase $\rhoG$. This type of interpolation function was chosen to obtain consistency with the expression of the evaporative dilation rate, which is explained in Section~\ref{sec:level_set_transport}.

In Fig.~\ref{fig:effective_mat_props}, the distribution of the \replaced{viscosity (left) and effective density (right)}{effective density (left) and viscosity (right)} is illustrated over the interface region for increasing ratios of these parameters between the liquid and the gaseous phase (liquid phase parameters are taken for \tiSixFour{} as exemplary material). For comparison, in the left panel of Fig.~\ref{fig:effective_mat_props} the density distribution according to \myeqref{eq:rhoEff} and the one obtained by an arithmetic phase-fraction weighted average similar to \myeqref{eq:muEff} is shown. It can be seen that for the employed reciprocal interpolation function the influence of different density ratios becomes much more pronounced in the interfacial region compared to the (standard) arithmetic phase-fraction weighted average.

\subsubsection{Interface \replaced{source terms}{fluxes}}
\label{sec:interface_fluxes}

For a vaporizing incompressible two-phase-flow model, the two phases are coupled by \deleted{singular fluxes}\added{source terms} consisting of (i) the evaporative dilation rate and (ii) the surface tension force. The singular evaporative dilation rate is stated as
\begin{equation}
	\label{eq:evaporDilSingular}
	\func{\evaporDilationRateSingular}{\Bx,t}=\func{\mDot}{\Bx,t}\,\left(\frac{1}{\rhoL}-\frac{1}{\rhoG}\right)\func{\diracDelta}{\Bx,t}
\end{equation}
with the evaporative mass flux $\mDot$, which is a prescribed quantity in our setting arising from the underlying assumption of isothermal conditions in this contribution. The singularity at the interface may be imposed by the Dirac delta distribution
\begin{equation}
	\label{eq:diracDelta}
	\func{\diracDelta}{\Bx,t}=\begin{cases}
		1 \text{ on } \Gamma \\
		0 \text{ else}
	\end{cases}
\end{equation}
with support on the discrete/sharp interface $\Gamma$. Considering the weak form of \myeqref{eq:evaporDilSingular}\added{, e.g., }in a finite element context\added{,} with the test function $w$, the latter can formally be applied as a sharp model via
\begin{equation}
	\label{eq:weakFormSharp}
	\BiLi{w}{\evaporDilationRateSingular} = \left(w\,,\,\mDot\,\left(\frac{1}{\rhoL}-\frac{1}{\rhoG}\right)\right)_{\Gamma} 
\end{equation}
exploiting the property of the Dirac delta function $\int_\Omega f \diracDelta\,d\Omega= \int_\Gamma\, d\Gamma$ for an arbitrary function $f$.

Alternatively, in a regularized model considering a continuous surface flux in the sense of \cite{brackbill1992}, which is employed in the present work, the Dirac delta function $\diracDelta$ is approximated by a regularized, smooth function $\symDelta$ preserving the property \text{$\int_{-1}^{1}\symDelta\,d\phi=1$}. We calculate  $\symDelta$ from the smoothed Heaviside function \eqref{eq:heaviside}
\begin{equation}
	\label{eq:delta}
	\symDelta(\phi)=||\nabla H_\phi(\phi)||
\end{equation}
with support within the interface region $0< H_\phi <1$. This leads to a slightly modified expression for the weak form of \myeqref{eq:weakFormSharp}, which reads as
\begin{equation}
	\BiLi{w}{\evaporDilationRateSingular} \approx \Big(w\,,\,\underbrace{\mDot\,\left(\frac{1}{\rhoL}-\frac{1}{\rhoG}\right)\symDelta}_{\evaporDilationRate}\Big)_{\Omega} \,
\end{equation}
where the regularized representation of the evaporative dilation rate applied to the continuity equation~\eqref{eq:continuity} is introduced 
\begin{equation}
	\label{eq:evaporDil}
	\func{\evaporDilationRate}{\Bx,t}=\func{\mDot}{\Bx,t}\,\left(\frac{1}{\rhoL}-\frac{1}{\rhoG}\right)\func{\symDelta}{\phi(\Bx,t)}\,.
\end{equation}
It is important to note that the consideration of the  evaporative dilation rate 
\eqref{eq:evaporDil} in \myeqref{eq:continuity} results \emph{inherently} in an additional, evaporation-induced pressure at the interface, i.e., the evaporation-induced recoil pressure,
\begin{equation}
	\vapP{p}(\Bx,t) = \func{\mDot}{\Bx,t}^2\left(\frac{1}{\rhoL}- \frac{1}{\rhoG}\right)\,\func{\symDelta}{\phi(\Bx,t)}\,.
\end{equation}
\added{For instance, using the Hertz--Knudsen relation \cite{knight1979theoretical} to compute the evaporative mass flux yields results consistent with the recoil pressure model presented by \cite{anisimov1995instabilities}, a widely adopted approach in melt pool modeling \cite{khairallah2016laser}}.
Hence, the consideration of an extra term for the evaporation-induced recoil pressure in the momentum equation, as considered in e.g. \cite{khairallah2016laser}, is not necessary.
Instead, the recoil pressure results naturally from the velocity gradients across the interface as induced by the dilation rate $\evaporDilationRate$~\eqref{eq:evaporDil}, which is demonstrated in Section~\ref{sec3}.
 
Similarly to the evaporative dilation rate, the surface tension force is modeled as a continuous surface \replaced{force}{flux} in the sense of \cite{brackbill1992}. It is expressed as
	\begin{equation}
		\label{eq:surfaceTension}
		\surfaceTensionForce(\Bx,t)=\alpha\kappa(\phi(\Bx,t))\nGamma(\phi(\Bx,t))\func{\deltaRho}{\phi(\Bx,t)}
	\end{equation} 
	and is considered in the momentum equation (\myeqref{eq:momentum_balance}) with the surface tension coefficient $\alpha$, the interface normal vector $\nGamma$ and the interface mean curvature $\kappa$, the latter two computed from the level-set function as described in Appendix~\ref{app:level_set}.
	Here, $\deltaRho$ represents a density-scaled delta function similar to \cite{yokoi2014density}, adjusted for the employed reciprocal density interpolation according to \cite{much2023}:
	\begin{equation}
		\func{\deltaRho}{\phi} = \func{\symDelta}{\phi}\,\func{\rhoEff}{\phi}\,c_\rho\quad\text{ with }c_\rho = \frac{\gasP{\rho}-\liqP{\rho}}{\liqP{\rho}\gasP{\rho}\ln\left(\frac{\gasP{\rho}}{\liqP{\rho}}\right)} \text{ for } \gasP{\rho} > 0 \wedge \liqP{\rho} > 0\,. 
	\end{equation}
	 This ensures that the magnitude of the surface-tension-induced acceleration is well-balanced across the interface.

\subsection{Formulations of a consistent level-set transport velocity for a diffuse evaporation-induced velocity jump}
\label{sec:level_set_transport}

\subsubsection{Preliminaries}
\label{eq:uGamma_Preliminaries}

For phase change across the liquid-gaseous interface $\Gamma$, a key modeling aspect lies in an accurate expression for the level-set transport velocity $\uGamma$ in \myeqref{eq:transport} that (i) predicts the evaporated liquid mass accurately, (ii) is a continuous field and (iii) ideally is divergence-free at least in the near-interface region. The determination of $\uGamma$ is not straightforward since the fluid velocity exhibits a (smeared) discontinuity across the interface in the pursued diffuse one-fluid formulation of the presented modeling framework. 
\begin{figure}[tb!]
	\centering
	\includegraphics{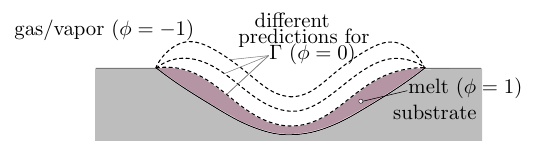}
	\caption{Schematic illustration of different predictions of the melt pool geometry, indicated by the dashed lines, obtained for different level-set transport velocities.}
	\label{fig:melt_pool_morphology}
\end{figure}

The importance of this quantity should be additionally highlighted for the example of melt pool dynamics of \pbfam{}. The exact prediction of the location of the melt pool surface is a crucial aspect, since it determines the morphology of the molten metal, as indicated in  Fig.~\ref{fig:melt_pool_morphology}. If the melt pool surface is incorrectly predicted, e.g., leading to a thinner layer of molten metal, the mass and the thermal mass of the melt pool \replaced{will change. Consequently,}{changes and consequently,} the dynamic behavior will be significantly different compared to the real melt pool morphology.


For now, let us consider a sharp interface description with a velocity jump from $\uL$ to $\uG$ present at the interface with $\uL$ being the velocity on the liquid side of the interface and $\uG$ the velocity on the gaseous side of the interface. Mass conservation across the interface according to the Rankine--Hugoniot condition  states  
\begin{equation}
\label{eq:rankineHugoniot}
\rhoL\left(\uGamma-\uL\right)\cdot\nGamma = 
\rhoG\left(\uGamma-\uG\right)\cdot\nGamma \equiv \mDot
\end{equation}
considering a reference frame moving with the interface \cite{tryggvason2011direct}\replaced{. \myeqref{eq:rankineHugoniot} implies that the velocity jump from the liquid to the gas phase occurs in the normal direction to the interface. The velocity component tangential to the interface is continuous. The variable }{, where} $\nGamma$ is the interface normal vector (pointing inside $\OmegaL$) and $\uGamma$ is the interface transport velocity, which is our quantity of interest. By rearrangement of the Rankine--Hugoniot condition, an expression for the transport velocity of the discrete interface is obtained
\begin{equation}
	\label{eq:uGammaSharp}
	\uGamma = \uL + \frac{\mDot}{\rhoL}\nGamma \qquad \text{ or } \qquad
	\uGamma = \uG + \frac{\mDot}{\rhoG}\nGamma\,\,\qquad\text{on }\Gamma.
\end{equation} 
The direct evaluation of \myeqref{eq:uGammaSharp} is not possible in our diffuse framework due to the smearing of the velocity discontinuity across the interface region. 

In the following, we discuss three different formulations for the computation of the level-set
transport velocity field in a narrow band around the interface, applicable to a diffuse phase-change framework.

\subsubsection{Variant 1: Divergence-free continuous level-set transport velocity based on a consistent density distribution}
\label{sec:uGammaContinuous}

In this section, we derive a continuous level-set transport velocity, suitable for diffuse jump conditions and flat or slightly curved interfaces, and an associated consistent interpolation function of the density between the two phases. If the reader is only interested in the final result, we recommend skipping the following paragraphs and continuing  before \myeqref{eq:transport_vel_local_continuous}.

We depart from the mass conservation equation
\begin{align}
	\label{eq:continuity2}
	\frac{D\rho}{Dt} + \rho\,\diver{\Bu} = 0
\end{align}
with the material time derivative of the density \added{$D\rho/Dt$} and the fluid velocity $\boldsymbol{u}$. For the sake of brevity, we denote the effective density as $\rho$ in this section. We assume that the density is a function of the smoothed Heaviside function $H_\phi(\phi)$ \eqref{eq:heaviside}. Thus, the material time derivative of the density is obtained by applying the chain rule
\begin{equation}
	\label{eq:rhoChainRule}
	\frac{D\rho}{Dt}=\fracPartial{\rho}{H_\phi}\,\fracPartial{H_\phi}{\phi}\,
	\frac{D\phi}{Dt}\,.
\end{equation}
By inserting the continuity equation (\myeqref{eq:continuity}) together with \myeqref{eq:rhoChainRule} into \myeqref{eq:continuity2} results in
\begin{equation}
	\label{eq:rhoChainRule2}
	\fracPartial{\rho}{H_\phi}\,\fracPartial{H_\phi}{\phi}\,
	\underbrace{
		\left(\fracPartial{\phi}{t}+\Bu\nabla\phi\right)
	}_{\frac{D\phi}{Dt}}
	+\rho\,\evaporDilationRate = 0\,.
\end{equation}

After rearrangement, consisting of the insertion of the defined evaporative dilation rate (\myeqref{eq:evaporDil}) \added{into \myeqref{eq:rhoChainRule2}}
\begin{equation}
	\added{
	\label{eq:rhoChainRule3}
	\fracPartial{\rho}{H_\phi}\,\fracPartial{H_\phi}{\phi}\,
		\left(\fracPartial{\phi}{t}+\Bu\nabla\phi\right)
		+\rho\,\mDot\,\left(\frac{1}{\rhoL}-\frac{1}{\rhoG}\right)\symDelta = 0\,
	}
\end{equation}
and taking into account the equality $\symDelta= \added{\partial H_\phi/\partial\,\phi}\, (\nabla\phi\cdot\nGamma)$ \added{in \myeqref{eq:rhoChainRule3}}, \added{we obtain:}
\begin{align}
	\added{
	\label{eq:rhoChainRule4}
	\underbrace{\fracPartial{\rho}{H_\phi}}_{\rho'}\,\fracPartial{H_\phi}{\phi}\,
\left(\fracPartial{\phi}{t}+\Bu\nabla\phi\right)+\rho\,\mDot\,\underbrace{\left(\frac{1}{\rhoL}-\frac{1}{\rhoG}\right)}_{c_1}
	\fracPartial{H_\phi}{\phi}\nabla\phi\cdot\nGamma
	 = 0\,.
	}
\end{align}
\added{Next, we cancel out $\fracPartialText{H_\phi}{\phi}$ from \myeqref{eq:rhoChainRule4} and divide by $\rho'=\added{\partial\rho/\partial\phi}$, arriving at the transformed form of }\myeqref{eq:rhoChainRule2} \added{for}
 \deleted{can be transformed into} the transport equation of the level set
\begin{equation}
	\fracPartial{\phi}{t} + \underbrace{\left(\Bu+c_1\mDot\,\frac{\rho}{\rho'}\nGamma\right)}_{:=\uGamma}\cdot\nabla\phi = 0\,
	\label{eq:uGammaContinuous_transport}
\end{equation}
with the definition of the transport velocity $\uGamma$, the interface normal vector $\nGamma$ and the abbreviation\deleted{s} $c_1=\added{1/\rhoG-1/\rhoL}$ \deleted{and $\rho'=\added{\partial\rho/\partial\phi}$}. At this point, it is not clear, which interpolation rule should be chosen to describe the smooth evolution of the density across the interface. We derive a suitable density interpolation function in the following. In a first step, in order to keep the level-set profile constant as the interface moves and to avoid artificial deformation of the level set field, we enforce that the interface velocity should be divergence-free 
\begin{equation}
	\diver\uGamma = 0\,.
\end{equation}
We apply the chain rule to compute the divergence of the transport velocity defined in \myeqref{eq:uGammaContinuous_transport}\replaced{. For this purpose, we assume}{
 under the assumption of} a potentially varying evaporation flux $\mDot$ over the finite interface region, consider the definition of the evaporative dilation rate \eqref{eq:evaporDil} and employ the
 definition of the interface curvature, $\diver\nGamma=\kappa$, which yields:
\begin{align}
	\diver\uGamma = 
	\underbrace{c_1\,\mDot\,\symDelta}_{\diver\boldsymbol{u}}
	+ c_1\,\frac{\rho}{\rho'}\nGamma\cdot\nabla\mDot
	+ c_1\underbrace{\mDot\,\frac{\rho}{\rho'}\kappa}_{\approx 0} 
	+c_1\,\mDot\,\left(1-\frac{\rho\,\rho''}{\rho'^2}\right)\nGamma\cdot\nabla\phi \equiv0\,.
\end{align}
Here, the abbreviation $\rho''=\added{\fracPartialSecText{\rho}{\phi}}$ is introduced.
The expression $\mDot\added{\kappa\,\rho/\rho'}$ may be neglected if the interface thickness, influenced by $\epsilon$, is considerably smaller than the curvature radius $R_\kappa=1/\kappa$, i.e., when approaching the limit case of a flat interface ($\epsilon/R_\kappa=0$). Note that in this context the ratio $\rho/\rho^{\prime}$ scales with the interface thickness parameter $\epsilon$. With this approximation and after division by the non-zero term $c_1$ and exploiting $\symDelta=\added{\partial H_\phi/\partial \phi}\,\nGamma\cdot\nabla\phi$, this results in the simplified form of the divergence-free condition of the transport velocity
\begin{align}
	0 &= \nGamma \cdot \left(\frac{\rho}{\rho'}\nabla\mDot+\mDot\,\left(\mDot+1-\frac{\rho\,\rho''}{\rho'^2}\right)\fracPartial{H_\phi}{\phi}\,\nabla\phi\right)\,.
	\label{eq:transport_velocity_pde}
\end{align}
	In \myeqref{eq:transport_velocity_pde}, the scalar product of the interface normal vector ($\nGamma$) and the gradient terms ($\nabla \phi$ and $\nabla\mDot$) represent an extraction of the component of the gradient in the interface normal direction. Therefore, we introduce a local coordinate system, in which the $x$-direction is aligned with the interface normal direction $\nGamma$, which allows us to express \myeqref{eq:transport_velocity_pde} as
\begin{equation}
	-\frac{\frac{d\mDot}{dx}}{\mDot} = \frac{dH_\phi}{dx} \left(\frac{2\rho'}{\rho}-\frac{\rho''}{\rho'}\right)\,,
\end{equation}
representing a linear differential equation.
The latter can be solved analytically by separation of variables 
and subsequent integration over the interface thickness $x$ with $-t/2\leq x\leq t/2$. Considering $H_\phi(-t/2)=0$ and $\rho(-t/2)=\rho_g$, this results in
\begin{align}
	C\,\int_{-t/2}^{x} \mDot\,\symDelta\,dx &= \frac{1}{\rho_g} - \frac{1}{\rho(x)}\,.
	\label{eq:transport_velocity_no_C}
\end{align}
Determination of the integration constant $C$ follows from  integration over the interface thickness $-t/2\leq x\leq t/2$ with $\rho(t/2)=\rho_l$
\begin{align}
	C &= \frac{1}{\underbrace{\int_{-t/2}^{t/2} \mDot\,\symDelta\,dx}_{\dot{\bar{m}}}}\left(\frac{1}{\rhoG} - \frac{1}{\rhoL}\right)\,.
	\label{eq:transport_velocity_integration_constant}
\end{align}
Finally, inserting \myeqref{eq:transport_velocity_integration_constant} into \myeqref{eq:transport_velocity_no_C} yields an expression for the density distribution over the interface that ensures a divergence-free transport velocity 
\begin{equation}
	\frac{1}{\rho} = 
	\frac{1}{\rhoG} - \frac{1}{\dot{\bar{m}}}\left(\frac{1}{\rhoG} - \frac{1}{\rhoL}\right)\,\int_{-t/2}^{x} \mDot\,\symDelta\,dx\,.
	\label{eq:reciprocalDensityNew}
\end{equation}
By evaluation of \myeqref{eq:reciprocalDensityNew} for a spatially constant evaporative mass flux in interface thickness direction ($\Bn_\Gamma \left(\Bn_\Gamma \cdot \nabla \mDot\right)=\B0$), the relations $\dot{\bar{m}}=\mDot$ and $\int_{-t/2}^{x} \mDot\,\symDelta\,dx=\mDot\,H_\phi$  hold. For this special case, the definition of the density distribution \eqref{eq:reciprocalDensityNew} reduces to
\begin{equation}
	\label{eq:reciprocalDensity}
	\frac{1}{\rho} = 
	\frac{H_\phi}{\rhoL} + 
	\frac{1-H_\phi}{\rhoG}\,.
\end{equation}
It represents the employed interpolation function used for the effective density of our diffuse framework given in \myeqref{eq:rhoEff}, for which we have demonstrated a mathematically consistent derivation to achieve a diverence-free level-set transport velocity. In this derivation, two main assumptions have been made: 1) the interface thickness has to be small compared to the interface curvature radius; 2) the variation of the evaporative mass flux has to be small across the interface thickness. Both assumptions can be justified if the interface thickness is chosen small enough. 

We conclude the central result of the derivation above with the definition of the level-set transport velocity 	\eqref{eq:uGammaContinuous_transport} and the reciprocal interpolation of the density \eqref{eq:reciprocalDensity}. The first approach for computing the level-set transport velocity of the diffuse model is defined by
\begin{equation}
	\uGamma^{(\text{V1})}(\Bx) = \Bu(\Bx) + \frac{\mDot}{\rho(\Bx)}\,\nGamma(\Bx) \qquad \text{ for }\Bx \text{ in }\Omega\,.
	\label{eq:transport_vel_local_continuous}
\end{equation}
It is referred to as \emph{variant 1} in the following. 
This equation modifies the local fluid velocity $\Bu(\Bx)$ by an evaporation-dependent contribution, considering only local field quantities making it attractive for a finite element framework. 

\subsubsection{Variant 2 and variant 3: Level-set transport velocity based on extended velocity fields}
\label{sec:uGammaHybrid}

The previously presented \emph{variant 1}~\eqref{eq:transport_vel_local_continuous} for computing the level-set transport velocity is particularly suited for flat or slightly curved interfaces. However, in certain practical applications such as \pbfam{}, a very thin interface thickness is required to achieve an acceptable accuracy owing to the fundamental assumptions of \emph{variant 1} (a small interface thickness compared to the curvature radius and a minimal variation of the evaporative mass flux across the interface). This necessitates an extremely fine spatial discretization, which leads to a significant increase in computational cost. Inspired by the sharp model equations \eqref{eq:uGammaSharp}, we propose two alternative variants to compute the transport velocity in the diffuse model which are more accurate for highly curved interfaces. We exploit extension algorithms which are similarly found in ghost fluid methods \cite{tanguy2007level}. 

\emph{Variant 2} considers an extension of the velocity from the liquid end of the interface region, i.e., from $\liqP{\Bx}$:
\begin{equation}
	\uGamma^{(\text{V2})}(\Bx) = \Bu(\liqP{\Bx}(\Bx) ) + \frac{\mDot}{\rhoL}\,\nGamma(\Bx)
	\qquad
	\text{ for }\Bx \text{ in }\Omega\,.
	\label{eq:transport_vel_extension_liquid}
\end{equation}
The liquid end of the interface region $\liqP{\Bx}$ is defined as the projection of a point $\Bx$ along the interface normal to the level-set isocontour where $H_\phi(\phi)$ attains 1, see \myeqref{eq:heaviside}.

In contrast, \emph{variant 3} considers an extension of the fluid velocity from the gaseous end of the interface region, i.e., from $\gasP{\Bx}$:
\begin{equation}
	\uGamma^{(\text{V3})}(\Bx) = \Bu(\gasP{\Bx}(\Bx) ) + \frac{\mDot}{\rhoG}\,\nGamma(\Bx)
	\qquad
	 \text{ for }\Bx \text{ in }\Omega\,.
	\label{eq:transport_vel_extension_gas}
\end{equation}
The gaseous end of the interface region $\gasP{\Bx}$ is defined as the projection of a point $\Bx$ along the interface normal to the level-set isocontour where $H_\phi(\phi)$ attains 0, see \myeqref{eq:heaviside}.

The two presented variants above require an extrapolation algorithm for determining $\Bu(\gasP{\Bx})$ or $\Bu(\liqP{\Bx})$ in a narrow band around the interface.  
	In \autoref{app:cpp}, algorithmic details are provided, along with an illustrative demonstration and verification based on the well-known two-phase flow benchmark example of a rising bubble. In the extension algorithm, we utilize closest point projection, as suggested in \cite{henri2022geometrical}: In the first step, corresponding to the current location $\Bx$, points at the liquid or gaseous end of the interface region, i.e., $\gasP{\Bx}(\Bx)$ or $\liqP{\Bx}(\Bx)$, defined as the closest point located at the level-set-isocontours
 \begin{equation}
 	\phi(\liqP{\Bx})=\phi(d(\Bx)=3\,\epsilon)\stackrel{\myeqref{eq:initial_phi}}{=}\tanh{(1.5)}
 	\end{equation} 
 	and
 	\begin{equation}
 	\phi(\gasP{\Bx})=\phi(d(\Bx)=-3\,\epsilon)\stackrel{\myeqref{eq:initial_phi}}{=}\tanh{(-1.5)}\,,
 \end{equation} 
 are computed. In the second step, the fluid velocities $\Bu(\liqP{\Bx}(\Bx) )$  or $\Bu(\gasP{\Bx}(\Bx) )$ are evaluated at the projection points $\gasP{\Bx}$ and $\liqP{\Bx}$, which can be used in \myeqrefsa{eq:transport_vel_extension_liquid}{eq:transport_vel_extension_gas}  to compute the level-set transport velocity. 

\subsubsection{Evaluation of the proposed level-set transport velocity approaches based on analytical benchmark examples}
\label{sec:uGamma_evaluation}

\begin{figure}[htbp!]
	\includegraphics{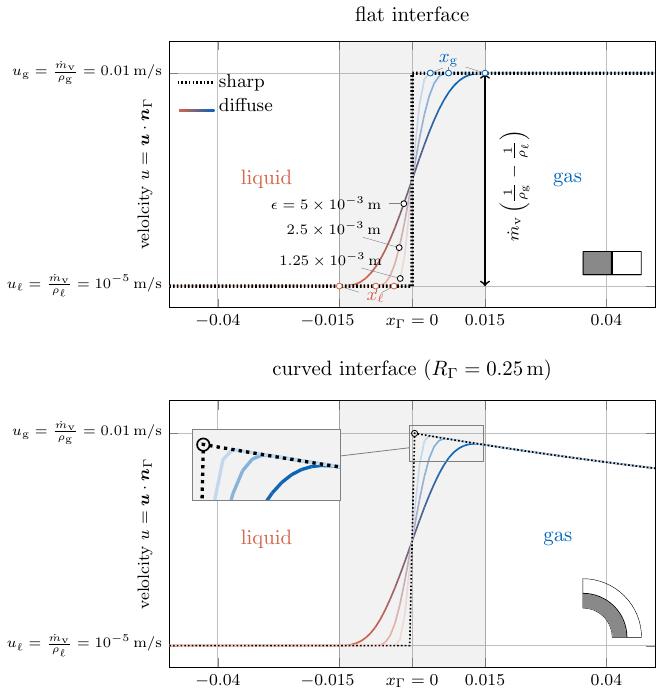}
	\caption{Comparison of sharp and diffuse models based on the analytical solution for the fluid velocity normal to the interface of a flat (top) and an axisymmetric curved (bottom) interface subject to evaporation. The parameters are chosen as $\mDot=\SI{0.01}{kg/m^2s}, \rhoL=\SI{1000}{kg/m^3}$, $\rhoG=\SI{1}{kg/m^3}$. For the curved interface, the diffuse model results in a slightly lower peak velocity compared to the sharp model through the inherent diffusion of the velocity over the curved interface zone.}
	\label{fig:transport_velocity_profiles_velocity}
\end{figure}

\begin{figure}[htbp!]
	\centering
	\includegraphics{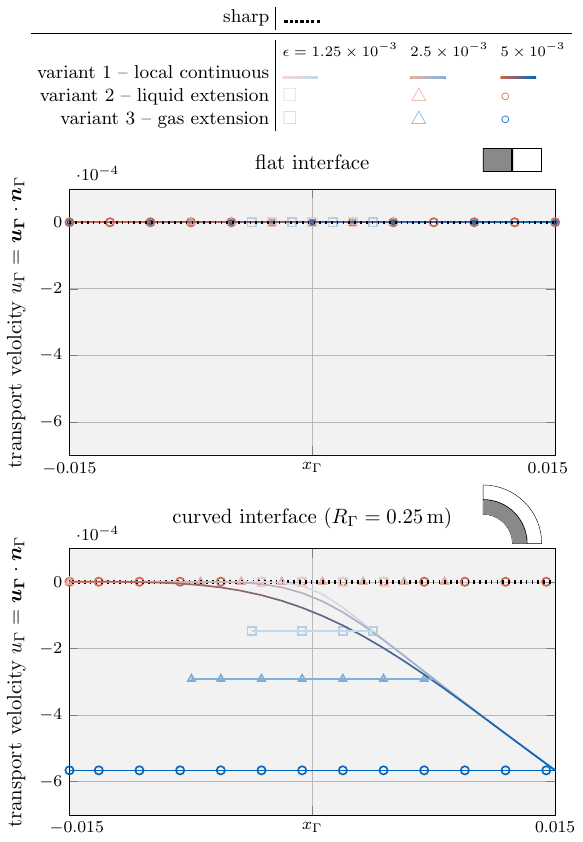}
	\caption{Evaluation of diffuse models based on the analytical solution for the level-set transport velocity of a flat (top) and an axisymmetric curved (bottom) interface subject to evaporation (cf. Fig.~\ref{fig:transport_velocity_profiles_velocity}):	The considered \emph{variants 1-3} yield identical results corresponding to the sharp reference solution for the flat interface. For the curved interface, only \emph{variant 2}, i.e., considering an extension velocity from the liquid end of the interface zone, yields a good approximation of the exact transport velocity for finite values of the interface thickness.}
	\label{fig:benchmark_transport_velocity}
\end{figure}

After having introduced three formulations of a level-set transport velocity suitable for a diffuse framework in Sections~\ref{sec:uGammaContinuous} and \ref{sec:uGammaHybrid}, we evaluate their strengths and weaknesses based on two simple yet illustrative analytical benchmark cases consisting of (i) a flat interface (cf. Fig.~\ref{fig:flat_interface}) and (ii) an axisymmetric curved interface (cf. left panel of Fig.~\ref{fig:const_law}).  At the interface liquid material evaporates with a spatially and temporally constant evaporative mass flux. Simultaneously, the evaporated volume is compensated by a prescribed inflow velocity on the liquid side of the interface to yield a spatially fixed interface location. Thus, we expect the computed level-set transport velocity $\uGamma$ to be zero.  For the two examples, analytical solutions exist for the velocity and the pressure field for both the diffuse and the sharp model, which are derived in \autoref{app:evapor_stefan} and \autoref{app:evapor_shell}, respectively.

In Fig.~\ref{fig:transport_velocity_profiles_velocity}, the resulting  profiles of the velocity component normal to the interface are depicted for the flat (top) and the axisymmetric curved interface (bottom). The black, dashed curves represent the velocity profiles derived for a sharp model --- representing the exact reference solution (cf. to \myeqref{eq:stefan_sharp_velocity} for the flat interface and \myeqref{eq:shell_u_analytical_sharp} for the curved interface). The colored solid curves correspond to the solutions obtained for the diffuse model for different values of the interface thickness parameter $\epsilon$ according to \myeqrefsa{eq:stefan_diffuse_velocity}{eq:shell_u_analytical}. For the curved interface, the diffuse model predicts a slightly lower peak velocity compared to the sharp model due to the inherent diffusion of the velocity over the curved interface zone, as also discussed in \cite{lee2017direct}.
For both geometries, the solution of the diffuse model tends to converge to the reference solution as the interface thickness decreases ($\epsilon\rightarrow0$) and thus is considered mathematically consistent. 

 The resulting level-set transport velocities for the two investigated geometries and different interface thickness parameters are shown in Fig.~\ref{fig:benchmark_transport_velocity}. We recall that the exact solution for the level-set transport velocity is by construction zero.

As depicted in the top panel of Fig.~\ref{fig:benchmark_transport_velocity}, for the flat interface, the transport velocity obtained by using \emph{variant 1}, \emph{variant 2} or \emph{variant 3} all agree with the reference solution as expected.
 
 However, when considering the curved interface, illustrated in in Fig.~\ref{fig:benchmark_transport_velocity} bottom, the transport velocities predicted by the diffuse model for \emph{variant 1} and \emph{variant 3} significantly deviate from the reference solution\replaced{. This deviation is particularly notable}{, particularly} at the critical location of the discrete interface, i.e., $\phi=0$\added{, where the level-set transport velocity plays a key role in determining the overall evaporated mass}. Furthermore, the solution according to \emph{variant~1} exhibits a substantial variation across the interface, leading to a violation of the initial assumption of a local divergence-free condition of the transport velocity. 
 It should be emphasized that, due to the analytical nature of the problem, spatial or temporal discretization errors play do not play a role. Consequently, the error can be attributed to the assumption made in Section~\ref{sec:uGammaContinuous} regarding a small ratio between the interface thickness and the curvature radius.
 Despite these discrepancies, it is important to note that the solution remains mathematically consistent, i.e., the error decreases as the interface thickness or the curvature approaches zero. 
 Nevertheless, for realistic values of \added{the} interface thickness\replaced{, the deviation between the transport velocity of the diffuse model in \emph{variant 1} and the sharp model can be significant. This is especially true when dealing with high velocity jumps and curved geometries, which is demonstrated in numerical benchmark examples in Section~\ref{sec3}.}{and particularly when dealing with high velocity jumps, the deviation between the transport velocity of the diffuse model according to \emph{variant 1} and the sharp model can be significant for curved geometries, which is demonstrated in numerical benchmark examples in Section~\ref{sec3}}.

\emph{Variant 2} tends to exhibit a better accuracy than \emph{variant 3} for modeling evaporation. \replaced{This is expected because the transport velocity closely resembles the velocity at the liquid end of the interface region, which differs significantly from the velocity at the gas end. The evaporation-induced velocity difference results from a significant density ratio and/or a large evaporative mass flux. }{This is expected because at high evaporation-induced velocity differences between the liquid and gas phases, resulting from a significant density ratio and/or large evaporative mass flux, the transport velocity closely resembles the velocity of the liquid end of the interface region.}

As a final remark, although the extension algorithm makes \emph{variant~2} and \emph{variant~3} computationally more expensive compared to the local nature of \emph{variant 1}, the resulting transport velocity is constant across the interface region by construction. This is advantageous for the level-set transport, potentially reducing the need for frequent reinitialization steps~\cite{coquerelle2016fourth}. 

\subsection{Constitutive relation for incompressible viscous flow with diffuse phase change}
\label{sec:const_law}
\subsubsection{Stokes' constitutive relation}
For modeling incompressible viscous flow, the Stokes' constitutive relation
\begin{equation}
	\label{eq:stokesLaw}
	\boldsymbol{\sigma} =-p\,\mathcal{I}+\underbrace{2\,\eff{\mu}\,\boldsymbol{\varepsilon}}_{\viscousStress}
\end{equation}
 is frequently employed, where $\boldsymbol{\sigma}$ is the Cauchy stress tensor, $p$ is the pressure, $\mathcal{I}$ is the second-order identity tensor, $\eff{\mu}$ is the effective dynamic viscosity and
$\boldsymbol{\varepsilon}$ the rate-of-deformation tensor according to
\begin{equation}
	\label{eq:rate_of_deformation}
	\boldsymbol{\varepsilon}= \frac{1}{2}\left(\,\nabla\Bu +\left(\nabla\Bu\right)^\top\right)\,.
\end{equation} 

\subsubsection{A \replaced{corrected}{modified} viscous stress formulation}
\label{sec:modified_viscous_stress}

For incompressible flow without phase change, \added{the rate-of-deformation tensor \added{according to \myeqref{eq:rate_of_deformation}} is purely deviatoric due to the divergence constraint, i.e,  $\func{\Tr}{\boldsymbol{\varepsilon}}\equiv\diver\Bu=0$.} \added{As a consequence,} the viscous stress \added{tensor} $\viscousStress\added{=2\,\eff{\mu}\,\boldsymbol{\varepsilon}}$ \replaced{is}{ represents a} purely deviatoric\added{.} \deleted{stress tensor} \deleted{since the rate-of-deformation tensor \added{according to \myeqref{eq:rate_of_deformation}} is purely deviatoric due to the divergence constraint, i.e,  $\func{\Tr}{\boldsymbol{\varepsilon}}\equiv\diver\Bu=0$.} However, in presence of phase change, the diffuse velocity jump \added{in normal direction to the interface}, introduced in \myeqref{eq:continuity} through the evaporative dilation rate \eqref{eq:evaporDil}, yields an intentional violation of the \replaced{incompressibility}{divergence-free} condition in the interface zone, i.e. $\diver\Bu=\func{\Tr}{\boldsymbol{\varepsilon}}\neq0$ for $\{\Bx\in\Omega~\vert~0<H(\Bx)<1\}$. As a consequence, \replaced{the evaporative dilation rate contributes }{contributions to the divergence of the velocity field as caused by the evaporative dilation rate would contribute} to the
rate-of-deformation tensor \added{according to} \myeqref{eq:rate_of_deformation}. According to \myeqref{eq:stokesLaw}, this would result in an evaporation-induced contribution to the viscous stress, which is deemed to be not physically meaningful but a pure\added{ly numerical} \replaced{artifact}{consequence} of the diffuse interface approximation. 

As a remedy, we propose in the following to \replaced{correct}{modify} the rate-of-deformation tensor by neglecting the volumetric deformation caused by the diffuse evaporative dilation rate in the evaluation of viscous stresses. \added{Specifically, }
\deleted{In order to overcome the mentioned shortcomings of the standard Stokes' constitutive relation described in the previous section,} we propose to \replaced{introduce a correction term into the }{compute a modified }rate-of-deformation tensor
\begin{equation}
	\mathbb{\varepsilon}^{(\text{mod})}=\mathbb{\varepsilon}-\underbrace{\Tr\left(\mathbb{\varepsilon}\right)\,\nGamma\otimes\nGamma}_{\mathbb{\varepsilon}^{(v)}}\,.
	\label{eq:modified_rate_of_deformation}
\end{equation}
and use it in \myeqref{eq:stokesLaw} via
\begin{equation}
	\label{eq:modified_viscous_stress}
	\viscousStress= 2\,\eff{\mu}\,\mathbb{\varepsilon}^{(\text{mod})}\,.
\end{equation} 
Thereby, the non-physical evaporation-induced volumetric deformation of the interface region\added{, denoted as $\mathbb{\varepsilon}^{(v)}$,} is subtracted. It is by definition a deviatoric tensor (i.e., $\func{\Tr}{\mathbb{\varepsilon}^{\text{(mod)}}}=0$) and \replaced{accounts for the fact}{considers} that evaporative deformation only affects the \added{strain rate} component in interface normal direction \added{(resulting from the evaporation-induced velocity jump in normal direction)}. The evaporation-induced rate-of-deformation $\mathbb{\varepsilon}^{(v)}$ is only non-zero in the interface zone, where the evaporative dilation rate $\tilde{v}^{(lg)}\neq0$ and thus $\func{\Tr}{\mathbb{\varepsilon}}\neq0$ holds. 
Using this expression within the momentum equation \eqref{eq:momentum_balance} leads to successful elimination of spurious pressure artifacts in the interface region for viscous flows with evaporation, as demonstrated based on an analytical example in Section~\ref{sec:evapor_shell_analytical} and several numerical examples in Section~\ref{sec3}.

\subsubsection{Analytical demonstration example: Evaporating circular shell}
\label{sec:evapor_shell_analytical}
\begin{figure}[tb!]
	\centering
	\includegraphics{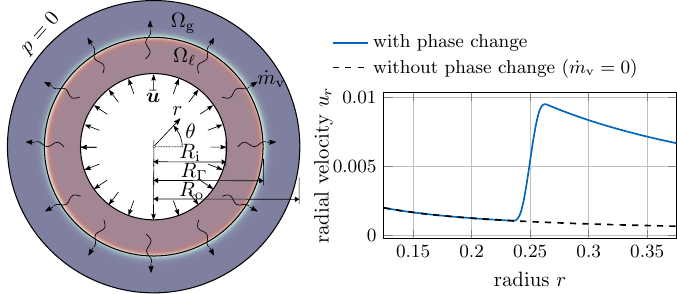}
	\caption{Circular shell subject to an axisymmetric evaporative mass flux $\mDot$: (left) geometry and boundary conditions; (right) analytical solution for the radial velocity component and comparison with the solution obtained \emph{without} phase change. The parameters are chosen as (SI units): $R_\*i=0.125$, $R_\Gamma=0.25$, $R_\*o=0.375$, $\rhoL=10$, $\rhoG=1$, $\mDot=0.01$, $\epsilon=\num{5e-3}$. A low density ratio is utilized on purpose to retain fine details in the region where $r<0.25$.}
	\label{fig:const_law}
\end{figure}

\noindent For an illustrative derivation of the proposed \replaced{corrected}{modified} viscous stress formulation in \myeqrefsa{eq:modified_rate_of_deformation}{eq:modified_viscous_stress} and without losing generality, we consider a circular shell under axisymmetric conditions as depicted in the left panel of Fig.~\ref{fig:const_law}, parametrized in 2D by the radius $r$ and the angular coordinate~$\theta$. The shell has an inner radius of $R_\*i$ and an outer radius of $R_\*o$.  
At the interface liquid material evaporates with a spatially and temporally constant evaporative mass flux~$\mDot$. Simultaneously, the evaporated volume is compensated by a prescribed inflow velocity on the liquid side of the interface
\begin{equation}
	u_r(r=R_\*i)=\bar{u}=\frac{\mDot}{\rhoL}\,\frac{R_\Gamma}{R_\*i}\,
\end{equation}
 to yield a spatially fixed interface location.
Considering the axisymmetry of the system, the velocity field and other field quantities do not depend on the angular coordinate $\theta$. Furthermore, the circumferential velocity $u_\theta(r,\theta)$ is zero throughout the domain. The only non-zero velocity component is the radial velocity component $u_r(r)$, which can be expressed in terms of the cylindrical coordinate system as:
\begin{equation}
	\Bu = \begin{bmatrix}u_r(r) \\ 0\end{bmatrix}\,.
\end{equation} 
For this example, an analytical solution for the radial velocity (and the pressure) is derived considering the present diffuse model, detailed  in \autoref{app:evapor_shell}:
\begin{equation}
	\label{eq:evapor_shell_radial_velocity}
	\evaporShellVelocity\,.
\end{equation}
It is illustrated in the right panel of Fig.~\ref{fig:const_law} for exemplary parameter values. The rate-of-deformation tensor \eqref{eq:rate_of_deformation} computed from this velocity field and expressed in cylindrical coordinates reads as
\begin{equation}
	\boldsymbol{\varepsilon} = 
	\begin{bmatrix}
		\fracPartial{u_r}{r} & \frac{1}{2}\left(\cancelto{0}{\fracPartial{u_\theta}{r}} - \cancelto{0}{\frac{u_\theta}{r}} + \frac{1}{r} \, \cancelto{0}{\fracPartial{u_r}{\theta}}\right) \\
		\text{sym.} & \frac{1}{r} \, \cancelto{0}{\fracPartial{u_\theta}{\theta}} + \frac{u_r}{r} 
	\end{bmatrix} =  
\begin{bmatrix}
	\fracPartial{u_r}{r} & 0 \\
	0 & \frac{u_r}{r}
\end{bmatrix}\,.
\label{eq:2d_example_eps}
\end{equation}
It can be seen that insertion of \myeqref{eq:evapor_shell_radial_velocity} into the rate-of-deformation tensor \eqref{eq:2d_example_eps} yields a radial normal strain rate component $\varepsilon_{rr}=\added{\partial u_r/\partial r}$ which differs from $\varepsilon_{\theta\theta}=\added{u_r/r}$. Thus, the volumetric strain rate results to $\func{\Tr}{\boldsymbol{\varepsilon}}=\added{\partial u_r/r+u_r/r}\neq0$. Consideration of the latter in \myeqref{eq:stokesLaw} would induce a non-physical, evaporation-induced viscous stress.

In contrast, evaluation of \myeqref{eq:modified_rate_of_deformation} for the analytical solution provided in \myeqref{eq:evapor_shell_radial_velocity} yields
\begin{equation}
\evaporModStrain =
\begin{bmatrix}
	\fracPartial{u_r}{r} & 0 \\
	0 & \frac{u_r}{r}
\end{bmatrix} - 
\left(\fracPartial{u_r}{r}+\frac{u_r}{r}\right)\,
\begin{bmatrix}
	1 \\
	0 
\end{bmatrix} \,\cdot\begin{bmatrix}
	1 &
	0 
\end{bmatrix} = 
\begin{bmatrix}
	- \frac{u_r}{r} & 0 \\
	0 &  \frac{u_r}{r}
\end{bmatrix},
\label{eq:2d_example_eps_eval}
\end{equation}
implying the desired purely deviatoric rate-of-deformation tensor, i.e., $\Tr(\evaporModStrain)=0$. Hence, by using the \replaced{corrected rate-of-deformation tensor to compute}{modified definition of} viscous stresses \eqref{eq:modified_rate_of_deformation}-\eqref{eq:modified_viscous_stress}, artificial evaporation-induced viscous stress contributions are reduced.

\begin{remark}
Recalling the considered example of the circular shell  (cf. right panel of Fig.~\ref{fig:const_law}) for incompressible two-phase flow without phase change, the analytical solution for the radial velocity component can be obtained from an analytical solution of the continuity equation as 
\begin{equation}
	u_r(r) =\frac{R_i}{r}\, \bar{u}.
\end{equation}
It is illustrated in the right panel of Fig.~\ref{fig:const_law} as the black, dashed line. 
Here, the rate-of-deformation tensor \eqref{eq:2d_example_eps} is calculated as 
\begin{equation}
	\boldsymbol{\varepsilon} = 
\begin{bmatrix}
	-\bar{u}\frac{R_i}{r^2}& 0 \\
	0 & \bar{u}\frac{R_i}{r^2}
\end{bmatrix}
\equiv
\begin{bmatrix}
	-\frac{u_r}{r}& 0 \\
	0 & \frac{u_r}{r}
\end{bmatrix}
\end{equation}
which is equal to \myeqref{eq:2d_example_eps_eval} and represents a purely deviatoric tensor.
\end{remark}

\subsection{Numerical framework}
The governing partial differential equations, i.e., \myeqrefs{eq:continuity}{eq:momentum_balance} and \myeqref{eq:transport}, as well as the additional equations for the level-set framework consisting of the reinitialization~\eqref{eq:reinitialization}, the filtered normal \eqref{eq:normal_vector_regularized}  and the filtered curvature \eqref{eq:curvature_regularized} are discretized in space using continuous finite elements based on Lagrange polynomials as test and trial functions. The resulting weak form for  \myeqrefs{eq:continuity}{eq:momentum_balance} and \myeqref{eq:transport} and additional notes on the discretization are presented in Appendix~\ref{app:weak_form}.
Finite element discretizations of transport terms, such as present in the governing equations, would typically require stabilization schemes at higher Reynolds numbers. We employ no \added{distinct }stabilization since the Reynolds numbers considered in this publication are moderate and potential oscillations in the level set field are flattened by the reinitialization\added{, where diffusion in the direction normal to the interface is employed (see \myeqref{eq:reinitialization})}.
The polynomial degree $k$
of the test and trial functions for the velocity field is $k_u=2$ while it is $k_p=1$ for the pressure field to ensure inf-sup stability. For the level-set field, we consider $k_\phi=1$. For the transport and reinitialization equation of the level-set field, the filtered normal vector and curvature calculation, we employ a refined mesh by subdividing it $\nsub$ times, in the spirit of \cite{Kronbichler18multiphase}. If not stated otherwise, we choose $\nsub=2$ leading to a level-set mesh to be a factor of two finer compared to the one of the Navier--Stokes equations. In order to avoid a mismatch in pressure space with the level-set space, we employ an interpolation of the level-set function onto the pressure space before evaluating the surface tension force \cite{zahedi2012spurious,Kronbichler18multiphase}.
For evaluating the integrals of the weak form, we consider numerical integration by evaluation at $(k_i+1)^\*{dim}$, with $i \in \{p, u, \phi\}$, Gaussian quadrature points.

For time integration, \allowbreak (semi-)implicit time stepping schemes are used. The coupled system of equations is solved based on operator splitting considering a weakly partitioned solution scheme, introducing an explicit (time lag) scheme between the
equations as outlined in Algorithm~\ref{algo:meltpooldg}. 
Thus, each of the fields is propagated fully implicitly, but the coupling terms, i.e., evaporative dilation rate and surface tension force are treated explicitly, which introduces a time-step limit. For computing the latter, we consider the capillary time-step limit according to \cite{brackbill1992}
\begin{equation}
	\Delta t_{\max} \leq \sqrt{\frac{\left(\liqP{\rho}+\gasP{\rho}\right)\,l_{\min}^{3}}{4\pi\alpha}}
\end{equation}
where $l_{\min}$ is the minimum edge length. It is noted that the time step limit could also be affected by the explicit treatment of the evaporative dilation rate, but in the absence of a detailed study of the latter, we estimate it empirically by trial and error in the following studies.

To allow for a high spatial resolution of the interface region, adaptive meshing schemes are considered. \added{In every time step, we assess whether refinement is necessary. If a cell located within 3.5 layers of the interface (i.e., if $-\log\left({\underset{K}{\max}{|\nabla \phi|\,\epsilon}}\right)<3.5$ holds) is not at the maximum refinement level, adaptive mesh refinement is performed. In such cases, we follow the remeshing strategy presented in \cite{Kronbichler18multiphase}: all cells within four layers of the interface (i.e., if $-\log\left({\underset{K}{\max}{|\nabla \phi|\,\epsilon}}\right)<4$ holds) or seven layers biased towards the flow direction (i.e., if $-\log\left({\underset{K}{\max}{|\nabla \phi|\,\epsilon}}\right)-\Delta t \left(\Bu\cdot\nabla\phi\right)/\left(\epsilon\,|\nabla \phi|\right)<7$ holds) are refined. 
The second criterion introduces an additional layer of roughly three cells in the downstream direction, which extends the duration for a valid mesh and consequently lowers the need for frequent remeshing.
}

To reduce the time for the matrix-vector product within the iterative solvers for the linear systems of equations, highly efficient matrix-free algorithms \added{as presented in \cite{kronbichler2012generic, Kronbichler2019}} are used for each field. \added{Matrix-free operator evaluation ensures high	node-level performance, in line with current trends in exascale finite-element algorithms described in \cite{kolev2021efficient}. This approach allows for the matrix-free evaluation of the operator action $\By = A\Bx$, computing the integrals underlying a finite-element discretization on the fly.	Specifically, it involves a loop over all cells, applying the element stiffness matrix on a vector restricted to the unknowns (degrees of freedom) of the cell \cite{kronbichler2012generic}, i.e., $\By=A(\Bx)=\sum_i R_i^T A_i R_i$.	For example for quadratic finite elements and systems of partial differential equations such as the system matrix of the incompressible Navier--Stokes equations linearized by a Newton method, the matrix-free kernels can be up to ten times as fast as matrix-based kernels \cite{Kronbichler18multiphase}.}

The \replaced{frameworks for adaptive mesh refinement and matrix-free operator evaluation}{latter two} are used from the open-source finite element package \texttt{deal.II} \cite{arndt2022deal} together with available parallelized MPI-based implementations using domain decomposition. In addition, we use and extend the open-source incompressible Navier--Stokes solver \texttt{adaflo} \cite{Kronbichler18multiphase}. As outlined in Algorithm~\ref{algo:meltpooldg}, for this purpose, we provide variable material properties and additional \replaced{right-hand side terms}{fluxes} to the Navier--Stokes solver.


For the solution of the linear equation systems, we use iterative solvers based on preconditioned Krylov subspace methods, i.e., the conjugate-gradient (CG) solver for symmetric systems and the generalized minimal residual method (GMRES) solver for non-symmetric systems \cite{saad2003iterative}. A summary on the linear solver settings can be found in \autoref{tab:solver_settings}. \replaced{Solving the fully coupled block system of the Navier--Stokes equations involves a saddle point structure \cite{benzi2005numerical}. To adress this, we employ  a  block-triangular preconditioner with an incomplete LU decomposition (ILU) for the velocity block and with the Cahout--Chabard approximation \cite{cahouet1988some} of the Schur complement as described in \cite{Kronbichler18multiphase}.}{As described in \cite{Kronbichler18multiphase}, for solving the fully coupled block system of the Navier--Stokes equations, which is of a saddle point structure \cite{benzi2005numerical}, we employ  a  block-triangular preconditioner with an incomplete LU decomposition (ILU) for the velocity block and with the Cahout--Chabard approximation \cite{cahouet1988some} of the Schur complement.}


\begin{algorithm}[H]
	\caption{Overall solution algorithm of the incompressible two-phase flow with evaporation framework.}\label{algo:meltpooldg}
	$\Bu\gets\Bu^{(0)},p\gets p^{(0)},\phi\gets \phi^{(0)}$ \Comment*[r]{set initial conditions at $t=t^{(0)}$}
	\For{$t\gets t^{\mathrm{(start)}}$ \KwTo $t^{\mathrm{(end)}}$}{
		\tcc{\underline{step 1:~compute predictors of the primary variables}}
		$\Bu, p, \phi\gets \texttt{extrapolate}(\Bu^{(n)}, \; p^{(n)}, \; \phi^{(n)},\;\Bu^{(n-1)}, \; p^{(n-1)}, \; \phi^{(n-1)}, \; ...)$\;
		\tcc{\underline{step 2:~compute level-set and geometric quantities}}
		$\phi,\;\nGamma,\;\kappa \gets \texttt{level\_set\_solver}(\phi,\;\uGamma, \mDot)$  (cf. Algorithm\,\ref{algo:ls})\;
		\tcc{\underline{step 3:~solve incompressible Navier--Stokes equations}}
		$\eff{\rho}\gets\rho(\phi), \eff{\mu}\gets\mu(\phi)$\Comment*[r]{compute effective material properties}
		$\evaporDilationRate \gets \evaporDilationRate (\mDot, \phi)$~\eqref{eq:evaporDil},\
		$\surfaceTensionForce \gets \surfaceTensionForce(\kappa,\nGamma,\phi)~
		\eqref{eq:surfaceTension}$ \\\hfill\Comment{compute interface \replaced{contributions}{fluxes}: evap. dilation, surface tension}
		$\Bu,p \gets \texttt{navier\_stokes\_solver}(\eff{\rho},\eff{\mu}, \surfaceTensionForce, \evaporDilationRate)$ (cf. algorithm in \cite{Kronbichler18multiphase})
	}
\end{algorithm}

\begin{table}[htbp!]
	\small
	\caption{Default solver settings; either the absolute tolerance ($\normltwo{\BR}<$ \ATOL) or the relative tolerance ( $\normltwo{\BR}/\normltwo{\BR^{(0)}}<$ \RTOL) need to be fulfilled. $\BR$ represents the residual of the current iteration and $\BR^{(0)}$ the initial residual.}
	\label{tab:solver_settings}
	\begin{tabular}{p{3.1cm}|cccc|cc}
		\toprule
		& \multicolumn{4}{c}{\emph{linear solver}} & \multicolumn{2}{c}{\emph{nonlinear solver}}                                                \\
		subproblem                                                   & type                              & preconditioner                       & \ATOL      & \RTOL      & \ATOL      & \RTOL \\
		\toprule
		Navier--Stokes equations \eqref{eq:continuity}-\eqref{eq:momentum_balance} & GMRES                             & ILU+Schur                                  & --         & \num{e-4}  & \num{e-10} & --    \\
		\midrule
		level-set advection \eqref{eq:transport}                                & GMRES                             & diagonal                             & \num{e-20} & \num{e-12} & --         & --    \\
		\midrule
		reinitialization \eqref{eq:reinitialization} & \multirow{3}{*}{CG} & \multirow{3}{*}{diagonal}                             & \multirow{3}{*}{\num{e-20}} & \multirow{3}{*}{\num{e-12}} &    \multirow{3}{*}{--}        & \multirow{3}{*}{--}   \\
		normal vector \eqref{eq:normal_vector_regularized} & & & & & & \\
		curvature \eqref{eq:curvature_regularized}  & & & & & & \\
		\bottomrule
	\end{tabular}
\end{table}

\section{Results}\label{sec3}
In the following, several benchmark examples are computed to evaluate the strengths and weaknesses of the diffuse framework for two-phase flow with evaporative phase change presented in Section~\ref{sec2}. If units are omitted in this section, they are assumed to correspond to SI standards, i.e., kg, m, s, K. As stated in the introduction, the focus of this contribution is to accurately predict the movement of the liquid surface for rapid evaporation. 
\replaced{
The analytical study in Section~\ref{sec:uGamma_evaluation} showed that \emph{variant 1} and \emph{variant 2} are the most promising for modeling interface movement under evaporation. Therefore, we exclude \emph{variant 3} from the subsequent numerical study to keep the study concise.
}{
Since it turned out from the analytical study regarding the level set transport velocity in Section~\ref{sec:uGamma_evaluation} that \emph{variant 1} and \emph{variant 2} are the most promising for modeling the interface movement subject to evaporation, we drop \emph{variant 3} for the subsequent numerical study to keep the study concise.} For the evaporative mass flux $\mDot$ (\si{kg/(m^2\,s)}), we prescribe an analytical function to mimic isothermal conditions. For every investigated example, an analytical solution exists for verification. 

\subsection{One-dimensional phase change}
\label{sec:1D_phase_change}

\begin{figure}[bt!]
	\centering
	\includegraphics{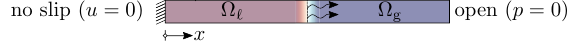}
	\caption{Illustration of the one-dimensional phase change problem.}
	\label{fig:stefans_problem_1d}
\end{figure}
\noindent
In this example, the behavior of a flat liquid surface subject to evaporation is analyzed. Thereto, a one-dimensional (1D) domain is considered with a prescribed spatially and temporally constant evaporation flux of $\mDot=\num{0.01}$, illustrated in Fig.~\ref{fig:stefans_problem_1d}. A similar example was considered e.g. in  \cite{lee2017direct}.
The domain $\Omega=x\in\left[0,1\right]$ is occupied with a liquid (left half) and a gaseous phase (right half), characterized by an initial position of the discrete interface at $x(\phi=0, t=0)\equiv x_\Gamma^{(0)}=0.5$ and the interface thickness parameter $\epsilon=0.02$. The fluid is initially at rest ($u^{(0)}=0$). Homogeneous Dirichlet boundary conditions for the velocity along the left domain boundary ($u(x=0)=0$) and an outlet boundary condition is assumed along the right domain boundary ($p(x=1)=0$). A uniform mesh with an element length of approx. 0.008 is employed. Considering a refined mesh for the level-set framework by subdividing it $\nsub=2$ times, this results in a resolution of the interface region by approx. 30 elements for the level-set field. The simulation is performed for the time period $0\leq t \leq 1$ with a constant time step size of $\num{5e-4}$. The parameters for the phase densities are specified as $\liqP{\rho}=1$ and $\gasP{\rho}=\num{e-3}$. Gravity forces are neglected.

	The analytical solution for this example is described in Appendix~\ref{app:evapor_stefan} for both the sharp and the diffuse model. For discussing the influence of the viscous stress tensor (cf. Section~\ref{sec:modified_viscous_stress}), in the following, we present simulation results considering two different rheology types of fluids, i.e., a quasi-inviscid fluid and a viscous Newtonian fluid.

\subsubsection{Quasi-inviscid fluid}
\begin{figure}[tb!]
	\includegraphics{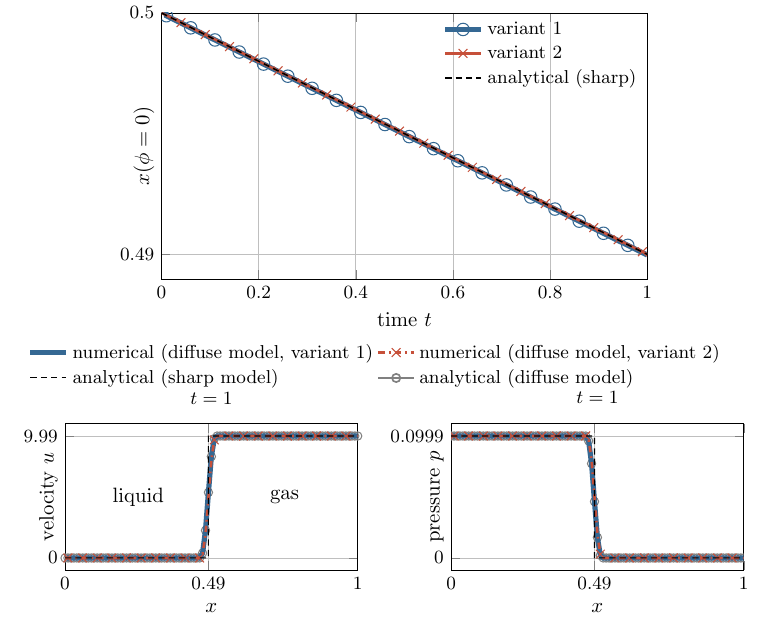}
	\caption{One-dimensional phase change problem subject to a spatially and constant evaporation flux (cf. Fig.~\ref{fig:stefans_problem_1d}) for a \emph{quasi-inviscid} fluid: (top) temporal movement of the discrete interface ($x(\phi=0)$); Both considered \emph{variants 1 and 2} for computing the level-set transport velocity yield identical results corresponding to the sharp reference solution; (bottom left) velocity and  (bottom right) pressure at the end of the simulation. The perfect agreement between the numerical results, the analytical solution of the diffuse model and the one of the sharp model outside the interface region verifies the numerical framework.}
	\label{fig:stefan_1d_inviscid_xGamma}
\end{figure}

\begin{figure}[tb!]
\centering
\includegraphics{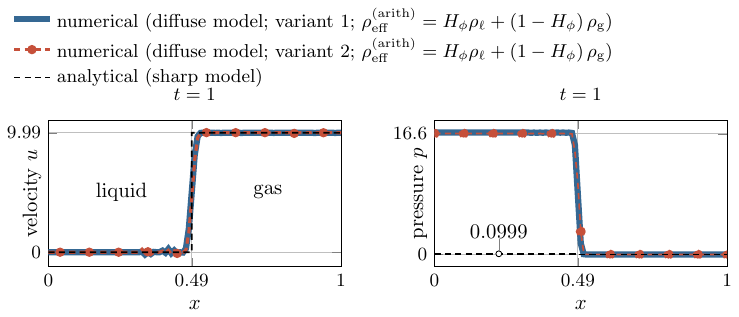} 
\caption{One-dimensional phase change problem subject to a spatially and constant evaporation flux (cf. Fig.~\ref{fig:stefans_problem_1d}) for a \emph{quasi-inviscid} fluid: Numerical solution for the velocity (left) and the pressure (right) at the end of the simulation, considering the \emph{effective density  in the momentum equation \eqref{eq:momentum_balance} as an arithmetic phase-weighted average}  $\rhoEff^{(\text{arith})} = H_\phi\liqP{\rho}+(1-H_\phi)\,\gasP{\rho}$. Due to the inconsistent interpolation function of the density between the two phases used for demonstration purposes, there is a significant discrepancy between the numerically predicted pressure and the reference solution in the liquid domain.
	}
\label{fig:stefan_1d_inviscid_profiles_density}
\end{figure}

First, we consider a quasi-inviscid fluid (realized by setting $\muEff=\num{e-10}$), for which the \deleted{standard} Stokes' law \added{without correction term} \eqref{eq:stokesLaw} holds and the modification of the viscous stress proposed in Section~\ref{sec:const_law} is not needed. 
According to Fig.~\ref{fig:stefan_1d_inviscid_xGamma}, the movement of the liquid surface is accurately modeled for both considered variants of the level-set transport velocity. The numerically predicted velocity and pressure, illustrated in the bottom panel of Fig.~\ref{fig:stefan_1d_inviscid_xGamma}, are in perfect agreement with the analytical solution of the diffuse model and coincide with the sharp model outside the interface region. It is stressed that the reciprocal density interpolation \eqref{eq:rhoEff} is mandatory for the diffuse framework to predict the correct solution for the pressure difference between the two phases, irrespective of the chosen level-set transport velocity variant. 
\replaced{For example, if the effective density in the momentum equation \eqref{eq:momentum_balance} were calculated according to an arithmetic phase-weighted average, the pressure would deviate significantly from the reference solution. This standard approach for two-phase flow without phase change would overestimate the pressure in the liquid phase by a factor of 166, as shown in Fig.~\ref{fig:stefan_1d_inviscid_profiles_density}.}{For example, if the effective density in the momentum equation \eqref{eq:momentum_balance} were calculated according to an arithmetic phase-weighted average, which is the standard approach for two-phase flow without phase change, this would result in a significant deviation of the predicted pressure compared to the reference solution, overestimating the pressure by a factor of 166 as shown in Fig.~\ref{fig:stefan_1d_inviscid_profiles_density}.}


\subsubsection{Viscous fluid}
\label{sec:stefans_problem_viscid}

\begin{figure}[tb!]
	\includegraphics{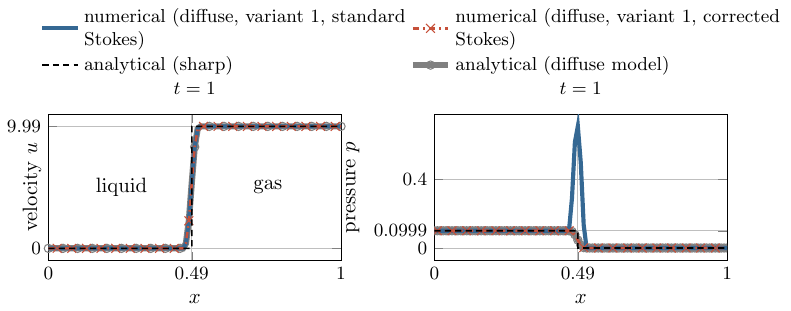}
	\caption{One-dimensional phase change problem subject to a spatially and constant evaporation flux (cf. Fig.~\ref{fig:stefans_problem_1d}) for a \emph{viscous} fluid: Numerical vs. analytical solution for the velocity (left) and the pressure (right) at the end of the simulation ($t=1$). The results for the velocity are not affected by the viscosity and resemble the one of Fig.~\ref{fig:stefan_1d_inviscid_xGamma}. However, the pressure profile obtained by \replaced{the}{a standard} Stokes' constitutive relation \added{using the uncorrected rate-of-deformation tensor} exhibits a significant peak in the interface zone (blue curve). The latter can be avoided by \replaced{using a correction term within the rate-of-deformation tensor}{a modified Stokes' relation} according to \myeqref{eq:modified_rate_of_deformation} (red curve).}
	\label{fig:stefan_1d_viscid_profiles}
\end{figure}

Next, we analyze a viscous fluid with a dynamic viscosity of $\liqP{\mu}=\gasP{\mu}=\num{e-3}$. We choose the viscosities to be equal between the phases in order to study the artificial evaporation-induced pressure jump in an isolated manner.
We expect viscosity not to have an influence on the results due to the 1D nature of the problem. For the present study, we consider two \replaced{variants}{types} \deleted{of constitutive formulations} for the calculation of the viscous stress tensor in our diffuse model: 
\replaced{the Stokes' constitutive relation for incompressible flow based on (i) the uncorrected rate-of-deformation tensor \eqref{eq:rate_of_deformation} and \eqref{eq:stokesLaw} (denoted as \emph{standard Stokes}) and (ii) the corrected rate-of-deformation tensor \myeqref{eq:modified_rate_of_deformation} and \myeqref{eq:modified_viscous_stress} (denoted as \emph{corrected Stokes}) taking into account a subtraction of the evaporation-induced deformation. }
{(i) the standard Stokes' constitutive relation for incompressible flow based on the standard rate-of-deformation tensor \eqref{eq:stokesLaw} (denoted as \emph{standard Stokes}) and (ii) the proposed modified version of the Stokes' constitutive relation \eqref{eq:modified_rate_of_deformation} (denoted as \emph{modified Stokes}) taking into account a subtraction of the evaporation-induced deformation. }
According to the left panel of Fig.~\ref{fig:stefan_1d_viscid_profiles}, viscosity has no influence on the velocity profile. However, by analyzing the pressure profile in the right panel of Fig.~\ref{fig:stefan_1d_viscid_profiles}, it becomes apparent that the result in the interface zone is manifested by a significant pressure elevation due to the contribution of the evaporation-induced volumetric strain-rate to the viscous stress using the \replaced{uncorrected rate-of-deformation tensor in the}{standard} Stokes relation. The latter is remedied by \replaced{adding the correction term to the rate-of-deformation tensor  \eqref{eq:modified_rate_of_deformation} and using it in the viscous stress relation \eqref{eq:modified_viscous_stress}.}{using the modified Stokes relation \eqref{eq:modified_rate_of_deformation}.} 

\begin{figure}[bt!]
	\includegraphics{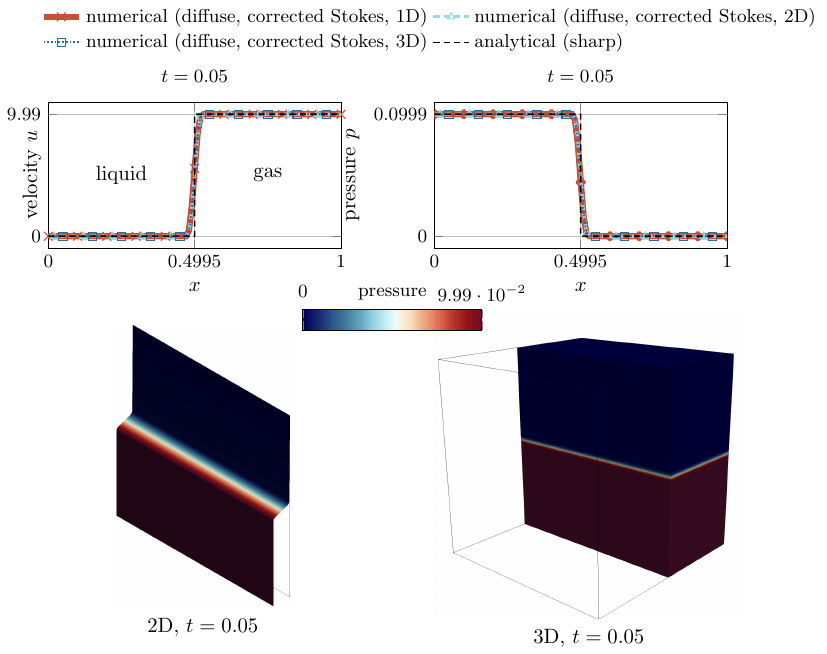}
	\caption{One-dimensional phase change problem subject to a spatially and constant evaporation flux (cf. Fig.~\ref{fig:stefans_problem_1d}) for a \emph{viscous} fluid considering \emph{variant 1}: Numerical vs. analytical solution for the velocity (left) and the pressure (right) at $t=0.05$. The results are computed considering a flat interface embedded in a 1D, 2D and 3D domain to investigate the influence of the spatial dimension on the pressure. For the 2D and 3D cases the values are evaluated along a line through the center that is normal to the interface. The agreement between the results confirms the general applicability of the viscous stress tensor \eqref{eq:modified_rate_of_deformation} to higher dimensions.}
	\label{fig:stefan_1d_2d_3d_viscid_profiles}
\end{figure}

Additional verification is performed by simulating the problem in 2D and 3D, shown in Figure~\ref{fig:stefan_1d_2d_3d_viscid_profiles}. It can be seen that the pressure, and for completeness the velocity, is identical to the purely one-dimensional case, which underlines the general applicability of the proposed modified Stokes' relation \eqref{eq:modified_rate_of_deformation} also to higher dimensions. \added{The total runtime for the 3D simulation (19496 degrees of freedom and 1000 time steps) was 1\,minute, using two cores of an AMD Ryzen Threadripper PRO 3995WX.}


\subsection{Evaporating droplet}
\label{sec:evaporating_droplet}

\begin{figure}[htbp!]
	\centering
	\includegraphics{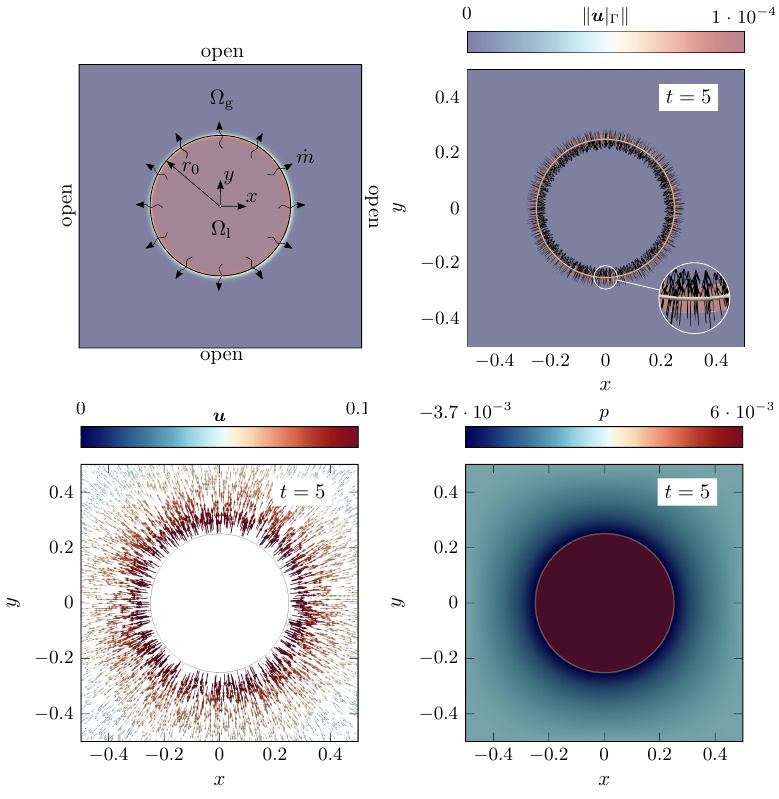}
	\caption{Evaporating droplet (\emph{quasi-inviscid},  \emph{variant 2}): (top left) problem setup; contour plots of (top right) the level-set transport velocity, (bottom left) the fluid velocity and (bottom right) the pressure, obtained at the final stage of the simulation.
	}
	\label{fig:evaporating_droplet}
\end{figure}

For verification of the proposed numerical framework for evaporation of highly curved surfaces, a circular droplet subject to a spatially and temporally constant evaporation flux $\mDot=\num{0.1}$ is simulated, illustrated in Fig.~\ref{fig:evaporating_droplet}\,(top left). A similar study was performed in \cite{lee2017direct,gibou2007level}. The domain $\Omega=[\num{-0.5},\num{0.5}]^2$ is occupied by a liquid droplet, characterized by the initial radius $r_0=\num{0.25}$ and the interface thickness parameter $\epsilon=\num{2e-3}$, embedded in a bulk vapor phase. The fluid is initially at rest ($\Bu^{(0)}=\boldsymbol{0}$). Along the domain boundary, outflow boundary conditions at zero pressure are assumed. In order to better resolve the interface domain, we employ adaptive mesh refinement with an element edge length between $\approx$ \num{0.0039} and \SI{0.0625}.
The simulation is performed for the time period $0\leq t \leq 5$ at a constant time step size of $\num{5e-3}$. The values for the phase densities are specified as $\liqP{\rho}=1000$ and $\gasP{\rho}=\num{1}$. The fluid is assumed to be quasi-inviscid. Surface tension and gravity forces are neglected for the sake of simplicity. 

The analytical solution for this example is derived from evaluating mass balance across the moving interface,  mentioned also in  \cite{lee2017direct}, resulting in the time derivative of the droplet radius $r$
\begin{equation}
	\frac{dr}{dt} = - \frac{\mDot}{\rho_l}\,.
\end{equation}

Similar to Section~\ref{sec:1D_phase_change}, for discussing the influence of the chosen type for the viscous stress tensor (cf. Section~\ref{sec:modified_viscous_stress}), we present simulation results considering a quasi-inviscid fluid and a viscous Newtonian fluid.

\begin{figure}[htbp!]
	\includegraphics{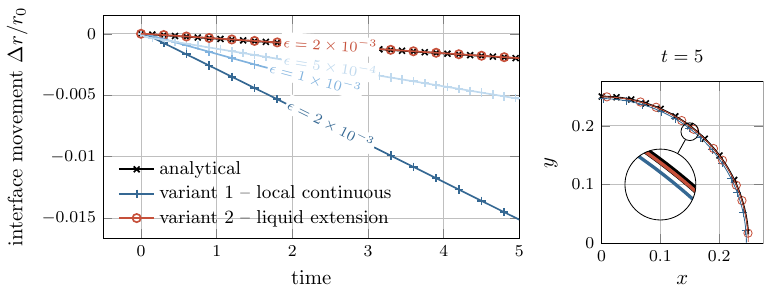}
\caption{Evaporating droplet (\emph{inviscid}): Relative movement of the interface over time considering \emph{variants 1} and \emph{2}. 
	It can be seen that \emph{variant 1}, i.e., the local modification of the fluid velocity, overestimates the interface movement and thus leads to violation of mass conservation while \emph{variant 2} is in excellent agreement with the analytical solution. 
}
\label{fig:evaporating_droplet_2}
\end{figure}

\begin{figure}[htbp!]  
\centering
\includegraphics{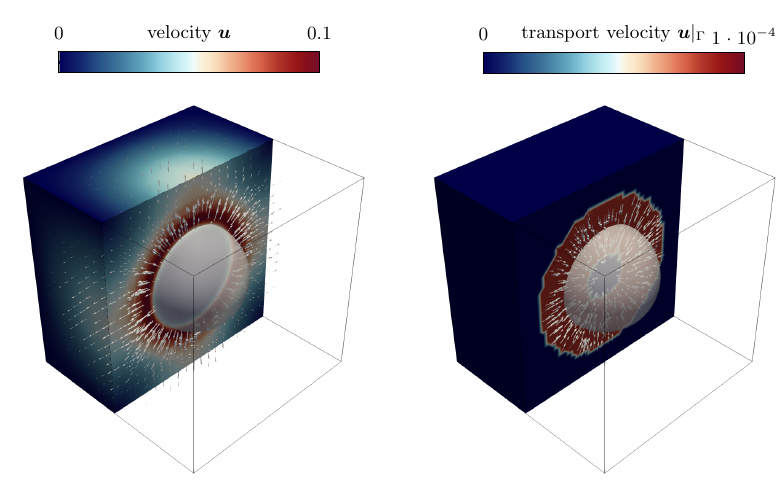}
\caption{Evaporating droplet (\emph{inviscid}, \emph{variant 2}): Results from a 3D simulation in the final stage of the simulation. The behavior resembles the one of the 2D simulation illustrated in Fig.~\ref{fig:evaporating_droplet}.}
\label{fig:evaporating_droplet_3D}
\end{figure}
	
\subsubsection{Quasi-inviscid fluid}
	
First, we consider a quasi-inviscid fluid (realized by setting $\muEff=\num{e-10}$), where the results obtained for a two-dimensional simulation are shown in Figs.~\ref{fig:evaporating_droplet}-\ref{fig:evaporating_droplet_2}.
In the left panel of Fig.~\ref{fig:evaporating_droplet_2}, the numerically predicted evolution of the relative movement of the interface is illustrated for \emph{variant 1} and \emph{variant 2} in comparison with the analytical solution. For the evaluation of the droplet radius from the numerical results, we performed an averaging over the droplet perimeter. Excellent agreement is obtained for \emph{variant 2}, while \emph{variant 1} overestimates the movement of the interface significantly\replaced{. This behavior}{, which} is in agreement with the analytical examples discussed in Section~\ref{sec:uGamma_evaluation} and shown in Figure~\ref{fig:benchmark_transport_velocity}. In addition, the zero-level-set isosurface at the final simulation time is shown in the right panel of Fig.~\ref{fig:evaporating_droplet_2}, where the overestimation of the droplet shrinkage becomes apparent.
 The resulting transport velocity according to \emph{variant 2} is illustrated in Fig.~\ref{fig:evaporating_droplet}\,(top right) for the final simulation time\replaced{. Here, }{, where} the \added{velocity} magnitude is constant over a narrow band around the interface and the vector points in radial direction into the droplet --- as expected. This investigation indicates that \emph{variant 2} is a promising candidate for accurate level-set transport in presence of evaporation for curved interfaces. For completeness, the velocity vectors and the pressure are shown in the bottom panels of Fig.~\ref{fig:evaporating_droplet}. It can be seen that the velocity is zero inside the droplet and increases significantly across the interface. This leads to an evaporation-induced pressure increase with a maximum value inside the droplet. 
 
For additional demonstration of the versatile applicability of the framework, the velocity field (left) and the transport velocity field (right) is shown for a 3D computation using \emph{variant 2} in Fig.~\ref{fig:evaporating_droplet_3D}. The results resemble the one of the 2D case.

\subsubsection{Viscous fluid}
\begin{figure}[htbp!]
	\includegraphics{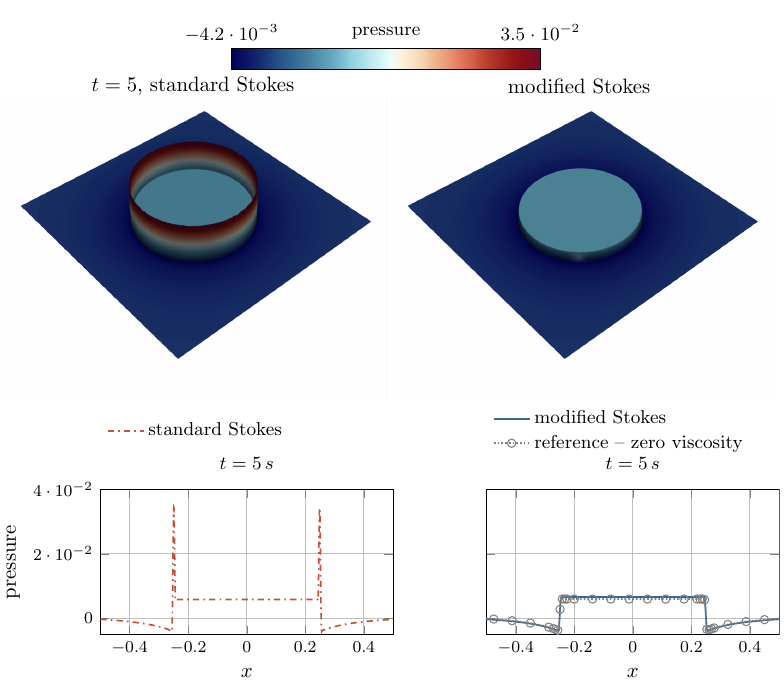}	
	\caption{Evaporating droplet (2D, \emph{viscid}, \emph{variant 2}): Pressure at the final simulation stage considering  the \deleted{standard} Stokes' constitutive relation (i)  \added{using the uncorrected rate-of-deformation tensor}~\eqref{eq:stokesLaw} (left column) and (ii) \added{using the corrected rate-of-deformation tensor} \eqref{eq:modified_viscous_stress} (right column). The first row shows contour plots of the pressure in a deformed state (pressure scale factor of 5). The second row shows pressure profile along the horizontal section through the center. It can be seen that the evaporation-induced  pressure elevation within the interface region can be avoided by using our proposed \replaced{evaporation-related correction of the rate-of-deformation tensor for computing viscous stress.}{modified Stokes' formulation.}}
	\label{fig:evaporating_droplet_2D_viscid_pressure}
\end{figure}

Next, we analyze a viscous fluid with a dynamic viscosity of $\liqP{\mu}=\gasP{\mu}=\num{e-3}$. We expect that viscosity should not have an influence on the results due to axisymmetry. 
For the present study, we consider two \replaced{variants}{types} \deleted{of constitutive formulations} for computing the stress tensor in our diffuse model: 
\replaced{the Stokes' constitutive relation for incompressible flow based on (i) the uncorrected rate-of-deformation tensor \eqref{eq:rate_of_deformation} and \eqref{eq:stokesLaw} (denoted as \emph{standard Stokes}) and (ii) the corrected rate-of-deformation tensor \myeqref{eq:modified_rate_of_deformation} and \eqref{eq:modified_viscous_stress} (denoted as \emph{corrected Stokes}) taking into account a subtraction of the evaporation-induced deformation. }
{(i) the Stokes' constitutive relation for incompressible flow based on the standard rate-of-deformation tensor \eqref{eq:stokesLaw} (denoted as \emph{standard Stokes}) and (ii) the modified version of the Stokes's constitutive relation taking into account a subtraction of the evaporation-induced deformation \added{proposed in }\eqref{eq:modified_rate_of_deformation} (denoted as \emph{modified Stokes}). }
Similar to the one-dimensional phase change case (cf. Section~\ref{sec:stefans_problem_viscid}) and according to Figure~\ref{fig:evaporating_droplet_2D_viscid_pressure} (left column) the interface region exhibits a significant pressure elevation without employing the evaporation correction of the deformation for the viscous stress. By using the \replaced{corrected rate-of-deformation tensor \eqref{eq:modified_rate_of_deformation} for computing viscous stress}{proposed modified formulation of the Stokes's constitutive relation  }, this is avoided as shown in Figure~\ref{fig:evaporating_droplet_2D_viscid_pressure} (right column).

\subsubsection{Remark on the computational effort}

\added{Table~\ref{tab:evapor_droplet_comp_effort} summarizes the computational effort split into the main tasks for the simulation of the evaporating droplet case, which represents the largest simulation of this contribution. It is evident that evaluating the level-set transport velocity in variant 2 consumes a substantial portion of the total computation time, especially in 3D. Thus, performance optimization of this task is necessary in view of large-scale simulations.}

\begin{table}[htbp]
	\caption{\added{Evaporating droplet (viscid): Distribution of the computational effort using 60 cores of AMD Ryzen Threadripper PRO 3995WX.}}
	\label{tab:evapor_droplet_comp_effort}
	\begin{tabular}{p{6cm}||c|c||c|c} 
		\toprule
		\multicolumn{5}{c}{Evaporating droplet, viscid}\\
		& \multicolumn{2}{c}{2D} & \multicolumn{2}{c}{3D}\\
		& variant 1 & variant 2 & variant 1 & variant 2 \\
		\midrule
		overall run time (60 cores) & \SI{7}{min} & \SI{10}{min} & \SI{6}{min} & \SI{1}{day}\\
		average DoFs ($\phi + p + \Bu$) & \num{447300} & \num{411977} &  \num{1890106} &  \num{2733955} \\
		total number of time steps & 1000 & 1000 &500 & 500 \\
		\midrule
		\midrule
		task & \multicolumn{4}{c}{relative effort} \\
		\midrule
		level set (advection, reinitialization, normal vector, curvature) &  44\,\% & 27\,\% & 28\,\%  &  $< 1\,\%$ \\
		level-set transport velocity (note: not optimized!) & $< 1\,\%$ & 31\,\% & 2\,\% & 99\,\%\\
		Navier-Stokes & 53\,\%& 40\,\% & 62\,\% &  $< 1\,\%$  \\
		adaptive meshing & $< 1$\,\%& $< 1\,\%$ & $< 1\,\%$ &  $< 1\,\%$ \\
		other (output, initial conditions, etc.) & $< 1$\,\% & $< 1$\,\% & $8\,\%$ & $< 1\,\%$ \\
		\bottomrule
	\end{tabular}
\end{table}

\subsection{Evaporating circular shell}
\label{sec:evaporating_circular_shell}

The examples shown previously are characterized by zero velocity in the liquid phase. In the following, we present a \replaced{new}{novel} benchmark example that allows us to evaluate the accuracy of our framework for curved surfaces in presence of fluid velocities in both phases --- the liquid and the vapor phase. This enables to mimic the typical situation of practically relevant problem types, such as melt pool dynamics of \pbfam{}. For the chosen setup we derive an analytical solution, which is presented in \autoref{app:evapor_shell}.

\begin{figure}[htbp!]
 	\includegraphics{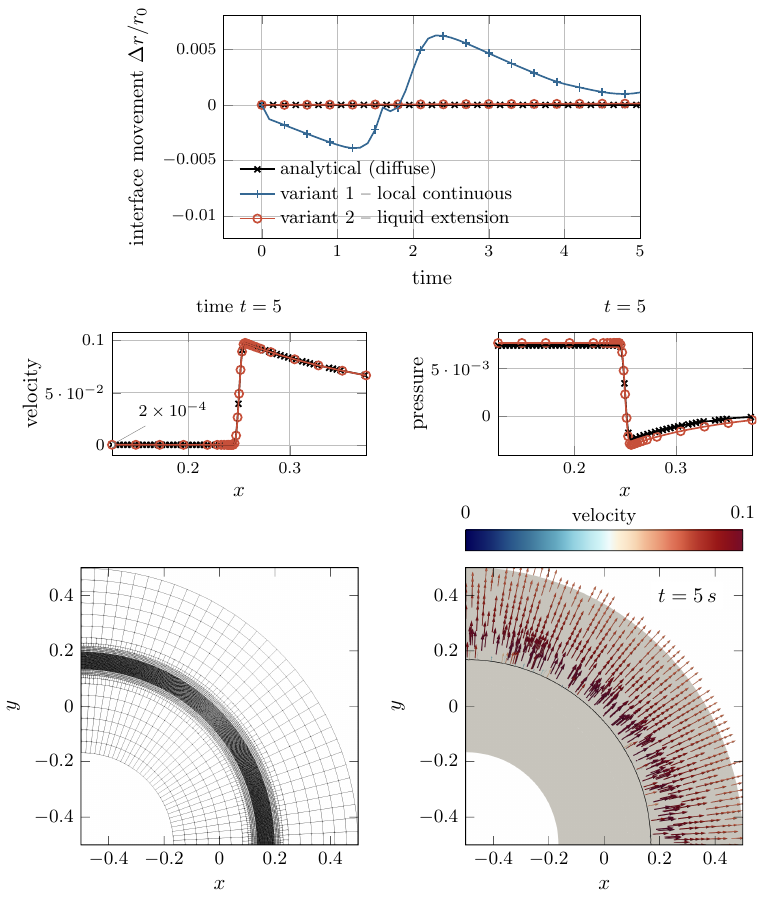}
 	\caption{Evaporating circular shell: (top) relative movement of the interface over time considering \emph{variant 1} and \emph{variant 2}; (center) velocity and pressure profile at the final simulation stage (only \emph{variant 2}); (bottom left) employed adaptively refined finite element mesh and (bottom right) resulting fluid velocity field shown as a vector plot (scale factor of 5) for \emph{variant 2}. Again, \emph{variant 2} yields a good agreement with the analytical solution.}
 	\label{fig:evapor_shell_2D_results}
 \end{figure}

We revisit the circular shell geometry described in Section~\ref{sec:evapor_shell_analytical} and illustrated in Fig.~\ref{fig:const_law} (left).  The domain $\Omega$ is described by a radius of the interior face $R_\*i=\num{0.125}$ and the exterior face $R_\*o=3\,R_\*i=\num{0.375}$. The initial liquid-vapor interface is positioned at $R_\Gamma=2\,R_\*i=\num{0.25}$.
The liquid surface is subject to a spatially and temporally constant evaporation flux $\mDot=\num{0.1}$. The inflow velocity at the interior boundary is chosen as  $\bar{u}=\mDot\,R_\Gamma/(\rho_l\,R_\*i)=\num{2e-5}$. This should balance the evaporated volume of the liquid phase and should prohibit the movement of the interface according to the analytical solution of the problem. The fluid is initially at rest ($\Bu^{(0)}=\boldsymbol{0}$). The initial level-set function is characterized by an interface thickness parameter $\epsilon=\num{2e-3}$. 
 Along the exterior domain boundary, outflow boundary conditions at zero pressure are assumed. In order to better resolve the interface domain, we employ adaptive mesh refinement with an element edge length between $\approx$ \num{3.068e-3} and \num{3.834e-4} in circumferential direction and $\approx$ \num{4.883e-4} and \num{3.906e-3} in radial direction (illustrated in  Figure~\ref{fig:evapor_shell_2D_results} bottom left). The simulation is performed for the time period $0\leq t \leq 1$ at a constant time step size of $\num{1e-3}$.  The material parameters comply with the evaporating droplet example of Section~\ref{sec:evaporating_droplet}.
 
 The results are shown in Figure~\ref{fig:evapor_shell_2D_results}. The relative interface movement for the investigated approaches of the level-set transport velocity \emph{variant 1} and \emph{variant 2} is plotted in the top panel of Figure~\ref{fig:evapor_shell_2D_results}. It can be seen that the results obtained with the level-set transport velocity according \emph{variant 2} are in good agreement with the analytical solution, while for \emph{variant 1} the undesirable motion of the interface is larger. The velocity and pressure profiles at the final simulation stage are depicted in the central panel of Figure~\ref{fig:evapor_shell_2D_results}\replaced{. The results}{, which} are in good agreement with the analytical solution for the diffuse model, presented in \myeqrefsa{eq:shell_u_analytical}{eq:shell_p_analytical}. This demonstrates the applicability of this method also to typical velocity scenarios for evaporative phase change, where the velocity in the liquid phase is non-zero but relatively small compared to the one in the vapor phase. For completeness, the velocity vectors are shown in the bottom right panel of Fig.~\ref{fig:evapor_shell_2D_results}. It should be noted that the velocity in the liquid part of the shell is so small (\num{2e-4}) compared to the one in the vapor part that there are no vectors visible in the liquid area. Again, it can be seen that the velocity increases significantly across the interface from the liquid to the gas phase.
 
 \added{ The total runtime for the simulation using variant 2 (126051 degrees of freedom and 1000 time steps) was 5\,minutes, using 60 cores of an AMD Ryzen Threadripper PRO 3995WX.}

\section{Conclusion}\label{sec4}
We have presented a \deleted{novel} mathematically consistent and robust diffuse-interface model for two-phase flow problems involving rapid evaporation. The model combines an incompressible Navier--Stokes solver with a conservative level-set formulation, and enhances it by a regularized representation of evaporation-induced discontinuities for ensuring robustness. The numerical discretization and high-performance solution approach utilizes a matrix-free adaptive finite element framework based on the open-source finite element library \texttt{deal.II}~\cite{arndt2022deal}, taking advantage of its adaptive mesh refinement and distributed point evaluation capabilities, as well as its matrix-free framework and a related incompressible Navier--Stokes solver~\cite{Kronbichler18multiphase}. To address the associated challenges of rapid evaporation, high density ratios, velocity jumps, and complex interface geometries including topological changes, we have made three major contributions to this research field.

 First, we have proposed mathematically consistent level-set transport velocity formulations particularly suitable for diffuse evaporation-induced velocity jump conditions, aiming at an accurate prediction of the evaporated mass.
 	Specifically, we have investigated two different variants based on an evaporation-dependent modification (i)~of the local fluid velocity and, alternatively, (ii) of the extension of the fluid velocity from the liquid or gas phase to the diffuse interface region via closest point projection. 
 	\replaced{
 		While approach~(ii) requires a higher numerical effort, it has been shown to offer greater accuracy. This is based on several analytical and numerical benchmarks. Specifically, the extension of the fluid velocity from the liquid phase, provides higher accuracy for a given interface thickness compared to approach~(i).}{
 	While the numerical effort is higher for approach~(ii), it has been demonstrated based on several analytical and numerical benchmarks that approach~(ii), especially considering the extension of the fluid velocity from the liquid phase, provides a higher accuracy for a given interface thickness compared to approach~(i).}
 	Approach~(i) requires a small interface thickness  to  curvature radius ratio, which is computationally expensive due to fine spatial discretization. Hence, we recommend using approach (ii) with liquid extension velocity for a better trade-off between accuracy and computational cost.
 	
 	Second, we show that accurate prediction of the evaporation-induced pressure jump requires a consistent, namely a reciprocal, density interpolation across the interface, which satisfies local mass conservation.
 	\replaced{Third, we have proposed a correction term for the Stokes-type constitutive relation in evaporating viscous two-phase flows. It neglects the contribution of the non-physical evaporation-induced volumetric deformation rate across the interface region to the viscous stress tensor.}
 	{Third, we have proposed a modified Stokes-type constitutive relation for evaporating viscous two-phase flows that neglects the contribution of the non-physical evaporation-induced volumetric deformation rate across the interface region to the viscous stress tensor.} This \deleted{novel} approach allows for the effective elimination of spurious pressure artifacts in the interface region, an issue that --- to the best of our knowledge --- has not been addressed in the literature.

In summary, this work has laid important groundwork 
 for the diffuse modeling of two-phase flows with rapid evaporation, which may be of interest for many types of engineering applications. 
 	We successfully verified our methods against various benchmarks, including scenarios with curved interfaces subject to rapid evaporation and high density contrast. In addition to well-established benchmark examples, we also proposed a \replaced{new}{novel} benchmark test including the derivation of an analytical solution. It represents a more general flow problem and is therefore closer to practical application scenarios than the aforementioned existing benchmarks. While this study focuses primarily on isothermal conditions to isolate evaporation-induced effects on the flow field, the extension to anisothermal conditions via incorporation of the heat transfer is possible and is part of our future work. As such, it will become an important building block of a high-fidelity thermal-multiphase flow model for the study of melt-vapor interactions in laser-based powder bed fusion of metals.

\section*{Declarations}
\bmhead{Author contributions}
MS and CM contributed to the derivation of model equations. MS was responsible for the specific code implementation and the numerical studies. PM and NM supported the
implementation. In addition, PM and MK contributed general-purpose functionality to this project via the \texttt{deal.II} library and the \texttt{adaflo} project. MS, CM, and WAW worked out the general conception of the proposed modeling approach. All authors participated in writing	and discussion of the manuscript.

\bmhead{Funding}
Magdalena Schreter-Fleischhacker received funding by the Austrian Science Fund (FWF) Schrödinger Fellowship (J4577). Christoph Meier gratefully
acknowledges the financial support from the European Research Council through the ERC Starting Grant ExcelAM
(project number: 101117579).

\bmhead{Acknowledgements}
\added{
The authors acknowledge collaboration with Bruno Blais, H\'{e}l\`{e}ne Papillon-Laroche as well as the \texttt{deal.II} community.
}

\bmhead{Competing interests}
The authors declare that they have no competing interests.

\bmhead{Availability of data and materials}
	The research code, numerical results and digital data obtained in this project are held on deployed servers that are
	backed up. The datasets used and/or analyzed during the current study are available from the corresponding author on
	reasonable request.
	
\begin{appendices}

\section{\added{Solution algorithm of the }level-set framework}\label{app:level_set}

\deleted{
Subsequent to solving the advection equation \eqref{eq:transport} of the level-set function $\phi$ at time~$t$, a reinitialization step \cite{olsson2007conservative} is performed to preserve the shape of the regularized indicator function as the interface moves. For this purpose, we solve}
	\deleted{
	$\fracPartial{\psi}{\tau}+\nabla\cdot\left(\frac{1-\psi^2}{2}\nGamma\right) =\epsilon\nabla\cdot((\nabla\psi\cdot\nGamma)\nGamma) \qquad  \text{in }\Omega\times[0,\tau]$
}
\deleted{
for the pseudo-time $\tau$ with initial condition $\psi(\Bx,\tau=0)=\phi(\Bx,t)$ and homogeneous Neumann boundary conditions $\nabla\psi\cdot\hat{\Bn}=0\text{ on }\partial\Omega\times[0,\tau]$
until steady state is obtained at $\tau=\tau^{(end)}$. Here, $\psi$ represents an auxiliary field, which is transferred to the level-set field, $\phi(\boldsymbol{x},t) = \psi(\boldsymbol{x},\tau=\tau^{(end)})$, at the end of the reinitialization pseudo-time stepping scheme. The parameter $\epsilon$ is the interface thickness parameter, and $\nGamma$ represents the interface normal vector, evaluated at time $t$ (or pseudo-time $\tau=0$) and assumed as constant over the pseudo-time. The determination of the latter is described below. For discretization in time, we employ a semi-implicit Euler time stepping scheme, considering an explicit scheme for the nonlinear compressive flux term $\left(\added{(1-\psi^2)/2}\nGamma\right)$ for obtaining a linear system of equations.
}

\deleted{
As proposed in \cite{olsson2007conservative}, the interface normal vector is computed from a projection step of the normalized level-set gradient 
$
	\bar{\Bn}_\Gamma=\frac{\nabla\phi}{|\nabla\phi|}\qquad\text{ in }\Omega\,$
to the level-set space
$	\nGamma-\eta_{n}\,h^2\Delta\nGamma=\bar{\Bn}_\Gamma\qquad\text{ in }\Omega\,$
subject to homogeneous Neumann boundary conditions $\nabla\nGamma\cdot\hat{\Bn}=\boldsymbol{0}\text{ on }\partial\Omega$. The filter parameter $\eta_{n}\,h^2$ is determined from the element edge length $h$ and the constant $\eta_n$ and represents the radius of nonlocal interaction. 
The mean curvature is defined as
$	\bar{\kappa}=-\nabla\cdot\nGamma\qquad\text{ in }\Omega\,.$
In order to avoid spurious high-frequency oscillations of the curvature, likewise to the projected interface normal vector, we compute a regularized curvature $\kappa$ as proposed in \cite{olsson2007conservative}
$	\kappa-\eta_{\kappa}\,h^2\Delta\kappa=\bar{\kappa}\qquad\text{ in }\Omega\,.$
with the filter parameter $\eta_{\kappa}\,h^2$ from the element edge length $h$ and a constant $\eta_{\kappa}$. 
We use homogeneous Neumann boundary conditions $\nabla\kappa\cdot\hat{\Bn}=\boldsymbol{0}\text{ on }\partial\Omega$. }

\deleted{Remark: For quadrilateral or hexahedral elements, we compute the element edge length $h$ as $h=\max(d)/\sqrt{dim}$, where $\max(d)$ is the largest diagonal of the element and $dim\in\{1,2,3\}$ is the considered dimension. The influence of the filter parameter was investigated in
\cite{zahedi2012spurious}\cite{cenanovic2020finite}. In our simulations we consider $\eta_{\kappa}=2$ and $\eta_{\boldsymbol{n}}=2$ as default values.
}

The overall solution algorithm for the level-set framework, consisting of the advection step, the reinitialization step and subsequent evaluation of geometric quantities of the interface, is summarized in Algorithm~\ref{algo:ls}.

\begin{algorithm}[H]
	\caption{Solve the level-set equation and compute filtered normal and curvature \texttt{level\_set\_solver}$(\phi,\;\boldsymbol{u},\; \mDot)$.}\label{algo:ls}
	\tcc{\underline{step 1:~compute level-set advection}}
	$\nGamma\gets\nGamma(\phi)$~\eqref{eq:normal_vector_regularized}  \Comment*[r]{update normal vector} 
		\tcc{perform fixed-point iteration to compute level set $\phi$}
	\While{$\Delta\phi>$\texttt{TOL}}{
		$\uGamma\gets\uGamma(\boldsymbol{u}, \mDot, \phi,\nGamma)$~\eqref{eq:transport_vel_local_continuous}/\eqref{eq:transport_vel_extension_liquid}/\eqref{eq:transport_vel_extension_gas}  \Comment*[r]{compute transport velocity}
		$\phi\gets\phi(\Bu_\Gamma)$~\eqref{eq:transport} \Comment*[r]{advect level set}
		$\nGamma\gets\nGamma(\phi)$~\eqref{eq:normal_vector_regularized}  \Comment*[r]{update normal vector}
	}
	\tcc{\underline{step 2:~perform level-set reinitialization}}
	\For{$\tau\gets \tau^{\text{(start)}}$ \KwTo $\tau^{\text{(end)}}$}{
		$\phi \gets \psi(\phi, \nGamma)$~\eqref{eq:reinitialization} \Comment*[r]{reinitialize level set with $\psi^{(0)}=\phi$}
	}
	\tcc{\underline{step 3:~compute geometric quantities}}
	$\nGamma\gets\nGamma(\phi), \kappa \gets \kappa(\phi, \nGamma)$~\eqref{eq:curvature_regularized}  \Comment*[r]{compute normal vector and curvature}
\end{algorithm}

\section{\added{Interpolation function for effective viscosity}}
\label{app:viscosity}

\begin{figure}[H]
	\centering
	\includegraphics[width=\textwidth]{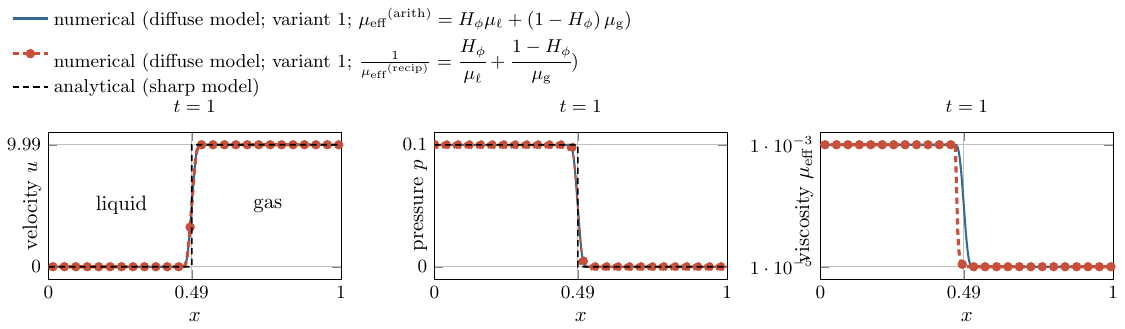}
	\caption{\added{One-dimensional phase change problem with phase-dependent viscosities of $\mu_\ell=\num{e-3}$ and $\mu_{\text{g}}=\num{e-5}$: Velocity, pressure and viscosity distribution at the final stage of the simulation considering the effective viscosity $\mu_{\text{eff}}$  as (i) an arithmetic phase-weighted average and (ii) a  reciprocal phase-weighted average. It can be seen that the pressure and velocity distribution is independent of the chosen interpolation function for the effective viscosity.}}
	\label{fig:1D_visc}
\end{figure}

\begin{figure}[H]
	\centering
	\includegraphics[width=\textwidth]{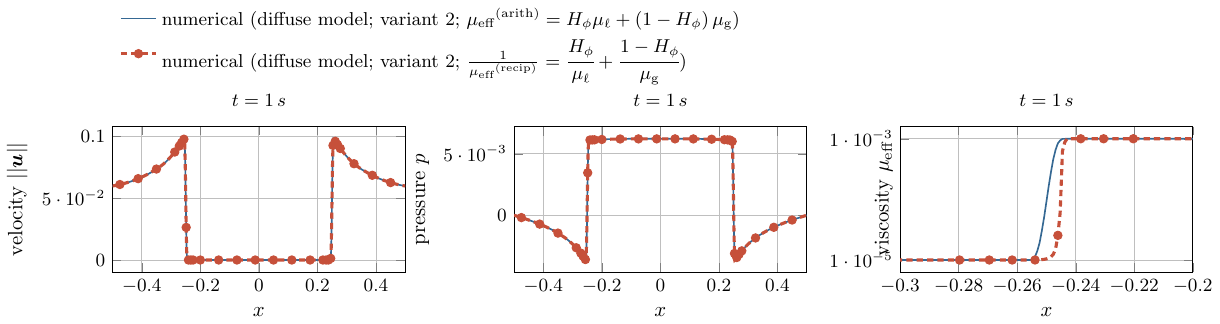}
	\caption{\added{Evaporating droplet problem with phase-dependent viscosities of $\mu_\ell=\num{e-3}$ and $\mu_{\text{g}}=\num{e-5}$: Velocity, pressure and viscosity distribution at the final stage of the simulation considering the effective viscosity $\mu_{\text{eff}}$  as (i) an arithmetic phase-weighted average and (ii) a  reciprocal phase-weighted average. Note the different scale on the $x$-axis for the viscosity distribution. It can be seen that the pressure and velocity distribution is independent of the chosen interpolation function for the effective viscosity.}
	\label{fig:2D_visc}}
\end{figure}

\added{
	To demonstrate that the choice of interpolation function for the effective viscosity is arbitrary, we compute the one-dimensional phase change (see Section~\ref{sec:1D_phase_change}) and evaporating droplet (Section~\ref{sec:evaporating_droplet}) problems for phase-dependent  viscosities of $\mu_\ell=\num{e-3}$ and $\mu_{\text{g}}=\num{e-5}$. 
	In Figures \ref{fig:1D_visc} and \ref{fig:2D_visc}, the velocity and pressure distribution are shown for two different types of effective viscosity interpolation functions, i.e., arithmetic and reciprocal phase-weighted average. 	
	It can be observed that the resulting velocity and pressure fields remain the same, regardless of the chosen effective viscosity interpolation function.}

\section{Extension of solution quantities from a level-set isosurface using closest point projection}
\label{app:cpp}

In the following, algorithmic aspects of performing a closest point projection from an arbitrary point inside the domain to a certain level-set isosurface are elaborated. In this work, this algorithm is used for computing the closest points to the liquid or gaseous ends of the interface region. At those points velocities are evaluated to compute extended velocity fields appearing in the models for the level-set transport velocity, i.e., \myeqrefsa{eq:transport_vel_extension_liquid}{eq:transport_vel_extension_gas}. Nevertheless, it could be also used to extend other quantities from certain level-set isosurfaces, e.g., to evaluate the mean curvature $\kappa$ at the zero-level-set isosurface and extend it over a narrow band to improve the accuracy of the continuum surface tension force model, similar to \cite{coquerelle2016fourth}.

\begin{figure}[tb!]
	\centering
	\includegraphics{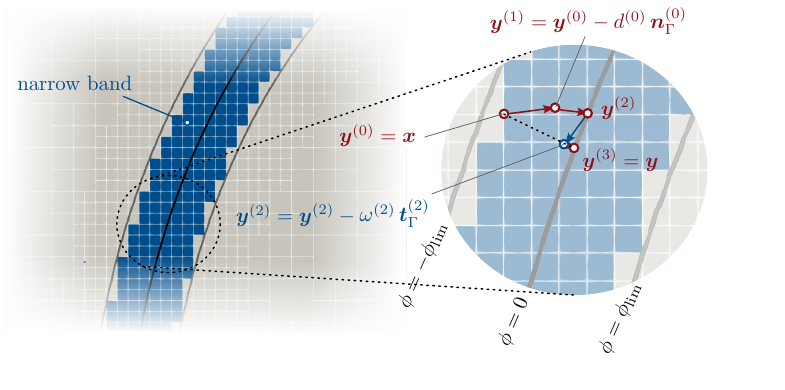}
	\caption{Sketch of the employed algorithm for closest point projection considering a narrow band (blue elements) around the zero-level-set isosurface, consisting of correction steps in interface normal direction (indicated by red colors) and one correction step in tangential direction (indicated by blue colors).}
	\label{fig:narrow_band}
\end{figure}
For the sake of demonstration and motivated by the example mentioned above, our isosurface of interest for performing a closest point projection is the discrete interface
\begin{equation}
	\Gamma = \{\By \in \Omega~|~\phi(\By)=0\}\,,
\end{equation}
represented by the zero-level-set isosurface.
The goal is to find for any point $\Bx$ of the domain the closest point $\By=\text{CP}(\Bx)$ on $\Gamma$ such that
$\forall \Bx \in \Omega$
\begin{equation}
	\label{eq:cpp}\By = \{ \text{CP}(\Bx) :\min_{\By\in\Gamma}(\|\Bx-\By \|) \wedge \vec{\Bx\By}\cdot\Bt_{(j),\Gamma}=0\} \quad  \text{ with } j \in 
	\begin{cases}
		\{\} \text{ for 1D} \\
		\{0\} \text{ for 2D} \\
		\{0, 1 \}  \text{ for 3D} 
	\end{cases}
\end{equation}
holds. The first criterion represents minimization of the distance and the second ensures that the local tangent plane described by the unit tangent vector(s) $\Bt_{(j),\Gamma}$ is orthogonal to the distance vector $\vec{\Bx\By}$. The tangential vectors are defined in 2D and 3D, respectively, as
\begin{align}
	\text{2D: }&\Bt_{(0)} \gets \begin{bmatrix}
		\Bn_{\Gamma,1} \\ 
		-\Bn_{\Gamma,0} 
	\end{bmatrix} \\
	\text{3D: }&
	\Bt_{(0)} \gets \Bv - \left(\Bv\cdot\Bn_\Gamma\right)\,\Bn_\Gamma 
	\qquad
	\Bt_{(1)} \gets \Bn_\Gamma \times \Bt_{(0)}
\end{align}
with $\Bv$ being an arbitrary unit vector that must not be parallel to $\Bn_\Gamma$. In practice, we set $\Bv$ to $\Be_x$ or $\Be_y$. 

For determining the closest point according to \myeqref{eq:cpp}, we implemented the algorithm similar to \cite{coquerelle2016fourth,henri2022geometrical}. First, we collect the support points of the finite element mesh in a narrow band around the target isosurface of the level-set function $\{\Bx \in \Omega~|~|\phi(\Bx)|<\phi_\*{lim}\}$ illustrated in Figure~\ref{fig:narrow_band}. Next, for each considered point $\Bx$ we perform a fixed-point iteration by performing a sequence of correction steps $k<k_{\max}$  to evaluate the closest point $\By$. We start with the initial guess  $\By^{(0)}=\Bx$. The computation consist of (i) a sequence of correction steps in normal direction 
\begin{equation}
	\By^{(k)}\gets\By^{(k)}-d\left(\By^{(k)}\right)\,\Bn_\Gamma\left(\By^{(k)}\right)
\end{equation} 
and if necessary (ii) one correction step in tangential direction
\begin{equation}
	\By^{(k+1)}\gets\By^{(k)}-\left(\overrightarrow{\Bx\By^{(k)}}\cdot  \Bt_{(j),\Gamma}\,\left(\By^{(k)}\right)\right)\,\Bt_{(j),\Gamma}\,\left(\By^{(k)}\right)
\end{equation}
until a certain tolerance for $\|\By^{(k+1)}-\By^{(k)}\|$ is reached.
Once the closest point $\By$ has been identified, it can be used to perform an extrapolation of solution values at the discrete interface to the narrow band interface region.

\begin{remark}
	Note that the point $\By^{(k)}$ may lie arbitrarily inside the computational domain $\Omega$ and may not necessarily comply with support points. Thus, the evaluation of the level-set function, the normal vector and the tangential vector(s) at this point, needed for the fixed-point iteration, comprises the following steps for a distributed finite element mesh among multiple processes: (1) identification of the process that owns the point; (2) identification of the attributed finite element and positions in the reference cell; (3) interpolation by means of shape functions. Subsequently, the values for the signed distance function, the normal vector and the tangential vector can be computed to perform the correction. This procedure is implemented in \text{deal.II}~\cite{arndt2022deal,schreter2023step87}.
\end{remark}

\begin{figure}[htbp!]
    \includegraphics{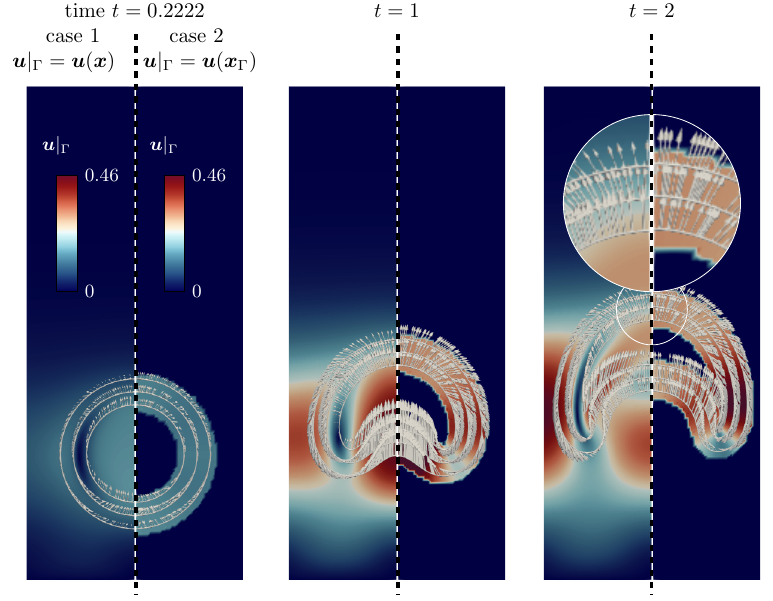}
	\caption{Snapshots from simulations of the rising bubble benchmark \cite{hysing2009quantitative} considering two different cases for computing the level-set transport velocity $\uGamma$: \emph{case 1} --- local evaluation of the fluid velocity ($\uGamma(\Bx)=\Bu(\Bx)$) indicated in the left part of one snapshot; \emph{case 2} --- extension fluid velocity from the interface ($\uGamma(\Bx)=\Bu(\Bx_\Gamma)$) indicated in the right part  of one snapshot; The domain $x,y \text{ in }\Omega=[0,1]\times[0,2]$ is discretized with quadrilateral elements with an adaptively refined mesh ranging between element edge length of $h_{\min}=\num{6.25e-3}$ and $h_{\max}=\num{0.025}$.}
	\label{fig:rising_bubble}
\end{figure}
To demonstrate and verify the capabilities of the closest point projection algorithm, we consider the well-known benchmark example of the rising of a bubble, presented in~\cite{hysing2009quantitative}. It should be noted that this is a pure two-phase flow problem without evaporation effects, i.e., $\mDot=0$. We determine the level-set transport velocity based on two approaches. \emph{Case 1} computes the transport velocity from the local fluid velocity, i.e., $\uGamma(\Bx)=\Bu(\Bx)$, which is the standard assumption for simulations without phase change. 
Alternatively, and similar to the approaches discussed within the presented two-phase flow with phase change framework in Section~\ref{sec2}, in \emph{case 2} we perform a closest point projection to the zero-level-set isosurface and subsequently extend the fluid velocity from the latter to a narrow band region to compute the level-set transport velocity, i.e.,  $\uGamma(\Bx)=\Bu(\Bx_\Gamma)$. The parameters (SI units) are chosen as $\liqP{\rho}=1$, $\gasP{\rho}=0.1$ , $\liqP{\mu}=0.01$, $\gasP{\mu}=0.001$, $\epsilon=0.01$. Gravity forces with $\boldsymbol{g}=[0, -9.81]^T$ and surface tension forces with $\sigma=0.001$ are considered. The simulation is performed for the time period $0\leq t \leq 1$ with a constant time step size of $\num{0.02}$. 

Snapshots from the simulation are shown in Fig.~\ref{fig:rising_bubble}, where the left half of each snapshot refers to \emph{case 1} and the right half to \emph{case~2}. It indicates the level-set isosurfaces at $\phi=\{-0.99, 0, 0.99\}$, and the color fields represent the computed level-set transport velocity. It can be seen that the resulting transport velocity according to \emph{case~1} is accompanied by a strong variation of the velocity across the interface region, which may lead to artificial deformation of the level-set field. This is not the case for \emph{case~2}, where the resulting level-set transport velocity remains constant across the interface thickness due to the employed extension algorithm.
By comparison of the two approaches, there is no apparent difference in the bubble shape, which underlines that the approximation of the level-set transport velocity $\uGamma\approx\Bu$ is perfectly valid for simulations without phase change.

\section{Analytical solution for the one-dimensional phase change problem for a sharp and diffuse model}
\label{app:evapor_stefan}

\begin{figure}[bt!]
	\centering
	\includegraphics{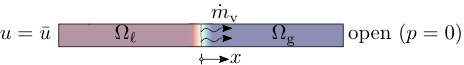}
	\caption{Illustration of the one-dimensional phase-change problem with inflow.}
	\label{fig:flat_interface}
	\end{figure}
\noindent
We consider a one-dimensional evaporative phase-change problem, representing a simplified version of the well-known Stefan's problem based on the assumption of isothermal conditions, illustrated in Fig.~\ref{fig:flat_interface}. We prescribe an inflow velocity on the liquid end and assume a zero pressure outlet on the gaseous end. 

\paragraph{Sharp model}
From the evaluation of the Rankine--Hugoniot conditions \eqref{eq:rankineHugoniot} and the continuity equation \eqref{eq:continuity} together with the Dirichlet boundary condition for the inflow velocity, the analytical solution for the velocity predicted by a sharp interface model, i.e., the exact solution, yields
\begin{equation}
	\label{eq:stefan_sharp_velocity}
	u(x,t) = \begin{cases} \bar{u} &\text{ for } x<x_\Gamma(t)  
		\\
		\bar{u}+	\mDot \left(\frac{1}{\rhoG} -\frac{1}{\rhoL}  \right) &\text{ for } x>x_\Gamma(t)  
	\end{cases}
\end{equation}
with the current position of the interface according to
\begin{equation}
	x_\Gamma(t) = x_\Gamma^{(0)}+\left(\bar{u}-\frac{\mDot}{\liqP{\rho}}\right)\,t
\end{equation}
and the interface transport velocity
\begin{equation}
	\label{eq:1DStefan_uGamma_sharp}
	u_\Gamma=\bar{u}-\frac{\mDot}{\liqP{\rho}} \qquad\text{ on } \Gamma\,.
\end{equation}
From \myeqref{eq:1DStefan_uGamma_sharp} it can be seen that if $\bar{u}$ is equal to  $\added{\mDot/\liqP{\rho}}$ the interface position remains static. By insertion of the velocity field into the momentum equation \eqref{eq:momentum_balance} and subsequent integration, considering the zero pressure outlet at the gaseous end, an analytical expression for the pressure for a sharp model is obtained as
\begin{equation}
	\label{eq:stefan_sharp_pressure}
	p(x,t) = 
	\begin{cases} 	
		\mDot^2 \left(\frac{1}{\rhoG} -\frac{1}{\rhoL}  \right)  &\text{ for } x<x_\Gamma(t) 
		\\
		0 &\text{ for } x>x_\Gamma(t) 
	\end{cases}\,.
\end{equation}

\paragraph{Diffuse model}
An analytical solution for the velocity in the diffuse-model case can be obtained from analytical integration of the continuity equation \eqref{eq:continuity}, specialized for the present use case and considering the inflow velocity at the liquid end, to
\begin{equation}
	\label{eq:stefan_diffuse_velocity}
	u(x,t) = 
	\bar{u}+\mDot \, (1-H_\phi(x,t)) \, \left(\frac{1}{\rhoG} -\frac{1}{\rhoL}  \right).
\end{equation}
By insertion of the velocity field into the momentum equation \eqref{eq:momentum_balance} and subsequent integration considering the zero pressure outlet at the gaseous end, an analytical expression for the pressure is obtained
\begin{equation}
	\label{eq:stefan_diffuse_pressure}
	p(x,t) = \mDot^2 \, H_\phi(x,t) \, \left(\frac{1}{\rhoG} -\frac{1}{\rhoL}\right)\,.
\end{equation}

\section{Analytical solution for the stationary evaporating circular shell}
\label{app:evapor_shell}

In the following, the analytical solution for the stationary evaporating circular shell, presented in Section~\ref{sec:evapor_shell_analytical} and illustrated in the left panel of Fig.~\ref{fig:const_law}, is derived. 

\paragraph{Diffuse model}

The continuity equation considering evaporative phase change \eqref{eq:continuity} and
specialized for stationary, axisymmetric conditions reads as
\begin{equation}\label{eq:conti_radial}
	\frac{d}{dr}(r\,u_r(r))= \mDot\left(\frac{1}{\rho_g}-\frac{1}{\rho_l}\right)\abs{\fracTotal{\Hphi}{r}}\,.
\end{equation}
It represents a first-order linear ordinary differential equation. If the evaporative mass flux $\mDot$, the interface position and accordingly the heaviside function $H_\phi$ is known, the right-hand side term of \myeqref{eq:conti_radial} is given.
According to the method of integrating factors, the integrating factor is determined as $I=\exp(\int \added{dr/r})=r$. If we multiply \myeqref{eq:conti_radial} by the integrating factor we obtain
\begin{equation}
u_r + \frac{d u_r}{dr}\,r = r\,\mDot\left(\frac{1}{\rho_g}-\frac{1}{\rho_l}\right)\abs{\fracTotal{\Hphi}{r}}\,
	\label{eq:conti_radial2}
\end{equation}
which can be rearranged to
\begin{equation}
	\frac{du_r}{dr} = \mDot\left(\frac{1}{\rho_g}-\frac{1}{\rho_l}\right)\abs{\fracTotal{\Hphi}{r}}\,-\frac{u_r}{r}\,.
\end{equation}
Integration of both sides of the equation with respect to $r$ gives
\begin{equation}
	r\,u_r(r) = \int r\,\mDot\left(\frac{1}{\rho_g}-\frac{1}{\rho_l}\right)\abs{\fracTotal{\Hphi}{r}}\,\mathrm{d}r + C\,.
\end{equation}
We integrate over the domain $R_\*i\leq r \leq R_\*o$ and determine the integration constant $C$ from the inflow boundary condition $u_r(r=R_\*i)=\bar{u}$
\begin{equation}
	C=R_\*i\,\bar{u} \,.
\end{equation}
Finally, the analytical solution for the radial velocity is obtained as
\begin{equation}
	\label{eq:shell_u_analytical}
	\evaporShellVelocity{}\,.
\end{equation}

The analytical solution for the pressure can be obtained from the momentum equation, where we consider the axisymmetric, stationary, inviscid case. It reads as
\begin{equation}
	\label{eq:momentum_equation}
	\rho u_r \fracTotal{u_r}{r} = - \fracTotal{p}{r}\,.
\end{equation}
The derivative of the radial velocity \eqref{eq:shell_u_analytical} with respect to $r$ reads as
\begin{align}
	\fracTotal{u_r}{r} &= -\bar{u}\,\frac{R_\*i}{r^2}  -
	\frac{1}{r^2} \, 	 \mDot\left(\frac{1}{\rho_g}-\frac{1}{\rho_l}\right) \left(r (H(r)-1)+\int_{R_\*i}^{r} H(r) \mathrm{d}r \right) \\
	\nonumber &+
	\frac{\mDot}{r}\left(\frac{1}{\rho_g}-\frac{1}{\rho_l}\right) \left((H(r)-1) + r\,\fracTotal{H}{r} + H(r)\right)\\ 
	\label{eq:du_dr}
	&= -\bar{u}\,\frac{R_\*i}{r^2}  + \mDot\left(\frac{1}{\rho_g}-\frac{1}{\rho_l}\right) \left(
	\frac{-1}{r^2} \, 	  \int_{R_\*i}^{r} H(r) \mathrm{d}r+
	\fracTotal{H}{r} + \frac{H(r)}{r}\right)\,.
\end{align}
Inserting \myeqrefsa{eq:du_dr}{eq:shell_u_analytical} into \myeqref{eq:momentum_equation} yields
\begin{align}
	\fracTotal{p}{r} = -	\rho u_r \left( \mDot\left(\frac{1}{\rho_g}-\frac{1}{\rho_l}\right)\abs{\fracTotal{H}{r}}\,-\frac{u_r}{r}\right)\,.
\end{align}
Integration over $r$ yields an analytical expression for the pressure considering the pressure boundary condition on the outer face $p(r=R_\*o)=0$:
\begin{align}
	\label{eq:shell_p_analytical}
	p(r) &= \int_{R_\*i}^{r} \left(-\rho u_r \left( \mDot\left(\frac{1}{\rho_g}-\frac{1}{\rho_l}\right)\abs{\frac{dH}{dr}}\,+\frac{u_r}{r}\right)\right) \mathrm{d}r \\
	\nonumber &+
	\int_{R_\*i}^{R_\*o} \left(\rho u_r \left( \mDot\left(\frac{1}{\rho_g}-\frac{1}{\rho_l}\right)\abs{\frac{dH}{dr}}\,-\frac{u_r}{r}\right)\right) \mathrm{d}r\,.
\end{align}
A \texttt{Python} script for evaluation of the radial velocity 
	\eqref{eq:shell_u_analytical} and the pressure 
	\eqref{eq:shell_p_analytical}, considering numerical integration, can be found in the supplementary materials.
	
\paragraph{Sharp model}	

As a reference, considering a sharp model, the analytical solution for the radial velocity component can be stated as
\begin{equation}
	\label{eq:shell_u_analytical_sharp}
	u_r(r) = \begin{cases} 
		\frac{R_\*i}{r}\,\bar{u}  &\text{ for } r<R_\Gamma\\
		\frac{R_\*i}{r}\,\bar{u} + \mDot\left(\frac{1}{\rhoG}-\frac{1}{\rhoL}\right)\,\frac{R_\Gamma}{r}  &\text{ for } r>R_\Gamma 
	\end{cases}\,.
\end{equation}

\section{Weak form and notes on the discretization of the governing equations}
\label{app:weak_form}

In the following, the spatial discretization by means of the finite element method of the main governing equations consisting of the Navier--Stokes equations \eqref{eq:continuity}-\eqref{eq:momentum_balance} and the level-set transport equation \eqref{eq:transport} is briefly summarized. For $L^2$-inner products we use the notation $(f,g)_\Omega=\int_\Omega f\,g\,dx$. For the sake of brevity, we omit the detailed description of the spatial and temporal discretization and refer interested readers to the literature.

\paragraph{Navier--Stokes equations}

\newcommand{\testU}{\delta\Bu}
\newcommand{\testP}{\delta p}
\newcommand{\testUh}{\delta\Bu^h}
\newcommand{\testPh}{\delta p^h}

The weak form of the Navier--Stokes equations \eqref{eq:continuity}-\eqref{eq:momentum_balance} is obtained by multiplication with weighting functions for the pressure and the velocity, denoted as $\testP$ and $\testU$. The solution space for the velocity is defined as
$\mathcal{S}_u = \{ \Bu \in \left(H^1(\Omega)\right)^{\text{dim}}~|~\Bu = \bar{\Bu} \text{ on } \boundaryDirU\}\,$
and for the corresponding weighting function as $\mathcal{V}_u = \{ \testU \in H^1(\Omega)~|~\testU = \boldsymbol{0} \text{ on } \boundaryDirU\}$. Here, $ H^1(\Omega)$ denotes the space of square integrable functions with square integrable derivatives. The solution space for the pressure and the corresponding weighting function is defined as
$\mathcal{S}_p =\mathcal{V}_p=L_2(\Omega)$ for inf-sup stability, where $L^2(\Omega)=H^0(\Omega)$ denotes the space of square integrable functions. Multiplication by the weighting functions, integration over the spatial domain, application of the divergence theorem and incorporation of the Dirichlet and Neumann boundary condition yields the weak form. Find $p \in \mathcal{S}_{p}$ and $u \in \mathcal{S}_{u}$ such that 
\begin{align}
	\label{eq:weakNavierStokes}
	\BiLi{\testP}{\diver\Bu} = 	\BiLi{\testP}{\tilde{v}^{(lg)}},                                               \nonumber    \\
	\BiLi{\delta \Bu}{\eff{\rho} \left(\fracPartial{\Bu}{t} + \left(\Bu\cdot\nabla\right) \Bu\right)} -
	\BiLi{\diver \testU}{p} + 		\BiLi{\nabla\testU}{\tau_{\mu}}  \\=	\BiLi{\testU}{\rhoEff\,\Bg + \surfaceTensionForce}
	+ \BiLiPure{\testU}{\bar{\Bt}}{\boundaryNeuU}\nonumber
	\\
	\quad \forall\,\{\testP, \testU\}\in\mathcal{V}_p\times\mathcal{V}_u \nonumber 
\end{align}
holds. The weak form \myeqref{eq:weakNavierStokes} is discretized in space based on a Bubnov-Galerkin ansatz for the discrete function values for the velocity
\begin{equation}
	\Bu^h(\Bx, t) = \sum_j N_{u,j}(\Bx)\,\Bu_j(t) \text{ and } 	\testUh(\Bx, t) = \sum_j N_{u,j}(\Bx)\,\testU_j(t)
\end{equation}
and the pressure
\begin{equation}
	p^h(\Bx, t) = \sum_j N_{p,j}(\Bx)\,p_j(t) \text{ and } 	\testPh(\Bx, t) = \sum_j N_{p,j}(\Bx)\,\testP_j(t)\,.
\end{equation}
Here, $N_{u,j}$ and $N_{p,j}$ are the shape functions and $\Bu_j$ and $p_j$ the nodal solution coefficients, for the velocity and the pressure. After insertion of the approximate solution for the velocity and the pressure into the weak form, the space-discrete problem is obtained. The latter is discretized in time using the backward differentiation formula BDF2. After that, the discrete system of equations is consistently linearized and solved using a full Newton-Raphson scheme as described in~\cite{Kronbichler18multiphase}.

\paragraph{Level-set advection equation}

\newcommand{\testPhi}{\delta\phi}
\newcommand{\testPhih}{\delta\phi^h}

The weak form of the level-set advection equation \eqref{eq:transport} is obtained by multiplication with the weighting functios for the level-set field, denoted as $\testPhi$. The solution space for the level-set is defined as
$\mathcal{S}_\phi = \{ \phi \in H^1(\Omega)~|~\phi = \bar{\phi} \text{ on } \partial\Omega_{\text{D},\phi}^{\text{inflow}}\}$ 
and for the corresponding weighting function as $\mathcal{V}_\phi =\{ \testPhi \in H^1(\Omega)~|~\testPhi = 0 \text{ on } \partial\Omega_{\text{D},\phi}^{\text{inflow}}\}$. 
Multiplication by the weighting function and integration over the spatial domain yields the weak form. Find $\phi \in \mathcal{S}_{\phi}$ such that 
\begin{equation}
	\label{eq:weakFormLevelSet}
	\BiLi{\testPhi}{\fracPartial{\phi}{t}+\uGamma\cdot\nabla\phi} = 0 \quad \forall \delta\phi\in\mathcal{V}_\phi
\end{equation}
holds.
The solution of the weak problem is discretized in space using a Bubnov-Galerkin ansatz for the level set
\begin{equation}
	\phi^h(\Bx, t) = \sum_j N_{\phi,j}(\Bx)\,\phi_j(t) \text{ and } 	\testPhih(\Bx, t) = \sum_j N_{\phi,j}(\Bx)\,\testPhi_j(t)\,.
\end{equation}
Here, $N_{\phi,j}$ are the shape functions and $\phi_j$ the nodal solution coefficients for the level set.  The temporal discretization is performed via the Crank--Nicolson time-integration scheme. It should be noted that due to the dependency of the level-set transport velocity $\uGamma$ on the level set in case of evaporation, as discussed in Section~\ref{sec:level_set_transport}, the resulting system of equations becomes nonlinear. Thus, we perform a fixed-point iteration for solving the nonlinear system of equations, as outlined in Algorithm~\ref{algo:ls}.


\section{\added{Convergence studies}}
\label{app:convergence_study}

\added{
We analyze the convergence behavior of the presented diffuse interface evaporating two-phase model with respect to the solution of a sharp interface evaporating two-phase model. As benchmarks, examples of one-dimensional phase change (cf. Section~\ref{sec:1D_phase_change}) and evaporating circular shell (cf. Section~\ref{sec:evaporating_circular_shell}) are used. For these, analytical solutions for the sharp interface model for both, pressure and velocity, are available. 
To measure the error, we employ the \mbox{$L^{2}$ norm}
}
\begin{equation} \label{eq:L2norm}
\added{	\LTwoNorm{(\bullet)}= \sqrt{\int_{\Omega} (\bullet)^{2}\,\diffd\Omega}\,.}
\end{equation}

\subsection{One-dimensional phase change}
\label{sec:convergence_1d}

\begin{figure}[tbp!]
\includegraphics[width=\textwidth]{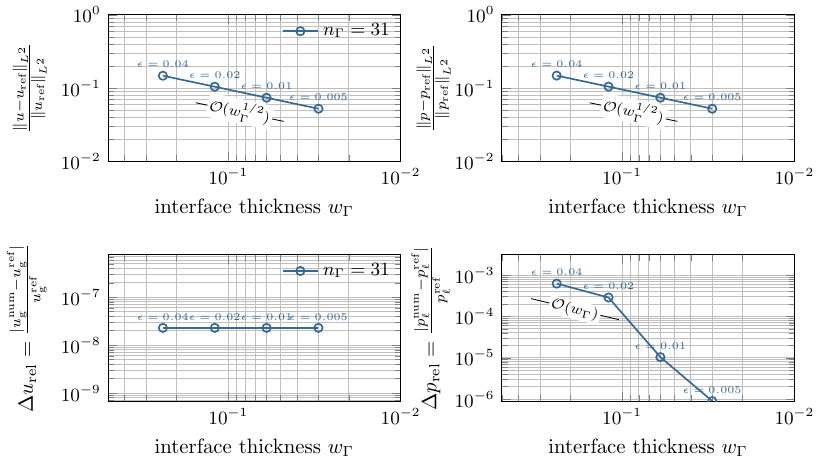}
\caption{\added{Convergence study for the one-dimensional phase change problem for a viscous fluid: velocity and pressure error against different interface thicknesses at $t=\num{1e-3}$ with time step size $\Delta t = \num{1e-6}$. We evaluate the velocity in the gas phase $\gasP{u}$ at $x=1$, and the pressure in the liquid phase $\liqP{u}$ at $x=0$.}}
\label{fig:onedim_error}
\end{figure}

\added{
We refer to the study of one-dimensional phase change of a viscid fluid presented in Section~\ref{sec:stefans_problem_viscid}.
Fig.~\ref{fig:onedim_error} shows the relative error of the velocity and the pressure with respect to the sharp interface reference solution according to \myeqrefsa{eq:stefan_sharp_velocity}{eq:stefan_sharp_pressure} for different values of the interface thickness $\interfaceThickness$. The interface thickness parameter $\epsilon$ is annotated.
The number of cells across the interface are chosen constant at $\numberOfElementsInInterface=31$, and the time step size is chosen as $\num{e-6}$. For all investigated interface thicknesses, the spatial and temporal
discretization error was checked to be negligible for that interface thickness resolution and time step size. Reducing the interface thickness while ensuring a sufficient mesh resolution leads to convergence to the sharp interface reference solution. 
The relative error in velocity and pressure measured in the $L^2$ norm (cf. top panel of Fig.~\ref{fig:onedim_error}) decreases with a convergence rate of order $\mathcal{O}(\interfaceThickness^{1/2})$ with respect to the interface thickness. This error is expected to be independent of the polynomial degree of the finite element function space. We believe that the observed behavior results from the discontinuity in velocity and pressure at the interface in the limit of zero interface thickness as will be discussed in Appendix~\ref{app:convergence_study_discussion}.  Therefore, to increase the accuracy, it is necessary to reduce the interface thickness $\interfaceThickness$.
}

\added{
Furthermore, we analyze the relative error in the difference of the pressure $\Delta p_{\text{rel}}$ and velocity $\Delta u_{\text{rel}}$ between the phases, presented in the bottom panel of Fig.~\ref{fig:onedim_error}. The relative velocity difference error has already reached a vanishingly small error in the order of the tolerance of the nonlinear solver, while the relative pressure difference error decreases with a convergence rate of order $\mathcal{O}(\interfaceThickness)$ and exhibits superconvergence with respect to the interface thickness. }

\added{For the chosen configuration in Section~\ref{sec:1D_phase_change} ($\epsilon=0.01$) the overall error in the velocity and pressure profile is still significant (7\,\%), which is attributed to the large value of the interface thickness. However, the pressure and velocity difference are computed with sufficient accuracy (0.001\,\%).
}

\subsection{\added{Evaporating circular shell}}
\label{sec:convergence_shell}

\begin{figure}[htbp!]
	\includegraphics[width=\textwidth]{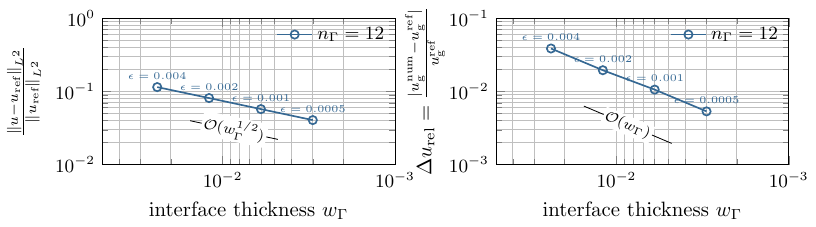}
	\caption{\added{Convergence study for the evaporating circular shell problem: 
		velocity and pressure error against different interface thicknesses at $t=0.1$ with time step size $\Delta t = \num{5e-4}$.
We evaluate the velocity in the gas phase $\gasP{u}$ at $r=0.375$, and the pressure in the liquid phase $\liqP{u}$ at $r=0.125$.	
}}
	\label{fig:evapor_shell_convergence}
\end{figure}

\added{
We refer to the study of the evaporating circular shell of a viscid fluid presented in Section~\ref{sec:evaporating_circular_shell}.
Fig.~\ref{fig:evapor_shell_convergence} shows the relative error of the velocity with respect to the sharp interface reference solution according to \myeqref{eq:shell_u_analytical_sharp} for different  values of the interface thickness $\interfaceThickness$. The interface thickness parameter $\epsilon$ is annotated.
The number of cells across the interface are chosen as $\numberOfElementsInInterface=12$, and the time step size is chosen constant at $\num{5e-4}$. For all investigated interface thicknesses, the spatial and temporal
discretization error was checked to be negligible for that interface thickness resolution and time step size. Reducing the interface thickness while ensuring a sufficient mesh resolution leads to convergence to the sharp interface reference solution. }

\added{
Similar to the one-dimensional phase change problem, the relative error in the radial velocity measured in the $L^2$ norm (cf. left panel of Fig.~\ref{fig:evapor_shell_convergence}) decreases with a convergence rate of order $\mathcal{O}(\interfaceThickness^{1/2})$ with respect to the interface thickness $\interfaceThickness$. 
 }

\added{
Furthermore, we analyze the relative error in the radial component of the velocity $\Delta u_{\text{rel}}$ in the gas phase away from the interface, presented in the right panel of Fig.~\ref{fig:evapor_shell_convergence}. It decreases with a convergence rate of order $\mathcal{O}(\interfaceThickness)$ with respect to the interface thickness. 
}

\added{
For the chosen configuration in Section~\ref{sec:1D_phase_change} ($\epsilon=0.002$) the overall error in the velocity and pressure profile is still significant (8\,\%), which is attributed to the large value of the interface thickness. The velocity in the gas phase is computed with higher accuracy (2\,\%).}

\newpage
\subsection{\added{Discussion}}
\label{app:convergence_study_discussion}

\added{For the cases discussed in Section~\ref{sec:convergence_1d} and \ref{sec:convergence_shell}, we observe a convergence order measured in the  $L^2$ norm  for the primary variables, pressure and velocity, of $\mathcal{O}(\interfaceThickness^{1/2})$. 
By comparison, in our previous study \cite{much2023}, we analyzed the accuracy of continuum surface flux models in approximating two-phase heat transfer under high material property ratios and extreme temperature gradients. In that work, we found that the relative temperature error, also measured in the $L^2$ norm,  for the case of interfacial heating resulting in a \emph{kink} in the temperature distribution at the interface converges to the sharp interface limit with a rate of $\mathcal{O}(\interfaceThickness)$. We attribute the lower convergence order of the $L^2$ error norm observed in the present cases to the presence of interface \emph{jumps} (instead of kinks) of the primary variables, pressure and velocity.
}

\added{For the relative error in the pressure and velocity away from the interface, our findings in Section~\ref{sec:convergence_1d} and \ref{sec:convergence_shell} yielded a convergence rate of $\mathcal{O}(\interfaceThickness)$. This is consistent with the work of Zahedi et al.~\cite{zahedi2012spurious}, who obtained the same convergence order for the pressure jump of a static capillary droplet.}

\end{appendices}


\begin{thebibliography}{76}
\ifx \bisbn   \undefined \def \bisbn  #1{ISBN #1}\fi
\ifx \binits  \undefined \def \binits#1{#1}\fi
\ifx \bauthor  \undefined \def \bauthor#1{#1}\fi
\ifx \batitle  \undefined \def \batitle#1{#1}\fi
\ifx \bjtitle  \undefined \def \bjtitle#1{#1}\fi
\ifx \bvolume  \undefined \def \bvolume#1{\textbf{#1}}\fi
\ifx \byear  \undefined \def \byear#1{#1}\fi
\ifx \bissue  \undefined \def \bissue#1{#1}\fi
\ifx \bfpage  \undefined \def \bfpage#1{#1}\fi
\ifx \blpage  \undefined \def \blpage #1{#1}\fi
\ifx \burl  \undefined \def \burl#1{\textsf{#1}}\fi
\ifx \doiurl  \undefined \def \doiurl#1{\url{https://doi.org/#1}}\fi
\ifx \betal  \undefined \def \betal{\textit{et al.}}\fi
\ifx \binstitute  \undefined \def \binstitute#1{#1}\fi
\ifx \binstitutionaled  \undefined \def \binstitutionaled#1{#1}\fi
\ifx \bctitle  \undefined \def \bctitle#1{#1}\fi
\ifx \beditor  \undefined \def \beditor#1{#1}\fi
\ifx \bpublisher  \undefined \def \bpublisher#1{#1}\fi
\ifx \bbtitle  \undefined \def \bbtitle#1{#1}\fi
\ifx \bedition  \undefined \def \bedition#1{#1}\fi
\ifx \bseriesno  \undefined \def \bseriesno#1{#1}\fi
\ifx \blocation  \undefined \def \blocation#1{#1}\fi
\ifx \bsertitle  \undefined \def \bsertitle#1{#1}\fi
\ifx \bsnm \undefined \def \bsnm#1{#1}\fi
\ifx \bsuffix \undefined \def \bsuffix#1{#1}\fi
\ifx \bparticle \undefined \def \bparticle#1{#1}\fi
\ifx \barticle \undefined \def \barticle#1{#1}\fi
\bibcommenthead
\ifx \bconfdate \undefined \def \bconfdate #1{#1}\fi
\ifx \botherref \undefined \def \botherref #1{#1}\fi
\ifx \url \undefined \def \url#1{\textsf{#1}}\fi
\ifx \bchapter \undefined \def \bchapter#1{#1}\fi
\ifx \bbook \undefined \def \bbook#1{#1}\fi
\ifx \bcomment \undefined \def \bcomment#1{#1}\fi
\ifx \oauthor \undefined \def \oauthor#1{#1}\fi
\ifx \citeauthoryear \undefined \def \citeauthoryear#1{#1}\fi
\ifx \endbibitem  \undefined \def \endbibitem {}\fi
\ifx \bconflocation  \undefined \def \bconflocation#1{#1}\fi
\ifx \arxivurl  \undefined \def \arxivurl#1{\textsf{#1}}\fi
\csname PreBibitemsHook\endcsname

\bibitem[\protect\citeauthoryear{Ly et~al.}{2017}]{ly2017metal}
\begin{barticle}
\bauthor{\bsnm{Ly}, \binits{S.}},
\bauthor{\bsnm{Rubenchik}, \binits{A.M.}},
\bauthor{\bsnm{Khairallah}, \binits{S.A.}},
\bauthor{\bsnm{Guss}, \binits{G.}},
\bauthor{\bsnm{Matthews}, \binits{M.J.}}:
\batitle{{Metal vapor micro-jet controls material redistribution in laser
  powder bed fusion additive manufacturing}}.
\bjtitle{Scientific Reports}
\bvolume{7}(\bissue{1}),
\bfpage{1}--\blpage{12}
(\byear{2017})
\doiurl{10.1038/s41598-017-04237-z}
\end{barticle}
\endbibitem

\bibitem[\protect\citeauthoryear{Kiss et~al.}{2019}]{kiss2019laser}
\begin{barticle}
\bauthor{\bsnm{Kiss}, \binits{A.M.}},
\bauthor{\bsnm{Fong}, \binits{A.Y.}},
\bauthor{\bsnm{Calta}, \binits{N.P.}},
\bauthor{\bsnm{Thampy}, \binits{V.}},
\bauthor{\bsnm{Martin}, \binits{A.A.}},
\bauthor{\bsnm{Depond}, \binits{P.J.}},
\bauthor{\bsnm{Wang}, \binits{J.}},
\bauthor{\bsnm{Matthews}, \binits{M.J.}},
\bauthor{\bsnm{Ott}, \binits{R.T.}},
\bauthor{\bsnm{Tassone}, \binits{C.J.}}, \betal:
\batitle{Laser-induced keyhole defect dynamics during metal additive
  manufacturing}.
\bjtitle{Advanced Engineering Materials}
\bvolume{21}(\bissue{10}),
\bfpage{1900455}
(\byear{2019})
\end{barticle}
\endbibitem

\bibitem[\protect\citeauthoryear{Cunningham
  et~al.}{2019}]{cunningham2019keyhole}
\begin{barticle}
\bauthor{\bsnm{Cunningham}, \binits{R.}},
\bauthor{\bsnm{Zhao}, \binits{C.}},
\bauthor{\bsnm{Parab}, \binits{N.}},
\bauthor{\bsnm{Kantzos}, \binits{C.}},
\bauthor{\bsnm{Pauza}, \binits{J.}},
\bauthor{\bsnm{Fezzaa}, \binits{K.}},
\bauthor{\bsnm{Sun}, \binits{T.}},
\bauthor{\bsnm{Rollett}, \binits{A.D.}}:
\batitle{{Keyhole threshold and morphology in laser melting revealed by
  ultrahigh-speed X-ray imaging}}.
\bjtitle{Science}
\bvolume{363}(\bissue{6429}),
\bfpage{849}--\blpage{852}
(\byear{2019})
\doiurl{10.1126/science.aav4687}
\end{barticle}
\endbibitem

\bibitem[\protect\citeauthoryear{Bitharas et~al.}{2022}]{bitharas2022interplay}
\begin{barticle}
\bauthor{\bsnm{Bitharas}, \binits{I.}},
\bauthor{\bsnm{Parab}, \binits{N.}},
\bauthor{\bsnm{Zhao}, \binits{C.}},
\bauthor{\bsnm{Sun}, \binits{T.}},
\bauthor{\bsnm{Rollett}, \binits{A.}},
\bauthor{\bsnm{Moore}, \binits{A.}}:
\batitle{The interplay between vapour, liquid, and solid phases in laser powder
  bed fusion}.
\bjtitle{Nature Communications}
\bvolume{13}(\bissue{1}),
\bfpage{2959}
(\byear{2022})
\doiurl{10.1038/s41467-022-30667-z}
\end{barticle}
\endbibitem

\bibitem[\protect\citeauthoryear{Khairallah et~al.}{2016}]{khairallah2016laser}
\begin{barticle}
\bauthor{\bsnm{Khairallah}, \binits{S.A.}},
\bauthor{\bsnm{Anderson}, \binits{A.T.}},
\bauthor{\bsnm{Rubenchik}, \binits{A.}},
\bauthor{\bsnm{King}, \binits{W.E.}}:
\batitle{{Laser powder-bed fusion additive manufacturing: Physics of complex
  melt flow and formation mechanisms of pores, spatter, and denudation zones}}.
\bjtitle{Acta Materialia}
\bvolume{108},
\bfpage{36}--\blpage{45}
(\byear{2016})
\doiurl{10.1016/j.actamat.2016.02.014}
\end{barticle}
\endbibitem

\bibitem[\protect\citeauthoryear{Chen and Yan}{2020}]{chen2020spattering}
\begin{barticle}
\bauthor{\bsnm{Chen}, \binits{H.}},
\bauthor{\bsnm{Yan}, \binits{W.}}:
\batitle{Spattering and denudation in laser powder bed fusion process:
  Multiphase flow modelling}.
\bjtitle{Acta Materialia}
\bvolume{196},
\bfpage{154}--\blpage{167}
(\byear{2020})
\doiurl{10.1016/j.actamat.2020.06.033}
\end{barticle}
\endbibitem

\bibitem[\protect\citeauthoryear{Meier et~al.}{2021}]{meier2020meltpool}
\begin{barticle}
\bauthor{\bsnm{Meier}, \binits{C.}},
\bauthor{\bsnm{Fuchs}, \binits{S.L.}},
\bauthor{\bsnm{Hart}, \binits{A.J.}},
\bauthor{\bsnm{Wall}, \binits{W.A.}}:
\batitle{A novel smoothed particle hydrodynamics formulation for
  thermo-capillary phase change problems with focus on metal additive
  manufacturing melt pool modeling}.
\bjtitle{Computer Methods in Applied Mechanics and Engineering}
\bvolume{381},
\bfpage{113812}
(\byear{2021})
\doiurl{10.1016/j.cma.2021.113812}
\end{barticle}
\endbibitem

\bibitem[\protect\citeauthoryear{Fuchs et~al.}{2022}]{fuchs2022versatile}
\begin{botherref}
\oauthor{\bsnm{Fuchs}, \binits{S.L.}},
\oauthor{\bsnm{Praegla}, \binits{P.M.}},
\oauthor{\bsnm{Cyron}, \binits{C.J.}},
\oauthor{\bsnm{Wall}, \binits{W.A.}},
\oauthor{\bsnm{Meier}, \binits{C.}}:
{A versatile SPH modeling framework for coupled microfluid-powder dynamics in
  additive manufacturing: binder jetting, material jetting, directed energy
  deposition and powder bed fusion}.
Engineering with Computers,
1--25
(2022)
\doiurl{10.1007/s00366-022-01724-4}
\end{botherref}
\endbibitem

\bibitem[\protect\citeauthoryear{Scardovelli and
  Zaleski}{1999}]{scardovelli1999direct}
\begin{barticle}
\bauthor{\bsnm{Scardovelli}, \binits{R.}},
\bauthor{\bsnm{Zaleski}, \binits{S.}}:
\batitle{Direct numerical simulation of free-surface and interfacial flow}.
\bjtitle{Annual Review of Fluid Mechanics}
\bvolume{31}(\bissue{1}),
\bfpage{567}--\blpage{603}
(\byear{1999})
\doiurl{10.1146/annurev.fluid.31.1.567}
\end{barticle}
\endbibitem

\bibitem[\protect\citeauthoryear{Tryggvason
  et~al.}{2011}]{tryggvason2011direct}
\begin{bbook}
\bauthor{\bsnm{Tryggvason}, \binits{G.}},
\bauthor{\bsnm{Scardovelli}, \binits{R.}},
\bauthor{\bsnm{Zaleski}, \binits{S.}}:
\bbtitle{Direct Numerical Simulations of Gas--liquid Multiphase Flows}.
\bpublisher{Cambridge University Press},
\blocation{Cambridge}
(\byear{2011}).
\doiurl{10.1017/CBO9780511975264}
\end{bbook}
\endbibitem

\bibitem[\protect\citeauthoryear{Hirt et~al.}{1974}]{hirt1974arbitrary}
\begin{barticle}
\bauthor{\bsnm{Hirt}, \binits{C.W.}},
\bauthor{\bsnm{Amsden}, \binits{A.A.}},
\bauthor{\bsnm{Cook}, \binits{J.}}:
\batitle{An arbitrary lagrangian-eulerian computing method for all flow
  speeds}.
\bjtitle{Journal of Computational Physics}
\bvolume{14}(\bissue{3}),
\bfpage{227}--\blpage{253}
(\byear{1974})
\doiurl{10.1016/0021-9991(74)90051-5}
\end{barticle}
\endbibitem

\bibitem[\protect\citeauthoryear{Tang}{2005}]{tang2005moving}
\begin{barticle}
\bauthor{\bsnm{Tang}, \binits{T.}}:
\batitle{Moving mesh methods for computational fluid dynamics}.
\bjtitle{Contemporary mathematics}
\bvolume{383}(\bissue{8}),
\bfpage{141}--\blpage{173}
(\byear{2005})
\end{barticle}
\endbibitem

\bibitem[\protect\citeauthoryear{Anderson et~al.}{1998}]{anderson1998diffuse}
\begin{barticle}
\bauthor{\bsnm{Anderson}, \binits{D.M.}},
\bauthor{\bsnm{McFadden}, \binits{G.B.}},
\bauthor{\bsnm{Wheeler}, \binits{A.A.}}:
\batitle{{Diffuse-interface methods in fluid mechanics}}.
\bjtitle{Annual Review of Fluid Mechanics}
\bvolume{30}(\bissue{1}),
\bfpage{139}--\blpage{165}
(\byear{1998})
\doiurl{10.1146/annurev.fluid.30.1.139}
\end{barticle}
\endbibitem

\bibitem[\protect\citeauthoryear{Anjos et~al.}{2014}]{anjos20143d}
\begin{barticle}
\bauthor{\bsnm{Anjos}, \binits{G.}},
\bauthor{\bsnm{Mangiavacchi}, \binits{N.}},
\bauthor{\bsnm{Borhani}, \binits{N.}},
\bauthor{\bsnm{Thome}, \binits{J.R.}}:
\batitle{{3D ALE finite-element method for two-phase flows with phase change}}.
\bjtitle{Heat Transfer Engineering}
\bvolume{35}(\bissue{5}),
\bfpage{537}--\blpage{547}
(\byear{2014})
\doiurl{10.1080/01457632.2013.833407}
\end{barticle}
\endbibitem

\bibitem[\protect\citeauthoryear{Jafari and Okutucu-Özyurt}{2016}]{jafari2016}
\begin{barticle}
\bauthor{\bsnm{Jafari}, \binits{R.}},
\bauthor{\bsnm{Okutucu-Özyurt}, \binits{T.}}:
\batitle{{3D numerical modeling of boiling in a microchannel by arbitrary
  Lagrangian–Eulerian (ALE) method}}.
\bjtitle{Applied Mathematics and Computation}
\bvolume{272},
\bfpage{593}--\blpage{603}
(\byear{2016})
\doiurl{10.1016/j.amc.2015.03.042}
\end{barticle}
\endbibitem

\bibitem[\protect\citeauthoryear{Gros et~al.}{2018}]{gros2018}
\begin{barticle}
\bauthor{\bsnm{Gros}, \binits{E.}},
\bauthor{\bsnm{Anjos}, \binits{G.}},
\bauthor{\bsnm{Thome}, \binits{J.}}:
\batitle{Moving mesh method for direct numerical simulation of two-phase flow
  with phase change}.
\bjtitle{Applied Mathematics and Computation}
\bvolume{339},
\bfpage{636}--\blpage{650}
(\byear{2018})
\doiurl{10.1016/j.amc.2018.07.052}
\end{barticle}
\endbibitem

\bibitem[\protect\citeauthoryear{Zhang et~al.}{2019}]{zhang2019locally}
\begin{barticle}
\bauthor{\bsnm{Zhang}, \binits{Y.}},
\bauthor{\bsnm{Chandra}, \binits{A.}},
\bauthor{\bsnm{Yang}, \binits{F.}},
\bauthor{\bsnm{Shams}, \binits{E.}},
\bauthor{\bsnm{Sahni}, \binits{O.}},
\bauthor{\bsnm{Shephard}, \binits{M.}},
\bauthor{\bsnm{Oberai}, \binits{A.A.}}:
\batitle{{A locally discontinuous ALE finite element formulation for
  compressible phase change problems}}.
\bjtitle{Journal of Computational Physics}
\bvolume{393},
\bfpage{438}--\blpage{464}
(\byear{2019})
\doiurl{10.1016/j.jcp.2019.04.039}
\end{barticle}
\endbibitem

\bibitem[\protect\citeauthoryear{Unverdi and
  Tryggvason}{1992}]{unverdi1992front}
\begin{botherref}
\oauthor{\bsnm{Unverdi}, \binits{S.O.}},
\oauthor{\bsnm{Tryggvason}, \binits{G.}}:
{A front-tracking method for viscous, incompressible, multi-fluid flows}.
Journal of Computational Physics,
25--37
(1992)
\doiurl{10.1016/0021-9991(92)90307-K}
\end{botherref}
\endbibitem

\bibitem[\protect\citeauthoryear{Hirt and Nichols}{1981}]{Hirt1981VolumeOF}
\begin{barticle}
\bauthor{\bsnm{Hirt}, \binits{C.W.}},
\bauthor{\bsnm{Nichols}, \binits{B.}}:
\batitle{{Volume of fluid (VOF) method for the dynamics of free boundaries}}.
\bjtitle{Journal of Computational Physics}
\bvolume{39}(\bissue{1}),
\bfpage{221}--\blpage{225}
(\byear{1981})
\doiurl{10.1016/0021-9991(81)90145-5}
\end{barticle}
\endbibitem

\bibitem[\protect\citeauthoryear{Osher and Sethian}{1988}]{Osher1988}
\begin{barticle}
\bauthor{\bsnm{Osher}, \binits{S.}},
\bauthor{\bsnm{Sethian}, \binits{J.}}:
\batitle{{Fronts propagating with curvature-dependent speed --- Algorithms
  based on Hamilton--Jacobi formulations}}.
\bjtitle{Journal of Computational Physics}
\bvolume{79}(\bissue{1}),
\bfpage{12}--\blpage{49}
(\byear{1988})
\doiurl{10.1016/0021-9991(88)90002-2}
\end{barticle}
\endbibitem

\bibitem[\protect\citeauthoryear{Sussman et~al.}{1994}]{sussman1994level}
\begin{barticle}
\bauthor{\bsnm{Sussman}, \binits{M.}},
\bauthor{\bsnm{Smereka}, \binits{P.}},
\bauthor{\bsnm{Osher}, \binits{S.}}:
\batitle{{A level set approach for computing solutions to incompressible
  two-phase flow}}.
\bjtitle{Journal of Computational Physics}
\bvolume{114}(\bissue{1}),
\bfpage{146}--\blpage{159}
(\byear{1994})
\doiurl{10.1006/jcph.1994.1155}
\end{barticle}
\endbibitem

\bibitem[\protect\citeauthoryear{Lowengrub and
  Truskinovsky}{1998}]{lowengrub1998quasi}
\begin{barticle}
\bauthor{\bsnm{Lowengrub}, \binits{J.}},
\bauthor{\bsnm{Truskinovsky}, \binits{L.}}:
\batitle{{Quasi-incompressible Cahn--Hilliard fluids and topological
  transitions}}.
\bjtitle{Proceedings of the Royal Society A}
\bvolume{454}(\bissue{1998}),
\bfpage{2617}--\blpage{2654}
(\byear{1998})
\doiurl{10.1098/rspa.1998.0273}
\end{barticle}
\endbibitem

\bibitem[\protect\citeauthoryear{Olsson et~al.}{2007}]{olsson2007conservative}
\begin{barticle}
\bauthor{\bsnm{Olsson}, \binits{E.}},
\bauthor{\bsnm{Kreiss}, \binits{G.}},
\bauthor{\bsnm{Zahedi}, \binits{S.}}:
\batitle{{A conservative level set method for two phase flow II}}.
\bjtitle{Journal of Computational Physics}
\bvolume{225}(\bissue{1}),
\bfpage{785}--\blpage{807}
(\byear{2007})
\doiurl{10.1016/j.jcp.2006.12.027}
\end{barticle}
\endbibitem

\bibitem[\protect\citeauthoryear{Popinet}{2009}]{popinet2009accurate}
\begin{barticle}
\bauthor{\bsnm{Popinet}, \binits{S.}}:
\batitle{{An accurate adaptive solver for surface-tension-driven interfacial
  flows}}.
\bjtitle{Journal of Computational Physics}
\bvolume{228}(\bissue{16}),
\bfpage{5838}--\blpage{5866}
(\byear{2009})
\doiurl{10.1016/j.jcp.2009.04.042}
\end{barticle}
\endbibitem

\bibitem[\protect\citeauthoryear{Jacqmin}{1999}]{jacqmin1999}
\begin{barticle}
\bauthor{\bsnm{Jacqmin}, \binits{D.}}:
\batitle{{Calculation of two-phase Navier--Stokes flows using phase-field
  modeling}}.
\bjtitle{Journal of Computational Physics}
\bvolume{155}(\bissue{1}),
\bfpage{96}--\blpage{127}
(\byear{1999})
\doiurl{10.1006/jcph.1999.6332}
\end{barticle}
\endbibitem

\bibitem[\protect\citeauthoryear{Chessa et~al.}{2002}]{chessa2002extended}
\begin{barticle}
\bauthor{\bsnm{Chessa}, \binits{J.}},
\bauthor{\bsnm{Smolinski}, \binits{P.}},
\bauthor{\bsnm{Belytschko}, \binits{T.}}:
\batitle{The extended finite element method (xfem) for solidification
  problems}.
\bjtitle{International Journal for Numerical Methods in Engineering}
\bvolume{53}(\bissue{8}),
\bfpage{1959}--\blpage{1977}
(\byear{2002})
\doiurl{10.1002/nme.386}
\end{barticle}
\endbibitem

\bibitem[\protect\citeauthoryear{Chessa and
  Belytschko}{2003}]{chessa2003enriched}
\begin{barticle}
\bauthor{\bsnm{Chessa}, \binits{J.}},
\bauthor{\bsnm{Belytschko}, \binits{T.}}:
\batitle{An enriched finite element method and level sets for axisymmetric
  two-phase flow with surface tension}.
\bjtitle{International Journal for Numerical Methods in Engineering}
\bvolume{58}(\bissue{13}),
\bfpage{2041}--\blpage{2064}
(\byear{2003})
\doiurl{10.1002/nme.946}
\end{barticle}
\endbibitem

\bibitem[\protect\citeauthoryear{Rasthofer
  et~al.}{2011}]{rasthofer2011extended}
\begin{barticle}
\bauthor{\bsnm{Rasthofer}, \binits{U.}},
\bauthor{\bsnm{Henke}, \binits{F.}},
\bauthor{\bsnm{Wall}, \binits{W.}},
\bauthor{\bsnm{Gravemeier}, \binits{V.}}:
\batitle{An extended residual-based variational multiscale method for two-phase
  flow including surface tension}.
\bjtitle{Computer Methods in Applied Mechanics and Engineering}
\bvolume{200}(\bissue{21-22}),
\bfpage{1866}--\blpage{1876}
(\byear{2011})
\doiurl{10.1016/j.cma.2011.02.004}
\end{barticle}
\endbibitem

\bibitem[\protect\citeauthoryear{Sauerland and
  Fries}{2013}]{sauerland2013stable}
\begin{barticle}
\bauthor{\bsnm{Sauerland}, \binits{H.}},
\bauthor{\bsnm{Fries}, \binits{T.-P.}}:
\batitle{{The stable XFEM for two-phase flows}}.
\bjtitle{Computers \& Fluids}
\bvolume{87},
\bfpage{41}--\blpage{49}
(\byear{2013})
\doiurl{10.1016/j.compfluid.2012.10.017}
\end{barticle}
\endbibitem

\bibitem[\protect\citeauthoryear{Schott et~al.}{2015}]{schottrasthofer2015face}
\begin{barticle}
\bauthor{\bsnm{Schott}, \binits{B.}},
\bauthor{\bsnm{Rasthofer}, \binits{U.}},
\bauthor{\bsnm{Gravemeier}, \binits{V.}},
\bauthor{\bsnm{Wall}, \binits{W.}}:
\batitle{{A face-oriented stabilized Nitsche-type extended variational
  multiscale method for incompressible two-phase flow}}.
\bjtitle{International Journal for Numerical Methods in Engineering}
\bvolume{104}(\bissue{7}),
\bfpage{721}--\blpage{748}
(\byear{2015})
\doiurl{10.1002/nme.4789}
\end{barticle}
\endbibitem

\bibitem[\protect\citeauthoryear{Hansbo et~al.}{2014}]{hansbo2014cut}
\begin{barticle}
\bauthor{\bsnm{Hansbo}, \binits{P.}},
\bauthor{\bsnm{Larson}, \binits{M.G.}},
\bauthor{\bsnm{Zahedi}, \binits{S.}}:
\batitle{{A cut finite element method for a Stokes interface problem}}.
\bjtitle{Applied Numerical Mathematics}
\bvolume{85},
\bfpage{90}--\blpage{114}
(\byear{2014})
\doiurl{10.1016/j.apnum.2014.06.009}
\end{barticle}
\endbibitem

\bibitem[\protect\citeauthoryear{Massing et~al.}{2018}]{massing2018stabilized}
\begin{barticle}
\bauthor{\bsnm{Massing}, \binits{A.}},
\bauthor{\bsnm{Schott}, \binits{B.}},
\bauthor{\bsnm{Wall}, \binits{W.A.}}:
\batitle{{A stabilized Nitsche cut finite element method for the Oseen
  problem}}.
\bjtitle{Computer Methods in Applied Mechanics and Engineering}
\bvolume{328},
\bfpage{262}--\blpage{300}
(\byear{2018})
\doiurl{10.1016/j.cma.2017.09.003}
\end{barticle}
\endbibitem

\bibitem[\protect\citeauthoryear{Claus and Kerfriden}{2019}]{claus2019cutfem}
\begin{barticle}
\bauthor{\bsnm{Claus}, \binits{S.}},
\bauthor{\bsnm{Kerfriden}, \binits{P.}}:
\batitle{A cutfem method for two-phase flow problems}.
\bjtitle{Computer Methods in Applied Mechanics and Engineering}
\bvolume{348},
\bfpage{185}--\blpage{206}
(\byear{2019})
\doiurl{10.1016/j.cma.2019.01.009}
\end{barticle}
\endbibitem

\bibitem[\protect\citeauthoryear{Frachon and Zahedi}{2019}]{frachon2019cut}
\begin{barticle}
\bauthor{\bsnm{Frachon}, \binits{T.}},
\bauthor{\bsnm{Zahedi}, \binits{S.}}:
\batitle{{A cut finite element method for incompressible two-phase
  Navier--Stokes flows}}.
\bjtitle{Journal of Computational Physics}
\bvolume{384},
\bfpage{77}--\blpage{98}
(\byear{2019})
\doiurl{10.1016/j.jcp.2019.01.028}
\end{barticle}
\endbibitem

\bibitem[\protect\citeauthoryear{Frachon and Zahedi}{2023}]{frachon2023cut}
\begin{barticle}
\bauthor{\bsnm{Frachon}, \binits{T.}},
\bauthor{\bsnm{Zahedi}, \binits{S.}}:
\batitle{A cut finite element method for two-phase flows with insoluble
  surfactants}.
\bjtitle{Journal of Computational Physics}
\bvolume{473},
\bfpage{111734}
(\byear{2023})
\doiurl{10.1016/j.jcp.2022.111734}
\end{barticle}
\endbibitem

\bibitem[\protect\citeauthoryear{Henneaux et~al.}{2023}]{henneaux2023}
\begin{barticle}
\bauthor{\bsnm{Henneaux}, \binits{D.}},
\bauthor{\bsnm{Schrooyen}, \binits{P.}},
\bauthor{\bsnm{Chatelain}, \binits{P.}},
\bauthor{\bsnm{Magin}, \binits{T.}}:
\batitle{{High-order enforcement of jumps conditions between compressible
  viscous phases: An extended interior penalty discontinuous Galerkin method
  for sharp interface simulation}}.
\bjtitle{Computer Methods in Applied Mechanics and Engineering}
\bvolume{415},
\bfpage{116215}
(\byear{2023})
\doiurl{10.1016/j.cma.2023.116215}
\end{barticle}
\endbibitem

\bibitem[\protect\citeauthoryear{Fechter and
  Munz}{2015}]{fechter2015discontinuous}
\begin{barticle}
\bauthor{\bsnm{Fechter}, \binits{S.}},
\bauthor{\bsnm{Munz}, \binits{C.-D.}}:
\batitle{{A discontinuous Galerkin-based sharp-interface method to simulate
  three-dimensional compressible two-phase flow}}.
\bjtitle{International Journal for Numerical Methods in Fluids}
\bvolume{78}(\bissue{7}),
\bfpage{413}--\blpage{435}
(\byear{2015})
\doiurl{10.1002/fld.4022}
\end{barticle}
\endbibitem

\bibitem[\protect\citeauthoryear{Fedkiw et~al.}{1999}]{fedkiw1999non}
\begin{barticle}
\bauthor{\bsnm{Fedkiw}, \binits{R.P.}},
\bauthor{\bsnm{Aslam}, \binits{T.}},
\bauthor{\bsnm{Merriman}, \binits{B.}},
\bauthor{\bsnm{Osher}, \binits{S.}}, \betal:
\batitle{{A non-oscillatory Eulerian approach to interfaces in multimaterial
  flows (the ghost fluid method)}}.
\bjtitle{Journal of Computational Physics}
\bvolume{152}(\bissue{2}),
\bfpage{457}--\blpage{492}
(\byear{1999})
\doiurl{10.1006/jcph.1999.6236}
\end{barticle}
\endbibitem

\bibitem[\protect\citeauthoryear{Fechter et~al.}{2018}]{fechter2018approximate}
\begin{barticle}
\bauthor{\bsnm{Fechter}, \binits{S.}},
\bauthor{\bsnm{Munz}, \binits{C.-D.}},
\bauthor{\bsnm{Rohde}, \binits{C.}},
\bauthor{\bsnm{Zeiler}, \binits{C.}}:
\batitle{{Approximate Riemann solver for compressible liquid vapor flow with
  phase transition and surface tension}}.
\bjtitle{Computers \& Fluids}
\bvolume{169},
\bfpage{169}--\blpage{185}
(\byear{2018})
\doiurl{10.1016/j.compfluid.2017.03.026}
\end{barticle}
\endbibitem

\bibitem[\protect\citeauthoryear{Lee and LeVeque}{2003}]{lee2003immersed}
\begin{barticle}
\bauthor{\bsnm{Lee}, \binits{L.}},
\bauthor{\bsnm{LeVeque}, \binits{R.J.}}:
\batitle{{An immersed interface method for incompressible Navier--Stokes
  equations}}.
\bjtitle{SIAM Journal on Scientific Computing}
\bvolume{25}(\bissue{3}),
\bfpage{832}--\blpage{856}
(\byear{2003})
\doiurl{10.1137/S1064827502414060}
\end{barticle}
\endbibitem

\bibitem[\protect\citeauthoryear{Burman et~al.}{2015}]{burman2015cutfem}
\begin{barticle}
\bauthor{\bsnm{Burman}, \binits{E.}},
\bauthor{\bsnm{Claus}, \binits{S.}},
\bauthor{\bsnm{Hansbo}, \binits{P.}},
\bauthor{\bsnm{Larson}, \binits{M.G.}},
\bauthor{\bsnm{Massing}, \binits{A.}}:
\batitle{{CutFEM: discretizing geometry and partial differential equations}}.
\bjtitle{International Journal for Numerical Methods in Engineering}
\bvolume{104}(\bissue{7}),
\bfpage{472}--\blpage{501}
(\byear{2015})
\doiurl{10.1002/nme.4823}
\end{barticle}
\endbibitem

\bibitem[\protect\citeauthoryear{Massing et~al.}{2015}]{massing2015nitsche}
\begin{barticle}
\bauthor{\bsnm{Massing}, \binits{A.}},
\bauthor{\bsnm{Larson}, \binits{M.}},
\bauthor{\bsnm{Logg}, \binits{A.}},
\bauthor{\bsnm{Rognes}, \binits{M.}}:
\batitle{A nitsche-based cut finite element method for a fluid-structure
  interaction problem}.
\bjtitle{Communications in Applied Mathematics and Computational Science}
\bvolume{10}(\bissue{2}),
\bfpage{97}--\blpage{120}
(\byear{2015})
\doiurl{10.2140/camcos.2015.10.97}
\end{barticle}
\endbibitem

\bibitem[\protect\citeauthoryear{Brackbill et~al.}{1992}]{brackbill1992}
\begin{barticle}
\bauthor{\bsnm{Brackbill}, \binits{J.U.}},
\bauthor{\bsnm{Kothe}, \binits{D.B.}},
\bauthor{\bsnm{Zemach}, \binits{C.}}:
\batitle{A continuum method for modeling surface tension}.
\bjtitle{Journal of Computational Physics}
\bvolume{100}(\bissue{2}),
\bfpage{335}--\blpage{354}
(\byear{1992})
\doiurl{10.1016/0021-9991(92)90240-y}
\end{barticle}
\endbibitem

\bibitem[\protect\citeauthoryear{Gibou et~al.}{2018}]{gibou2018review}
\begin{barticle}
\bauthor{\bsnm{Gibou}, \binits{F.}},
\bauthor{\bsnm{Fedkiw}, \binits{R.}},
\bauthor{\bsnm{Osher}, \binits{S.}}:
\batitle{{A review of level-set methods and some recent applications}}.
\bjtitle{Journal of Computational Physics}
\bvolume{353},
\bfpage{82}--\blpage{109}
(\byear{2018})
\doiurl{10.1016/j.jcp.2017.10.006}
\end{barticle}
\endbibitem

\bibitem[\protect\citeauthoryear{Meier et~al.}{2017}]{meier2017thermophysical}
\begin{botherref}
\oauthor{\bsnm{Meier}, \binits{C.}},
\oauthor{\bsnm{Penny}, \binits{R.W.}},
\oauthor{\bsnm{Zou}, \binits{Y.}},
\oauthor{\bsnm{Gibbs}, \binits{J.S.}},
\oauthor{\bsnm{Hart}, \binits{A.J.}}:
{Thermophysical phenomena in metal additive manufacturing by selective laser
  melting: fundamentals, modeling, simulation and experimentation}.
Annual Review of Heat Transfer
(1),
241--316
(2017)
\doiurl{10.1615/annualrevheattransfer.2018019042}
\end{botherref}
\endbibitem

\bibitem[\protect\citeauthoryear{Juric and
  Tryggvason}{1998}]{juric1998computations}
\begin{barticle}
\bauthor{\bsnm{Juric}, \binits{D.}},
\bauthor{\bsnm{Tryggvason}, \binits{G.}}:
\batitle{Computations of boiling flows}.
\bjtitle{International Journal of Multiphase Flow}
\bvolume{24}(\bissue{3}),
\bfpage{387}--\blpage{410}
(\byear{1998})
\doiurl{10.1016/S0301-9322(97)00050-5}
\end{barticle}
\endbibitem

\bibitem[\protect\citeauthoryear{Welch and Wilson}{2000}]{welch2000volume}
\begin{barticle}
\bauthor{\bsnm{Welch}, \binits{S.W.}},
\bauthor{\bsnm{Wilson}, \binits{J.}}:
\batitle{A volume of fluid based method for fluid flows with phase change}.
\bjtitle{Journal of Computational Physics}
\bvolume{160}(\bissue{2}),
\bfpage{662}--\blpage{682}
(\byear{2000})
\doiurl{10.1006/jcph.2000.6481}
\end{barticle}
\endbibitem

\bibitem[\protect\citeauthoryear{Nguyen et~al.}{2001}]{nguyen2001boundary}
\begin{barticle}
\bauthor{\bsnm{Nguyen}, \binits{D.Q.}},
\bauthor{\bsnm{Fedkiw}, \binits{R.P.}},
\bauthor{\bsnm{Kang}, \binits{M.}}:
\batitle{A boundary condition capturing method for incompressible flame
  discontinuities}.
\bjtitle{Journal of Computational Physics}
\bvolume{172}(\bissue{1}),
\bfpage{71}--\blpage{98}
(\byear{2001})
\doiurl{10.1006/jcph.2001.6812}
\end{barticle}
\endbibitem

\bibitem[\protect\citeauthoryear{Gibou et~al.}{2007}]{gibou2007level}
\begin{barticle}
\bauthor{\bsnm{Gibou}, \binits{F.}},
\bauthor{\bsnm{Chen}, \binits{L.}},
\bauthor{\bsnm{Nguyen}, \binits{D.}},
\bauthor{\bsnm{Banerjee}, \binits{S.}}:
\batitle{{A level set based sharp interface method for the multiphase
  incompressible Navier--Stokes equations with phase change}}.
\bjtitle{Journal of Computational Physics}
\bvolume{222}(\bissue{2}),
\bfpage{536}--\blpage{555}
(\byear{2007})
\doiurl{10.1016/j.jcp.2006.07.035}
\end{barticle}
\endbibitem

\bibitem[\protect\citeauthoryear{Son and Dhir}{2007}]{son2007level}
\begin{barticle}
\bauthor{\bsnm{Son}, \binits{G.}},
\bauthor{\bsnm{Dhir}, \binits{V.K.}}:
\batitle{A level set method for analysis of film boiling on an immersed solid
  surface}.
\bjtitle{Numerical Heat Transfer, Part B: Fundamentals}
\bvolume{52}(\bissue{2}),
\bfpage{153}--\blpage{177}
(\byear{2007})
\doiurl{10.1080/10407790701347720}
\end{barticle}
\endbibitem

\bibitem[\protect\citeauthoryear{Tanguy et~al.}{2007}]{tanguy2007level}
\begin{barticle}
\bauthor{\bsnm{Tanguy}, \binits{S.}},
\bauthor{\bsnm{M{\'e}nard}, \binits{T.}},
\bauthor{\bsnm{Berlemont}, \binits{A.}}:
\batitle{A level set method for vaporizing two-phase flows}.
\bjtitle{Journal of Computational Physics}
\bvolume{221}(\bissue{2}),
\bfpage{837}--\blpage{853}
(\byear{2007})
\doiurl{10.1016/j.jcp.2006.07.003}
\end{barticle}
\endbibitem

\bibitem[\protect\citeauthoryear{Hardt and Wondra}{2008}]{hardt2008evaporation}
\begin{barticle}
\bauthor{\bsnm{Hardt}, \binits{S.}},
\bauthor{\bsnm{Wondra}, \binits{F.}}:
\batitle{Evaporation model for interfacial flows based on a continuum-field
  representation of the source terms}.
\bjtitle{Journal of Computational Physics}
\bvolume{227}(\bissue{11}),
\bfpage{5871}--\blpage{5895}
(\byear{2008})
\doiurl{10.1016/j.jcp.2008.02.020}
\end{barticle}
\endbibitem

\bibitem[\protect\citeauthoryear{Lee et~al.}{2017}]{lee2017direct}
\begin{barticle}
\bauthor{\bsnm{Lee}, \binits{M.S.}},
\bauthor{\bsnm{Riaz}, \binits{A.}},
\bauthor{\bsnm{Aute}, \binits{V.}}:
\batitle{Direct numerical simulation of incompressible multiphase flow with
  phase change}.
\bjtitle{Journal of Computational Physics}
\bvolume{344},
\bfpage{381}--\blpage{418}
(\byear{2017})
\doiurl{10.1016/j.jcp.2017.04.073}
\end{barticle}
\endbibitem

\bibitem[\protect\citeauthoryear{Tanguy et~al.}{2014}]{tanguy2014benchmarks}
\begin{barticle}
\bauthor{\bsnm{Tanguy}, \binits{S.}},
\bauthor{\bsnm{Sagan}, \binits{M.}},
\bauthor{\bsnm{Lalanne}, \binits{B.}},
\bauthor{\bsnm{Couderc}, \binits{F.}},
\bauthor{\bsnm{Colin}, \binits{C.}}:
\batitle{Benchmarks and numerical methods for the simulation of boiling flows}.
\bjtitle{Journal of Computational Physics}
\bvolume{264},
\bfpage{1}--\blpage{22}
(\byear{2014})
\doiurl{10.1016/j.jcp.2014.01.014}
\end{barticle}
\endbibitem

\bibitem[\protect\citeauthoryear{Scapin et~al.}{2020}]{scapin2020volume}
\begin{barticle}
\bauthor{\bsnm{Scapin}, \binits{N.}},
\bauthor{\bsnm{Costa}, \binits{P.}},
\bauthor{\bsnm{Brandt}, \binits{L.}}:
\batitle{A volume-of-fluid method for interface-resolved simulations of
  phase-changing two-fluid flows}.
\bjtitle{Journal of Computational Physics}
\bvolume{407},
\bfpage{109251}
(\byear{2020})
\doiurl{10.1016/j.jcp.2020.109251}
\end{barticle}
\endbibitem

\bibitem[\protect\citeauthoryear{Arndt et~al.}{2023}]{arndt2022deal}
\begin{barticle}
\bauthor{\bsnm{Arndt}, \binits{D.}},
\bauthor{\bsnm{Bangerth}, \binits{W.}},
\bauthor{\bsnm{Bergbauer}, \binits{M.}},
\bauthor{\bsnm{Feder}, \binits{M.}},
\bauthor{\bsnm{Fehling}, \binits{M.}},
\bauthor{\bsnm{Heinz}, \binits{J.}},
\bauthor{\bsnm{Heister}, \binits{T.}},
\bauthor{\bsnm{Heltai}, \binits{L.}},
\bauthor{\bsnm{Kronbichler}, \binits{M.}},
\bauthor{\bsnm{Maier}, \binits{M.}}, \betal:
\batitle{{The deal. II library, version 9.5}}.
\bjtitle{Journal of Numerical Mathematics}
\bvolume{31}(\bissue{3}),
\bfpage{231}--\blpage{246}
(\byear{2023})
\doiurl{10.1515/jnma-2023-0089}
\end{barticle}
\endbibitem

\bibitem[\protect\citeauthoryear{Kronbichler and
  Kormann}{2012}]{kronbichler2012generic}
\begin{barticle}
\bauthor{\bsnm{Kronbichler}, \binits{M.}},
\bauthor{\bsnm{Kormann}, \binits{K.}}:
\batitle{A generic interface for parallel cell-based finite element operator
  application}.
\bjtitle{Computers \& Fluids}
\bvolume{63},
\bfpage{135}--\blpage{147}
(\byear{2012})
\doiurl{10.1016/j.compfluid.2012.04.012}
\end{barticle}
\endbibitem

\bibitem[\protect\citeauthoryear{Kronbichler
  et~al.}{2018}]{Kronbichler18multiphase}
\begin{barticle}
\bauthor{\bsnm{Kronbichler}, \binits{M.}},
\bauthor{\bsnm{Diagne}, \binits{A.}},
\bauthor{\bsnm{Holmgren}, \binits{H.}}:
\batitle{{A fast massively parallel two-phase flow solver for microfluidic chip
  simulation}}.
\bjtitle{International Journal of High Performance Computing Applications}
\bvolume{32}(\bissue{2}),
\bfpage{266}--\blpage{287}
(\byear{2018})
\doiurl{10.1177/1094342016671790}
\end{barticle}
\endbibitem

\bibitem[\protect\citeauthoryear{Proell et~al.}{2023}]{proell2023highly}
\begin{botherref}
\oauthor{\bsnm{Proell}, \binits{S.D.}},
\oauthor{\bsnm{Munch}, \binits{P.}},
\oauthor{\bsnm{Kronbichler}, \binits{M.}},
\oauthor{\bsnm{Wall}, \binits{W.A.}},
\oauthor{\bsnm{Meier}, \binits{C.}}:
A highly efficient computational approach for fast scan-resolved simulations of
  metal additive manufacturing processes on the scale of real parts.
Additive Manufacturing,
103921
(2023)
\doiurl{10.1016/j.addma.2023.103921}
\end{botherref}
\endbibitem

\bibitem[\protect\citeauthoryear{Munch et~al.}{2024}]{munch2023}
\begin{barticle}
\bauthor{\bsnm{Munch}, \binits{P.}},
\bauthor{\bsnm{Ivannikov}, \binits{V.}},
\bauthor{\bsnm{Cyron}, \binits{C.}},
\bauthor{\bsnm{Kronbichler}, \binits{M.}}:
\batitle{On the construction of an efﬁcient ﬁnite-element solver for
  phase-ﬁeld simulations of many-particle solid-state-sintering processes}.
\bjtitle{Computational Materials Science}
\bvolume{231},
\bfpage{112589}
(\byear{2024})
\doiurl{10.1016/j.commatsci.2023.112589}
\end{barticle}
\endbibitem

\bibitem[\protect\citeauthoryear{Zahedi et~al.}{2012}]{zahedi2012spurious}
\begin{barticle}
\bauthor{\bsnm{Zahedi}, \binits{S.}},
\bauthor{\bsnm{Kronbichler}, \binits{M.}},
\bauthor{\bsnm{Kreiss}, \binits{G.}}:
\batitle{Spurious currents in finite element based level set methods for
  two-phase flow}.
\bjtitle{International Journal for Numerical Methods in Fluids}
\bvolume{69}(\bissue{9}),
\bfpage{1433}--\blpage{1456}
(\byear{2012})
\doiurl{10.1002/fld.2643}
\end{barticle}
\endbibitem

\bibitem[\protect\citeauthoryear{Cenanovic et~al.}{2020}]{cenanovic2020finite}
\begin{barticle}
\bauthor{\bsnm{Cenanovic}, \binits{M.}},
\bauthor{\bsnm{Hansbo}, \binits{P.}},
\bauthor{\bsnm{Larson}, \binits{M.G.}}:
\batitle{Finite element procedures for computing normals and mean curvature on
  triangulated surfaces and their use for mesh refinement}.
\bjtitle{Computer Methods in Applied Mechanics and Engineering}
\bvolume{372},
\bfpage{113445}
(\byear{2020})
\doiurl{10.1016/j.cma.2020.113445}
\end{barticle}
\endbibitem

\bibitem[\protect\citeauthoryear{Peskin}{2002}]{peskin2002immersed}
\begin{barticle}
\bauthor{\bsnm{Peskin}, \binits{C.S.}}:
\batitle{The immersed boundary method}.
\bjtitle{Acta Numerica}
\bvolume{11},
\bfpage{479}--\blpage{517}
(\byear{2002})
\doiurl{10.1017/S0962492902000077}
\end{barticle}
\endbibitem

\bibitem[\protect\citeauthoryear{Knight}{1979}]{knight1979theoretical}
\begin{barticle}
\bauthor{\bsnm{Knight}, \binits{C.J.}}:
\batitle{{Theoretical modeling of rapid surface vaporization with back
  pressure}}.
\bjtitle{AIAA J}
\bvolume{17}(\bissue{5}),
\bfpage{519}--\blpage{523}
(\byear{1979})
\end{barticle}
\endbibitem

\bibitem[\protect\citeauthoryear{Anisimov and
  Khokhlov}{1995}]{anisimov1995instabilities}
\begin{bbook}
\bauthor{\bsnm{Anisimov}, \binits{S.I.}},
\bauthor{\bsnm{Khokhlov}, \binits{V.A.}}:
\bbtitle{Instabilities in Laser-Matter Interaction}.
\bpublisher{CRC Press},
\blocation{Boca Raton}
(\byear{1995})
\end{bbook}
\endbibitem

\bibitem[\protect\citeauthoryear{Yokoi}{2014}]{yokoi2014density}
\begin{barticle}
\bauthor{\bsnm{Yokoi}, \binits{K.}}:
\batitle{{A density-scaled continuum surface force model within a balanced
  force formulation}}.
\bjtitle{Journal of Computational Physics}
\bvolume{278}(\bissue{1}),
\bfpage{221}--\blpage{228}
(\byear{2014})
\doiurl{10.1016/j.jcp.2014.08.034}
\end{barticle}
\endbibitem

\bibitem[\protect\citeauthoryear{Much
  et~al.}{2024}]{much2023}
\begin{barticle}
\bauthor{\bsnm{Much}, \binits{N.}},
\bauthor{\bsnm{Schreter-Fleischhacker}, \binits{M.}},
\bauthor{\bsnm{Munch}, \binits{P.}},
\bauthor{\bsnm{Kronbichler}, \binits{M.}},
\bauthor{\bsnm{Meier}, \binits{C.}},
\bauthor{\bsnm{Wall}, \binits{W.A.}}:
\batitle{Improved accuracy of continuum surface flux models for metal additive
  manufacturing melt pool simulations}.
\bjtitle{Advanced Modeling and Simulation in Engineering Sciences}
\bvolume{11}(\bissue{16}),
\bfpage{1}--\blpage{40}
(\byear{2024})
\doiurl{10.1186/s40323-024-00270-6}
\end{barticle}
\endbibitem

\bibitem[\protect\citeauthoryear{Henri et~al.}{2022}]{henri2022geometrical}
\begin{barticle}
\bauthor{\bsnm{Henri}, \binits{F.}},
\bauthor{\bsnm{Coquerelle}, \binits{M.}},
\bauthor{\bsnm{Lubin}, \binits{P.}}:
\batitle{Geometrical level set reinitialization using closest point method and
  kink detection for thin filaments, topology changes and two-phase flows}.
\bjtitle{Journal of Computational Physics}
\bvolume{448},
\bfpage{110704}
(\byear{2022})
\doiurl{10.1016/j.jcp.2021.110704}
\end{barticle}
\endbibitem

\bibitem[\protect\citeauthoryear{Coquerelle and
  Glockner}{2016}]{coquerelle2016fourth}
\begin{barticle}
\bauthor{\bsnm{Coquerelle}, \binits{M.}},
\bauthor{\bsnm{Glockner}, \binits{S.}}:
\batitle{A fourth-order accurate curvature computation in a level set framework
  for two-phase flows subjected to surface tension forces}.
\bjtitle{Journal of Computational Physics}
\bvolume{305},
\bfpage{838}--\blpage{876}
(\byear{2016})
\doiurl{10.1016/j.jcp.2015.11.014}
\end{barticle}
\endbibitem

\bibitem[\protect\citeauthoryear{Kronbichler and
  Kormann}{2019}]{Kronbichler2019}
\begin{botherref}
\oauthor{\bsnm{Kronbichler}, \binits{M.}},
\oauthor{\bsnm{Kormann}, \binits{K.}}:
{Fast Matrix-Free Evaluation of Discontinuous Galerkin Finite Element
  Operators}.
ACM Trans. Math. Softw.
\textbf{45}(3)
29:\bfpage{1}--\blpage{40}
(2019)
\doiurl{10.1145/3325864}
\end{botherref}
\endbibitem

\bibitem[\protect\citeauthoryear{Kolev et~al.}{2021}]{kolev2021efficient}
\begin{barticle}
\bauthor{\bsnm{Kolev}, \binits{T.}},
\bauthor{\bsnm{Fischer}, \binits{P.}},
\bauthor{\bsnm{Min}, \binits{M.}},
\bauthor{\bsnm{Dongarra}, \binits{J.}},
\bauthor{\bsnm{Brown}, \binits{J.}},
\bauthor{\bsnm{Dobrev}, \binits{V.}},
\bauthor{\bsnm{Warburton}, \binits{T.}},
\bauthor{\bsnm{Tomov}, \binits{S.}},
\bauthor{\bsnm{Shephard}, \binits{M.S.}},
\bauthor{\bsnm{Abdelfattah}, \binits{A.}}, \betal:
\batitle{Efficient exascale discretizations: High-order finite element
  methods}.
\bjtitle{The International Journal of High Performance Computing Applications}
\bvolume{35}(\bissue{6}),
\bfpage{527}--\blpage{552}
(\byear{2021})
\end{barticle}
\endbibitem

\bibitem[\protect\citeauthoryear{Saad}{2003}]{saad2003iterative}
\begin{bbook}
\bauthor{\bsnm{Saad}, \binits{Y.}}:
\bbtitle{Iterative Methods for Sparse Linear Systems}.
\bpublisher{SIAM},
\blocation{Philadelphia}
(\byear{2003}).
\doiurl{10.1137/1.9780898718003}
\end{bbook}
\endbibitem

\bibitem[\protect\citeauthoryear{Benzi et~al.}{2005}]{benzi2005numerical}
\begin{barticle}
\bauthor{\bsnm{Benzi}, \binits{M.}},
\bauthor{\bsnm{Golub}, \binits{G.H.}},
\bauthor{\bsnm{Liesen}, \binits{J.}}:
\batitle{Numerical solution of saddle point problems}.
\bjtitle{Acta Numerica}
\bvolume{14},
\bfpage{1}--\blpage{137}
(\byear{2005})
\doiurl{10.1017/S0962492904000212}
\end{barticle}
\endbibitem

\bibitem[\protect\citeauthoryear{Cahouet and Chabard}{1988}]{cahouet1988some}
\begin{barticle}
\bauthor{\bsnm{Cahouet}, \binits{J.}},
\bauthor{\bsnm{Chabard}, \binits{J.-P.}}:
\batitle{{Some fast 3D finite element solvers for the generalized Stokes
  problem}}.
\bjtitle{International Journal for Numerical Methods in Fluids}
\bvolume{8}(\bissue{8}),
\bfpage{869}--\blpage{895}
(\byear{1988})
\doiurl{10.1002/fld.1650080802}
\end{barticle}
\endbibitem

\bibitem[\protect\citeauthoryear{Schreter-Fleischhacker and
  Munch}{2023}]{schreter2023step87}
\begin{botherref}
\oauthor{\bsnm{Schreter-Fleischhacker}, \binits{M.}},
\oauthor{\bsnm{Munch}, \binits{P.}}:
{The deal.II tutorial step-87: evaluation of finite element solutions at
  arbitrary points within a distributed mesh with application to two-phase
  flow}
(2023)
\doiurl{10.5281/zenodo.8411345}
\end{botherref}
\endbibitem

\bibitem[\protect\citeauthoryear{Hysing et~al.}{2009}]{hysing2009quantitative}
\begin{barticle}
\bauthor{\bsnm{Hysing}, \binits{S.-R.}},
\bauthor{\bsnm{Turek}, \binits{S.}},
\bauthor{\bsnm{Kuzmin}, \binits{D.}},
\bauthor{\bsnm{Parolini}, \binits{N.}},
\bauthor{\bsnm{Burman}, \binits{E.}},
\bauthor{\bsnm{Ganesan}, \binits{S.}},
\bauthor{\bsnm{Tobiska}, \binits{L.}}:
\batitle{{Quantitative benchmark computations of two-dimensional bubble
  dynamics}}.
\bjtitle{International Journal for Numerical Methods in Fluids}
\bvolume{60}(\bissue{11}),
\bfpage{1259}--\blpage{1288}
(\byear{2009})
\doiurl{10.1002/fld.1934}
\end{barticle}
\endbibitem

\end{thebibliography}

\end{document}